\documentclass{nature_plusfigure}
\usepackage{graphicx}
\usepackage{subcaption}
\usepackage{amsmath,amssymb,xcolor,url}
\usepackage{multirow}
\usepackage{longtable}
\usepackage{upgreek}
\usepackage{url}
\usepackage{hyperref}
\graphicspath{{./}{figures/}}
\usepackage{lineno}
\usepackage[none]{hyphenat}
\usepackage{multibib}
\newcites{New}{References}

\newcommand{\mn}{{Mon. Not. R. Astron. Soc.}}
\newcommand{\nar}{{New Astron. Reviews}}
\newcommand{\physrep}{{Phys. Reports}}
\newcommand{\mnras}{\mn}
\newcommand{\aj}{{"Astron. J."}}
\newcommand{\apj}{{Astrophys. J.}}
\newcommand{\apjl}{{Astrophys. J. Lett.}}
\newcommand{\apjs}{{Astrophys. J. Supp.}}

\newcommand{\aap}{{Astron. Astrophys.}}
\newcommand{\nat}{{Nature}}

\newcommand{\prd}{{Phys. Rev. D}}

\newcommand{\pasp}{{Pub. Ast. Soc. Pac.}}
\newcommand{\procspie}{Proc. SPIE}
\newcommand{\ssr}{Space Science Reviews}

\newcommand{\at}{AT2022cmc}

\def\mr{\mathrm}
\def\eps{\epsilon}
\def\epse{\epsilon_{\rm e}}
\def\epsB{\epsilon_{\rm B}}
\def\mp{m_{\rm p}}
\def\me{m_{\rm e}}
\def\nua{\nu_{\rm a}}
\def\ga{\gamma_{\rm a}}
\def\Eiso{E_{\rm iso}}
\def\kB{k_{\rm B}}
\newcommand{\lrb}[1]{\left({#1}\right)}

\usepackage[symbol]{footmisc}

\usepackage{cleveref}

\usepackage[none]{hyphenat}
\usepackage{multibib}
\newcites{New}{References}
\usepackage{lineno}
\usepackage{subcaption} % to add labels to panels

\title{A very luminous jet from the disruption of a star by a massive black hole}

\author{Igor Andreoni$^{1,2,3}$\footnote[1]{\label{NGF} Neil Gehrels Fellow.}~\footnote[2]{\label{authorship} These two authors contributed equally to this work.},
	Michael W. Coughlin$^{4 \dagger}$,Daniel A. Perley$^{5}$, Yuhan Yao$^{6}$, Wenbin Lu$^{7}$, S. Bradley Cenko$^{1,3}$, Harsh Kumar$^{8}$, Shreya Anand$^{6}$, Anna Y. Q. Ho$^{9,10,11}$, Mansi M. Kasliwal$^{6}$, Antonio de Ugarte Postigo$^{12}$, Ana Sagu{\'e}s-Carracedo$^{13}$, Steve Schulze$^{13}$, D. Alexander Kann$^{14}$, S. R. Kulkarni$^{6}$, Jesper Sollerman$^{15}$, Nial Tanvir$^{16}$, Armin Rest$^{17,18}$, Luca Izzo$^{19}$, Jean J. Somalwar$^{6}$, David L. Kaplan$^{20}$, Tom{\'a}s Ahumada$^{2}$, G. C. Anupama$^{21}$, Katie Auchettl$^{22,23,24}$, Sudhanshu Barway$^{21}$, Eric C. Bellm$^{25}$, Varun Bhalerao$^{8}$, Joshua S. Bloom$^{9,10}$, Michael Bremer$^{26}$, Mattia Bulla$^{15}$, Eric Burns$^{27}$, Sergio Campana$^{28}$, Poonam Chandra$^{29}$, Panos Charalampopoulos$^{30}$, Jeff Cooke$^{31,32}$, Valerio D'Elia$^{33}$, Kaustav Kashyap Das$^{6}$, Dougal Dobie$^{31,32}$, José Feliciano Agüí Fernández$^{14}$, James Freeburn$^{31,32}$, Cristoffer Fremling$^{6}$, Suvi Gezari$^{17}$, Simon Goode$^{31,32}$, Matthew Graham$^{6}$, Erica Hammerstein$^{2}$, Viraj R. Karambelkar$^{6}$, Charles D. Kilpatrick$^{34}$, Erik C. Kool$^{15}$, Melanie Krips$^{26}$, Russ R. Laher$^{35}$, Giorgos Leloudas$^{30}$, Andrew Levan$^{36}$, Michael J. Lundquist$^{37}$, Ashish~A.~Mahabal$^{6,38}$, Michael S. Medford$^{9,10}$, M. Coleman Miller$^{1,2}$, Anais Möller$^{31,32}$, Kunal Mooley$^{6}$, A. J. Nayana$^{39}$, Guy Nir$^{9}$, Peter T. H. Pang$^{40,41}$, Emmy Paraskeva$^{42,43,44,45}$, Richard A. Perley$^{46}$, Glen Petitpas$^{47}$, Miika Pursiainen$^{30}$, Vikram Ravi$^{6}$, Ryan Ridden-Harper$^{48}$, Reed Riddle$^{49}$, Mickael Rigault$^{50}$, Antonio C. Rodriguez$^{6}$, Ben Rusholme$^{35}$, Yashvi Sharma$^{6}$, I. A. Smith$^{51}$, Robert D. Stein$^{6}$, Christina Thöne$^{52}$, Aaron Tohuvavohu$^{53}$, Frank Valdes$^{54}$, Jan van Roestel$^{6}$, Susanna D. Vergani$^{55,56}$, Qinan Wang$^{17}$, Jielai Zhang$^{31,32}$ [affiliations can be found \hyperref[sec:affiliations]{after the references}]
	}

\newcommand{\arcmin}{$^{\prime}$}

\begin{document}

\maketitle

\begin{abstract}
Tidal disruption events (TDEs) are bursts of electromagnetic energy released when supermassive black holes (SMBHs) at the centers of galaxies violently disrupt a star that passes too close\cite{Rees1988}.
TDEs provide a new window to study accretion onto SMBHs; in some rare cases, this accretion leads to launching of a relativistic jet \cite{Bloom2011Sci, Burrows2011Nat, Levan2011Sci, Zauderer2011Nat, Cenko2012, Brown2015, Pasham2015, Yuan2016}, but the necessary conditions are not fully understood. The best studied jetted TDE to date is Swift J1644+57, which was discovered in  gamma-rays, but was too obscured by dust to be seen at optical wavelengths. Here we report the optical discovery of \at, a rapidly fading source at cosmological distance (redshift $z=1.19325$) whose unique lightcurve transitioned into a luminous plateau within days. Observations of a bright counterpart at other wavelengths, including X--rays, sub-millimeter, and radio, supports the interpretation of \at\ as a jetted TDE containing a synchrotron ``afterglow'', likely launched by a SMBH with spin $a\gtrsim 0.3$.  Using 4 years of Zwicky Transient Facility (ZTF\cite{Graham2019PASP}) survey data, we calculate a rate of $0.02 ^{+ 0.04 }_{- 0.01 }$ Gpc$^{-3}$ yr$^{-1}$ for on-axis jetted TDEs based on the luminous, fast-fading red component, thus providing a measurement complementary to the rates derived from X--ray and radio observations\cite{Sun2015ApJ}. Correcting for the beaming angle effects, this rate confirms that $\sim$1\% of TDEs have relativistic jets. Optical surveys can use \at\ as a prototype to unveil a population of jetted TDEs.
\end{abstract}

\begin{figure*}
\centering
    \begin{subfigure}[t]{0.67\textwidth}
         \centering
         \includegraphics[width=\textwidth]{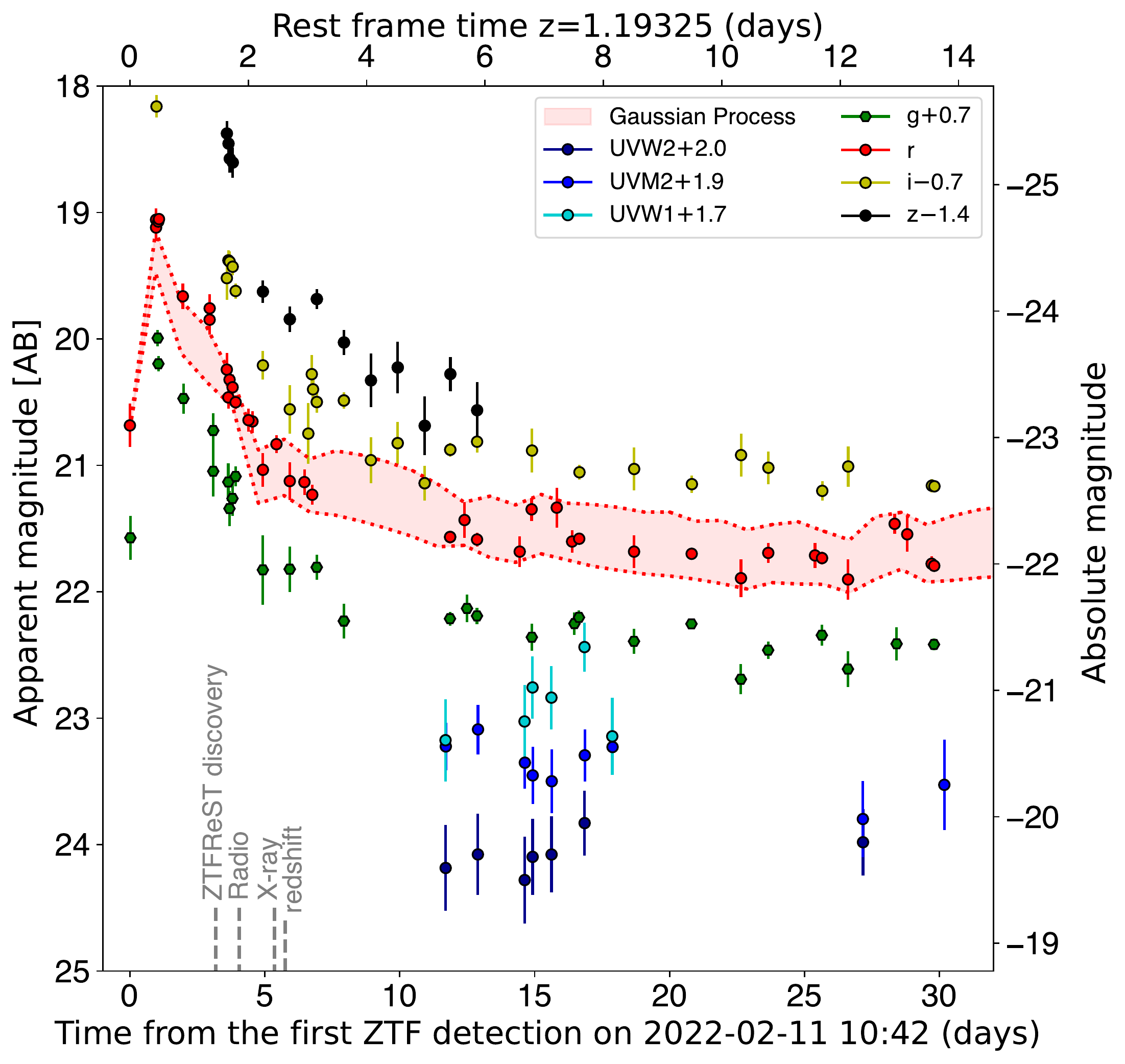}
        \caption{Optical and ultraviolet light curve}
     \end{subfigure}
    \begin{subfigure}[t]{0.32\textwidth}
         \centering
         \vspace{-3.9in}
         \includegraphics[width=0.9\textwidth]{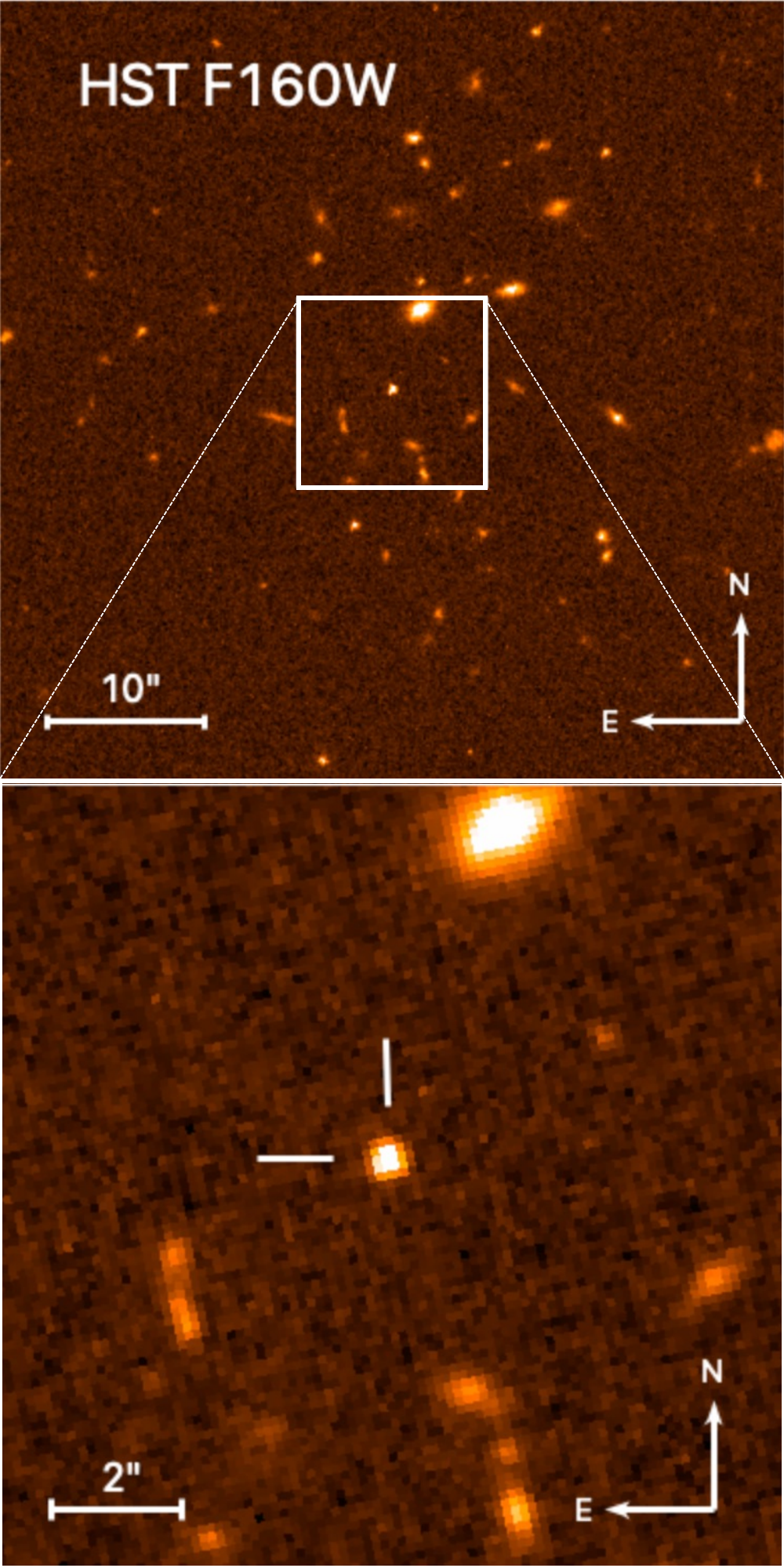}
         \caption{Hubble Space Telescope image}
     \end{subfigure}
\caption{\textbf{\at\ light curve and images in the near infrared, optical, and ultraviolet.}  (a) Apparent and absolute magnitudes show the fast evolution ($> 1$\,mag/day) at early times, the transition into a plateau, and the large luminosity of the transient in the optical in both phases. Magnitudes in this plot are corrected for Galactic extinction $E(B-V) = 0.01$\,mag\cite{Planck2014dust}.} A Gaussian process regression estimate is shown for the $r$-band data to guide the eye (the colored band represents the standard deviation from the central prediction).  More observations are available in $o, H, J, Ks, F606W, F106W$ bands all of which are reported, along with some upper limits in other bands, in \ref{table:photometry}. Error bars shown are $\pm 1~\sigma$. (b)
\at\ was clearly detected in Hubble Space Telescope (HST) images in $F606W$ (optical) and $F160W$ (near infrared) filters. 
A host galaxy likely underlies the bright transient and might be revealed by future observations from space.
 \label{fig:lc_optical}
\end{figure*}

On 2022 February 11 10:42:40 UTC,
ZTF detected a transient, ZTF22aaajecp (Fig.\,\ref{fig:lc_optical}),
located at right ascension $\alpha$ =
$13^{\rm{h}}34^{\rm{m}}43^{\rm{s}}.20232$
and declination $\delta$ = $+33^{\circ} 13' 00''.6565$ (equinox J2000, obtained via radio data analysis with uncertainty 0.01$''$, Methods section~\ref{Methods:VLA}) in its nightly cadenced survey. Our ``ZTFReST"\cite{AndreoniCough2021ztfrest} pipeline, using data obtained on the next two nights, flagged it to be atypical due to its rapid rise and fade (significantly faster than typical supernovae; Methods section \ref{Methods:Identification}--\ref{sec:comparison_transients}).

\begin{figure*}
\vspace{-0.7in}
\centering
    \begin{subfigure}[t]{0.47\textwidth}
         \centering
         \includegraphics[height=3.5in]{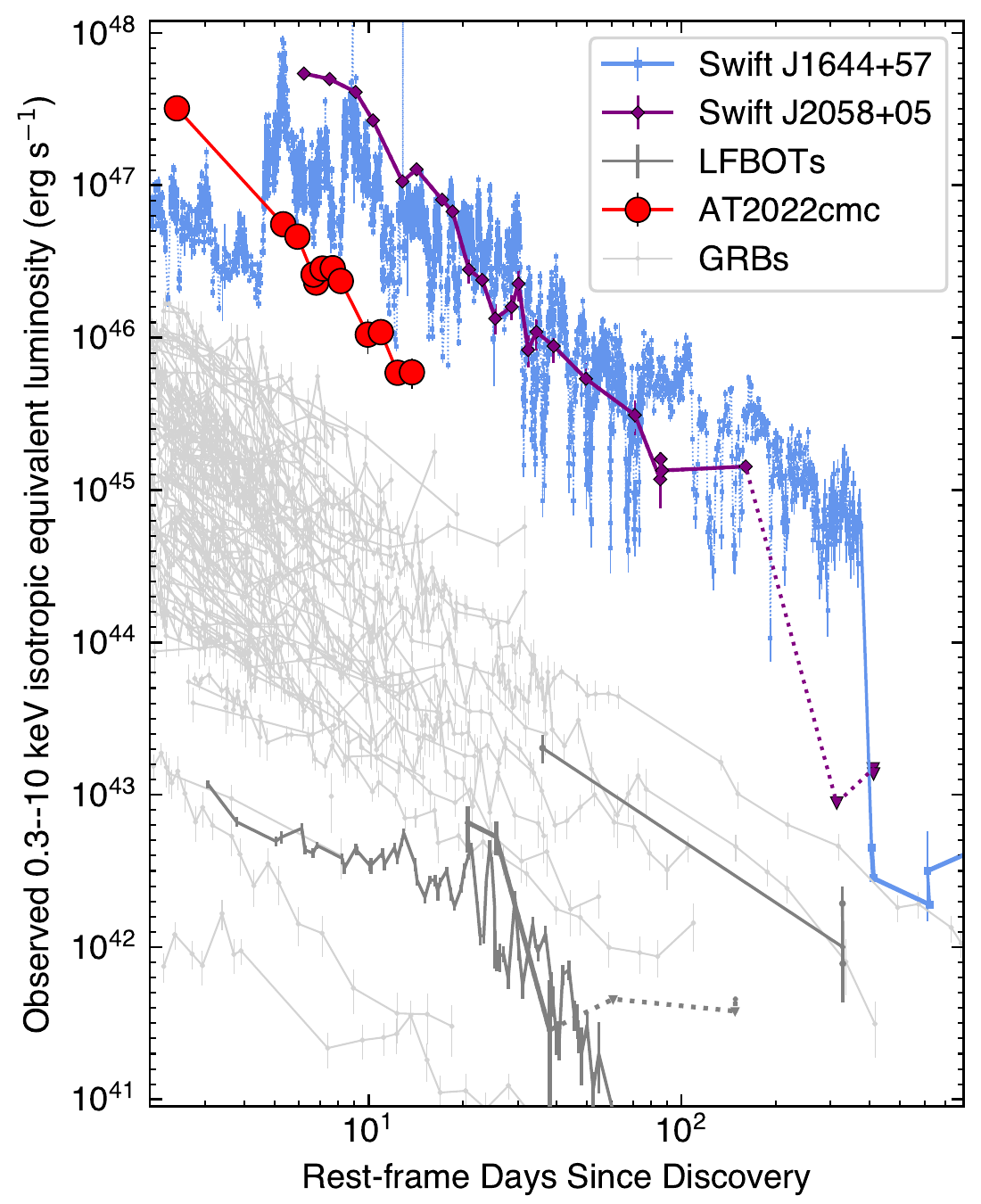}
         \caption{X--ray light curve}
         %\label{fig:}
     \end{subfigure}
    \begin{subfigure}[t]{0.49\textwidth}
         \centering
         \includegraphics[height=3.5in]{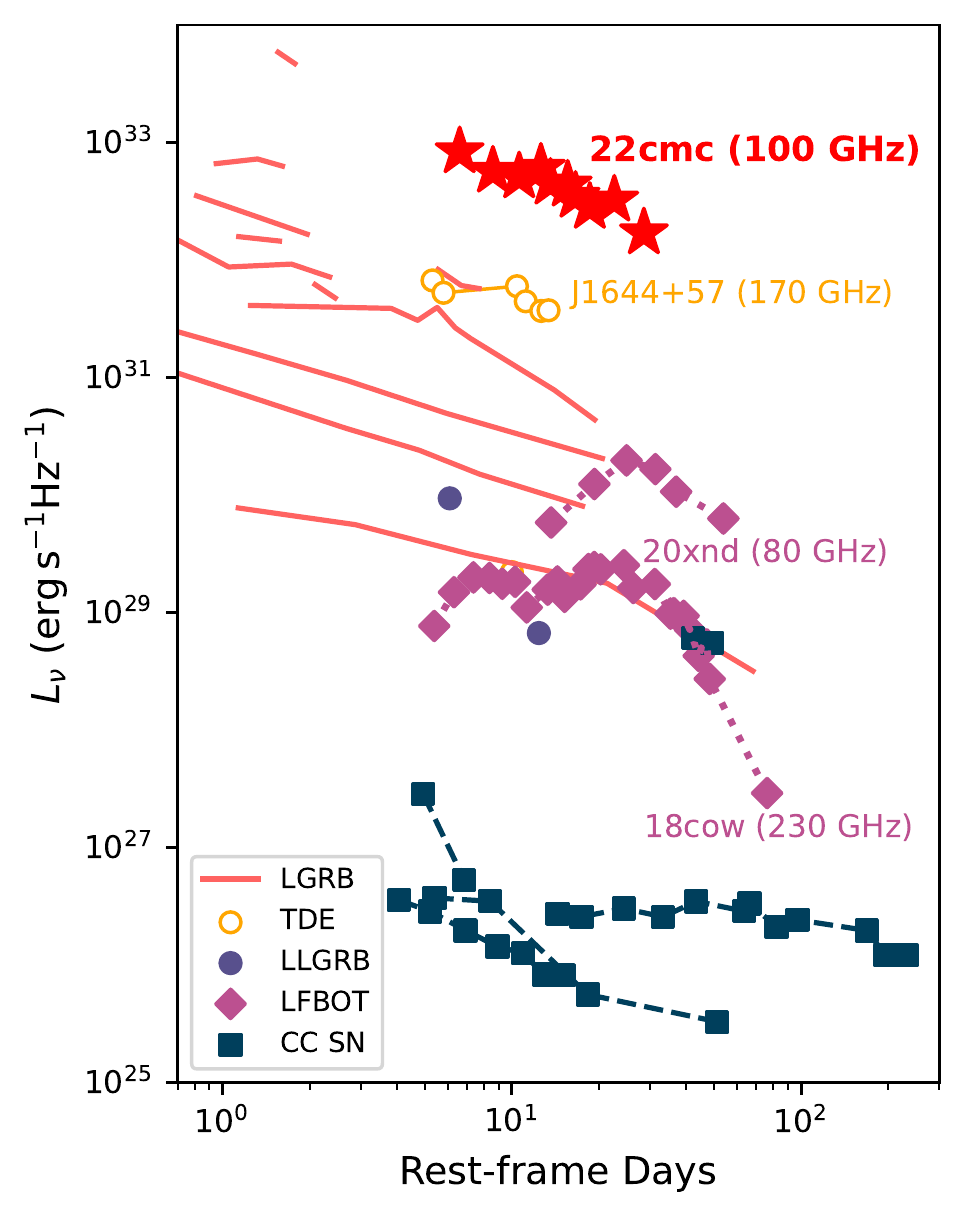}
         \caption{Millimeter light curve}
         %\label{fig:}
     \end{subfigure}
         \begin{subfigure}[b]{0.41\textwidth}
         \centering
         \includegraphics[width=\textwidth]{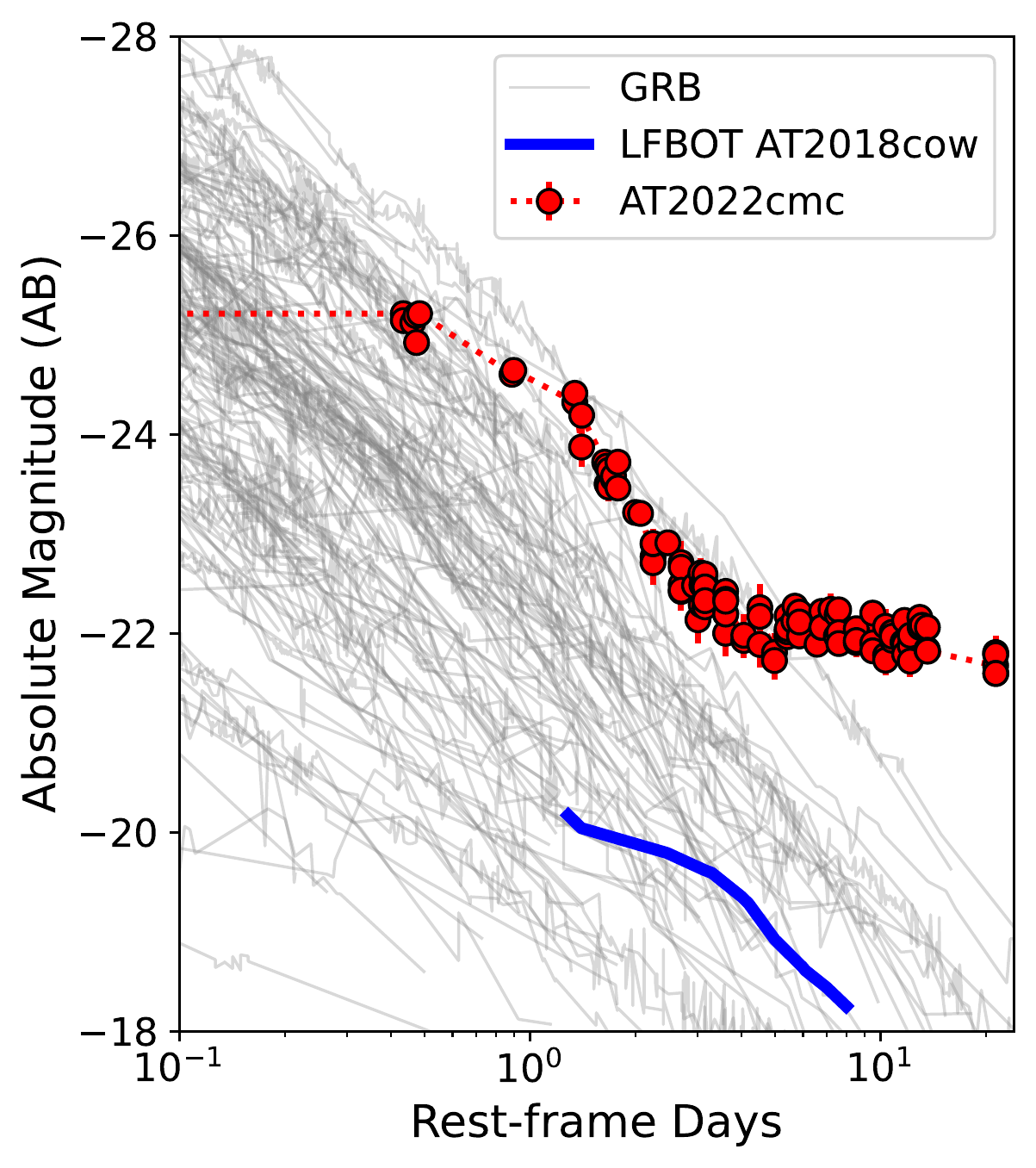}
         \caption{Optical light curve}
         \label{fig:}
     \end{subfigure}
         \begin{subfigure}[b]{0.5\textwidth}
         \centering
         \includegraphics[width=\textwidth]{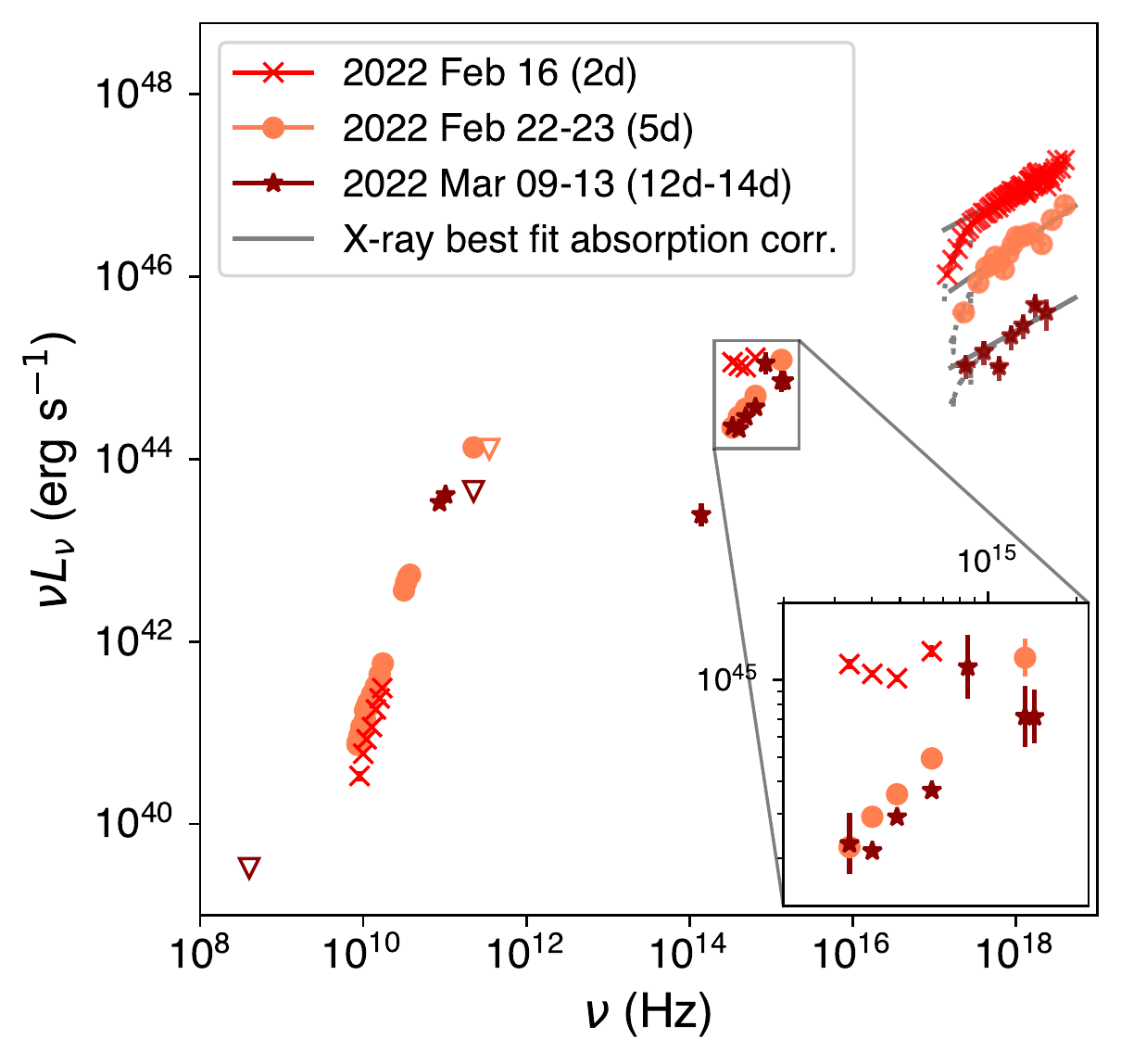}
         \caption{Spectral energy distribution}
         %\label{fig:}
     \end{subfigure}

\caption{\begin{small}
{ {\bf \at\ is among the most luminous extragalactic transients ever observed.} (a) Comparison between the X--ray observations of \at, the jetted TDE candidates Swift J1644+57 and Swift J2058+05, GRBs, and luminous fast blue optical transients (LFBOTs).  The onset time is here set to the first ZTF detection, but its true value is poorly constrained. (b) SMA millimeter light curve of \at\ compared to light curves of millimeter-bright cosmic explosions at similar frequencies (frequencies provided in the rest-frame): long-duration gamma-ray bursts (LGRBs), low-luminosity GRBs (LLGRBs), LFBOTs, core-collapse supernovae (CC SN), and TDEs.
(c) Comparison between the optical light curve of \at\, K-corrected to $r$-band (see Methods section~\ref{sec:comparison_transients}), the light curves of GRB afterglows, and the light curve of the prototypical LFBOT AT2018cow.
(d) Radio to X--ray spectral energy distribution (SED). A change in the shape of the SED is especially evident in the optical/UV between 2022 February 16 and March 09-13 (2\,d, 5\,d, and 12-14\,d in the rest frame from the first detection), suggesting a transition between two different emission components.}
 \end{small}} 
 \label{fig:sed_X}
\end{figure*}

\begin{figure*}[t]
 \centering
\includegraphics[width=5.5in]{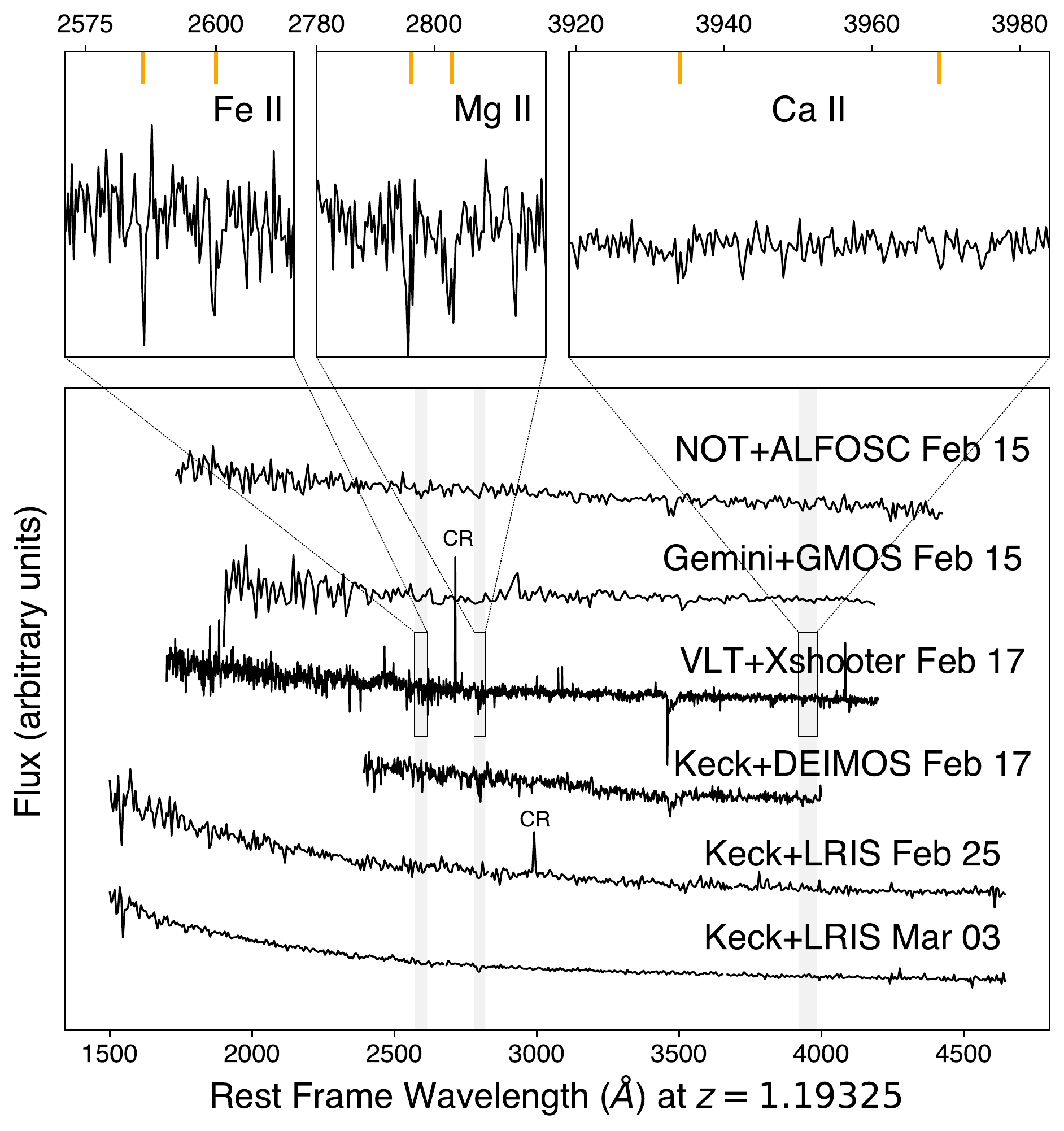}
  \caption{\textbf{Spectra at rest frame for redshift $z=1.19325$.} Optical spectra were acquired from the night after the identification of \at\ until several weeks afterwards. Observed wavelengths were corrected for the redshift by a multiplicative factor $(1+z)^{-1}$. Features in the VLT/X-shooter spectrum (top panels) enabled the redshift to be firmly established; orange bars mark the wavelengths of Fe II, Mg II, and Ca II lines. In the month since its first detection, the spectra of \at\ appear otherwise featureless. The absorption line around 3,500\AA\ is telluric (non astrophysical) and the apparent narrow emission features are cosmic rays (CR).}
 \label{fig:spectra}
\end{figure*}

We reported the source to the Transient Name Server, with assigned IAU name \at. Multi-wavelength observations were triggered, enabling the discovery of a bright counterpart in the X--rays\cite{Pasham2022GCN.31601NICERdetection}, with a 0.3--6\,keV flux of $(3.04 \pm 0.05)\times 10^{-11}$\,erg\,s$^{-1}$cm$^{-2}$ (Methods section~\ref{Methods:NICER}-\ref{Methods:Swift}), as well as counterparts in the decimeter\cite{Perley2022GCN.31592.VLAdetection} and sub-millimeter\cite{Perley2022GCN.31627.SMAdetection} bands (Fig.\,\ref{fig:sed_X}; Methods section~\ref{Methods:VLA}-\ref{Methods:JCMT}). 
The redshift of the transient, $z=1.19325\pm0.00024$ (luminosity distance $D_L = 8.444$\,Gpc assuming a Planck cosmology\cite{Planck2018}), was first secured by absorption lines in the spectrum obtained with the X-shooter instrument on the Very Large Telescope\cite{Tanvir2022GCN.31602.VLTredshift} (Fig.\,\ref{fig:spectra}; Methods section~\ref{sec: redshift}; Methods section~\ref{Methods:Xshooter}). This redshift measurement implies an absolute optical luminosity of  $M_i \simeq -25$\,mag (AB) for the observed peak. However, the host galaxy must be very faint (below $\sim$  24.5\,mag), as it was not found in deep archival images (Methods section~\ref{sec:host}). 

We undertook an intensive multi-wavelength monitoring program from radio to X--ray frequencies. 
Sub-millimeter and radio observations revealed a heavily self-absorbed radio spectrum  up to hundreds of GHz (Methods section~\ref{Methods:VLA}-\ref{Methods:JCMT}; \ref{fig:radio_SED}). The X--ray, radio, and submillimeter counterparts to \at\ are all among the most luminous identified to date for high-redshift transients (Fig.\,\ref{fig:sed_X}). The long-term evolution has shown a decline in X--ray luminosity, and a radio peak moving to lower frequencies.

The infrared/optical/ultraviolet light curve (Fig.\,\ref{fig:lc_optical}) revealed a red color and dramatic rise and decay for about four days post-discovery, before the evolution slowed and the color became bluer. Optical/IR spectra were acquired in both phases, but never showed broad features typically observed in explosive transients\cite{GalYam2017hsn}.

The exceptionally high isotropic-equivalent luminosity across wavelength, and rapid spectral and temporal evolution on sub-day timescales, mark \at\ as extremely unusual, even amongst the rapidly expanding ``zoo'' of astronomical transients (Fig.~\ref{fig:sed_X}), with $\sim 100$ new objects reported publicly per night.
In Methods section~\ref{sec:comparison_transients}, we compare \at\ with energetic transients, some well-known and others exotic. These include a kilonova arising from $r$-process element production in a compact binary merger, a luminous fast blue optical transient (LFBOT), which is a poorly-understood class likely related to stellar collapse to a black hole, and a $\gamma$-ray burst (GRB) arising due to the collapse of a star. Although astronomical surveys may not have sampled the full region of parameter space available to each class, we exclude an association between \at\ and these transient classes. 

The only remaining class of objects which can produce the observed optical, X--ray and submillimeter luminosities is a rare jetted TDE.
Space-based observatories performing searches in $\gamma$-rays and X--rays have disclosed a handful of TDEs with relativistic jets\cite{Brown2015}, the last one more than a decade ago. The best studied jetted TDE so far was Swift J1644+57\cite{Bloom2011Sci, Burrows2011Nat, Levan2011Sci, Zauderer2011Nat}, which showcased several exceptional characteristics: long-lived X--ray emission with variability on very short time scales ($\sim$\,100\,s), radio emission indicating a newly formed relativistic jet, and an origin in the nucleus of a galaxy. The near-infrared transient associated with Swift J1644+57 faded beyond the detection limit in $\sim 10$\,days\cite{Burrows2011Nat}. Unlike \at, no optical or ultraviolet transient was detected\cite{Burrows2011Nat}, though this was unsurprising given the large inferred host galaxy extinction\cite{Burrows2011Nat}. A direct comparison between the properties of \at\ and Swift J1644+57 is presented in Tab.~\ref{tab:comparisonJ1644}.

\begin{figure*}[t]
 \centering
\includegraphics[width=\textwidth]{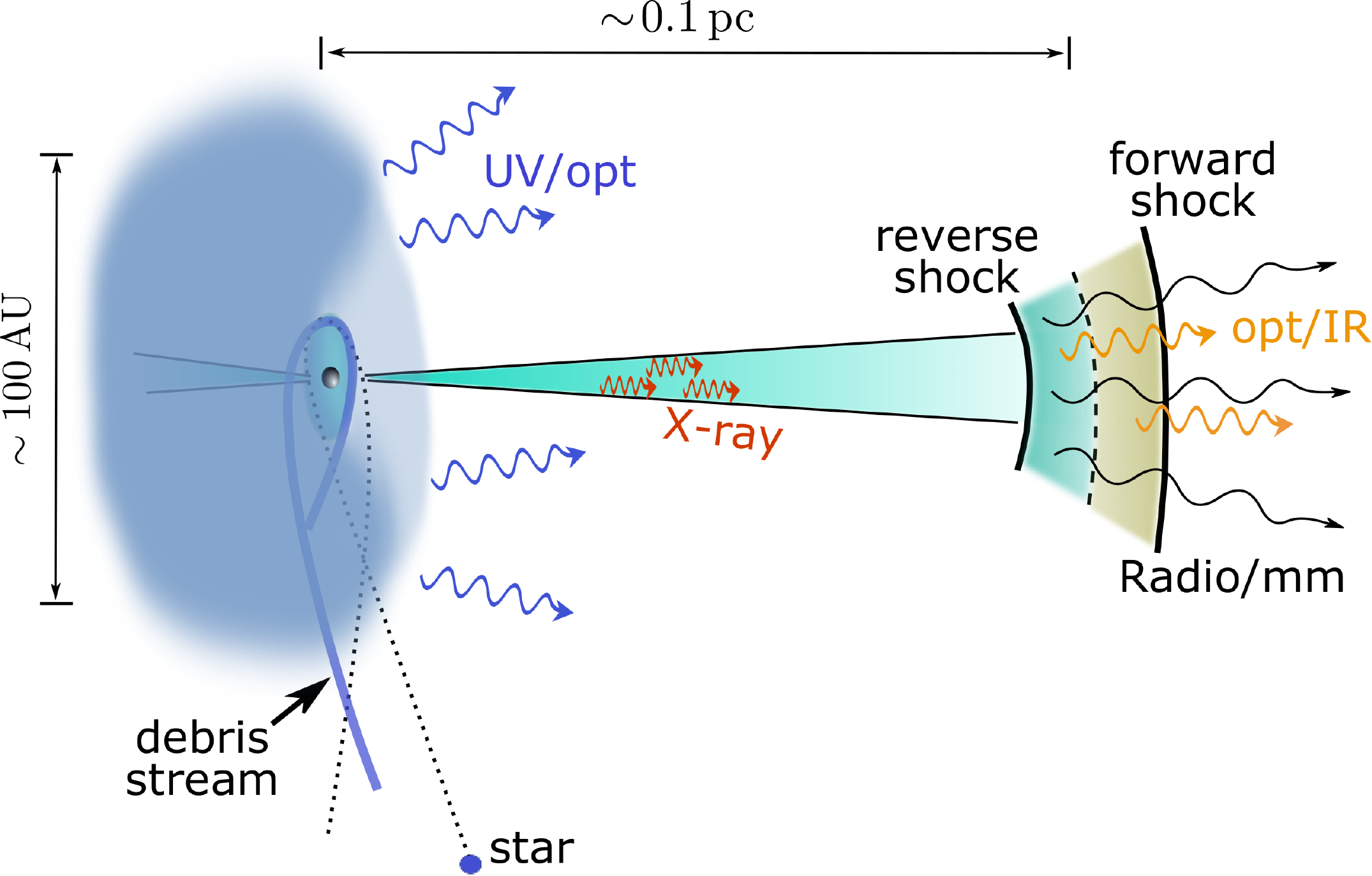}
  \caption{\textbf{Our interpretation of \at\ as a jetted TDE}. This cartoon picture offers a visual representation (not to scale) of the physical processes explained in the main text. Black dotted line: original geodesic of the star (note the general relativistic apsidal precession). Thick blue line: the stellar debris gas undergoing self-intersection. Thick blue envelope of size $\sim 100\rm\, AU$ (or $10^{15} \rm\, cm$): optically thick gas (likely an outflow) reprocessing the X--rays and extreme-UV emission from the accretion disk into the UV/optical band, as observed from other non-jetted TDEs. Light blue disk of size $\sim1\rm\, AU$ (of the order the tidal disruption radius): accretion disk near the black hole. Light blue cones: relativistic jets launched from the innermost regions of the disk. Shocks at a distance of $\sim0.1\rm\,pc$ (or $3 \times 10^{17}$\,cm) from the black hole: reverse shock dominates the radio/millimeter emission, and both reverse shock and forward shock contribute to the non-thermal optical/IR emission.
}
 \label{fig:sketch}
\end{figure*}

We now describe a possible explanation for \at, aided by the broad-brush picture shown in Fig.\,\ref{fig:sketch}. The event started when an ill-fated star approached the SMBH on a nearly parabolic trajectory and was ripped apart into a stream of gaseous debris. About half of the mass stayed bound to the black hole, underwent general relativistic apsidal precession as the gas fell back towards the pericenter, and then produced strong shocks at the self-crossing point \cite{lu20_self_intersection}. The shocked gas then circularized to form an accretion disk around the black hole whose rapid spin generated a pair of relativistic jets \cite{blandford77_BZ_jet}. The high X--ray luminosity (Fig.~\ref{fig:sed_X}a) and flux variability on a timescale of $t_{\rm var}\sim \rm hr$\cite{Pasham2022ATel15232variability, Yao2022ATel15230NuSTAR} suggest that the X--rays were generated by internal dissipation within the jet at a distance $ < 2t_{\rm var} \Gamma^2 c\sim 0.01\mathrm{\, pc}\, (t_{\rm var}/\mathrm{hr}) (\Gamma/10)^2$ from the black hole and that our line of sight was within the relativistic beaming cone of the jet, as was also the case for Swift J1644+57. Here, $\Gamma\sim 10$ is the jet Lorentz factor (as constrained by the radio spectrum, see Methods section \ref{sec:relativistic_motion}) and $c$ is the speed of light. The jet power of \at\ inferred from X--ray observations is consistent with being generated by the Penrose-Blandford-\.{Z}najek mechanism in a magnetically arrested disk\cite{Tchekhovskoy2014}.  Under this mechanism, we infer from the jet power that the SMBH is rapidly rotating with a spin parameter $a\gtrsim 0.3$ for \at\ and $a\gtrsim 0.7$ for Swift J1644+57. We conclude that a high spin is likely required to launched a relativistic jet.

The optical and ultraviolet observations revealed a fast-fading red ``flare'' ($\sim 1$ day) that transitioned quickly to a slow blue ``plateau'', enabling the study of two components generated by the tidal disruption: the relativistic jet and the thermal component from bound stellar debris accreting onto the black hole. 
The fast-fading red component can be explained as follows.
As the jet, which carried $10^{53}$ to $10^{54}\rm\, erg$ of isotropic-equivalent energy, propagated to large distances of $r_{\rm dec} \sim 0.2\rm\, pc$, it was significantly decelerated by driving a forward shock into the surrounding gas of hydrogen with number density of the order $1\rm\, {cm^{-3}}$ (see Methods). At the same time, a reverse shock was propagating into the jet material, similar to cosmological GRBs\cite{kumar15_GRB_review}. Electrons were accelerated to relativistic speeds by these shocks and then produced synchrotron emission at radio/millimeter to X--ray wavelengths. 
The bright millimeter emission was dominated by the reverse shock-heated electrons at early time before the reverse shock crossed the most energetic parts of the jet, but the forward shock emission dominated at later time.

The slowly-fading blue, thermal optical/UV emission was produced by the optically thick outflows from the self-crossing shock and the accretion disk\cite{lu20_self_intersection}, which can be responsible for the blue plateau observed for weeks after the initial flare. As is known from non-jetted TDEs, this gas component produces a blackbody-like spectrum with temperature $10^4$--$10^5\rm\, K$ and peak luminosity of $10^{44}$--$10^{45}\rm\,$ erg\,s$^{-1}$, consistent with our optical observations. The high rest-frame UV luminosity ($\sim 10^{45}\rm\, erg\,s^{-1}$) and blackbody temperature ($\sim 3\times 10^4\rm\, K$) of \at\ (see Methods section \ref{sec:lightcurve_modelling}) are likely due to a viewing angle close to the jet axis \cite{Dai18_viewing_angle}. 

Given the above properties, on balance we conclude that \at\ is most likely generated by (nearly) on-axis jetted relativistic material from the tidal disruption of a star by a massive black hole at the center of a galaxy with low dust extinction. This would make \at\ the furthest jetted TDE discovered to date and the only one for which it was possible to observe a complex optical light curve that transitions from a fast red component into a blue plateau. Our interpretation of a TDE naturally leads to a prediction that, if a host galaxy is eventually detected (e.g., with Hubble Space Telescope or James Webb Space Telescope), then the transient position should be astrometrically coincident with the nucleus and/or host light centroid. Under the TDE interpretation, since the jet is already on-going when the blue UV component is observed, this suggests that the disk formation occurs on a timescale shorter than the evolutionary time of the blue UV component, which is of the order of weeks in the rest frame. This provides important constraints on the highly uncertain hydrodynamics of the disk formation process \cite{bonnerot21_disk_formation}.

Besides Swift J1644+37, which triggered the Swift Burst Alert Telescope (BAT) onboard, two more jetted TDE candidates have been detected by BAT ground-based analysis with similar X--ray and radio properties: Swift J2058+05 \cite{Cenko2012,Pasham2015} and Swift J1112--82 \cite{Brown2015}. We find that $<5\%$ of GRBs like the one associated with Swift J1644+57 would result in a Swift/BAT onboard trigger if the source is placed at the same distance as AT2022cmc. 
Another jetted TDE\cite{Mattila2018Sci} was identified in the radio and infrared bands in the Arp 299 galaxy, but not in the optical and X--rays. Based on these, a jetted TDE rate of  ${\sim}0.03^{+0.04}_{-0.02}\,$Gpc$^{-3}$\,yr$^{-1}$ was obtained\cite{Sun2015ApJ}, which is small compared to the rate of non-jetted TDEs\cite{stone20_TDE_rate} ($\sim10^3\rm\, Gpc^{-3}\,yr^{-1}$). A major open question then is why apparently only a small fraction of TDEs launch jets\cite{deColle20_jet_TDE_review}. The solution to this question will likely shed light on the decades-old puzzle of jet launching from accreting SMBHs. However, a more complete survey of jetted TDEs is needed to pin down their event rate.

Using AT2022cmc's optical light curve and the ZTF survey footprint so far, we calculate an intrinsic rate of $0.02 ^{+ 0.04 }_{- 0.01 }$ Gpc$^{-3}$ yr$^{-1}$ for jetted TDEs oriented towards Earth, obtained independently of discoveries made by high-energy and radio surveys. 
This rate is consistent with previous estimates of the on-axis rate of jetted TDEs, which suggests that host galaxy extinction is often small. This results confirms that a very small fraction $\sim\! 10^{-2}(f_{\rm b}/10^{-2})^{-1}$ of TDEs launch relativistic jets with properties similar to AT2020cmc\cite{AlVe2020}, where $f_{\rm b}$ is the relativistic beaming factor (likely of the order $\Gamma^{-2}\sim 10^{-2}$). However, the connection between routinely-discovered TDEs and rare jetted TDEs remains unclear. Based on the observations of \at, we suggest that a connection exists between jetted TDEs and and the newly-identified class of luminous featureless TDEs\cite{Hammerstein2022arXiv} (Methods section~\ref{sec:featureless}), which could harbor relativistic jets, but might be observed off-axis. This hypothesis can be tested with future, deep follow-up observations in the radio and X--rays. If this connection is confirmed, it will offer a new way to study the system geometry and rapidly grow the known samples thanks to the high luminosity of these transients.

Previous work\cite{Levan2011Sci, Zauderer2011Nat, Cenko2012} presents prospects for radio and X--ray discovery of a population of jetted TDEs. Here we demonstrated that the discovery of such energetic phenomena not only has become accessible to the optical community, but optical may also be the best technique for discovery at the highest redshifts, which uniquely enables the study of distant quiescent SMBHs.
Future observations of \at-like systems will provide statistical samples required to understand dynamics of TDE jets, why some TDEs produce relativistic jets and others do not, and the degree of multi-messenger emission in jetted TDEs.

\newpage
%%%%%%%%%%%%%
% FIGURES 

\begin{table}
    \centering
    \begin{tabular}{lllll}
    \hline \hline
 Property & AT2022cmc & Sw J1644+57 & References \\
    \hline
 Redshift ($z$)&  1.19325 & 0.3534 & \cite{Tanvir2022GCN.31602.VLTredshift, Lundquist2022GCN.31612DEIMOS}, this work; \cite{Levan2011Sci} \\
 Gamma-ray burst & No & Yes & this work; \cite{Bloom2011Sci,  Burrows2011Nat} \\
 X--ray $L_{\rm iso}$ (0.3--6\,keV) & $2.4\times10^{47}$\,erg\,s$^{-1}$  & $\sim 3\times10^{48}$\,erg\,s$^{-1}$ & \cite{Pasham2022GCN.31601NICERdetection}, this work; \cite{Bloom2011Sci, Burrows2011Nat, Levan2011Sci}  \\
 X--ray hour timescale variability & Yes & Yes & \cite{Pasham2022ATel15232variability}; \cite{Bloom2011Sci, Burrows2011Nat, Levan2011Sci} \\
 Ultraviolet transient & Yes & No & this work; \cite{Burrows2011Nat, Levan2011Sci} \\
 Optical transient & Yes & No & this work, \cite{Burrows2011Nat, Levan2011Sci} \\
 Optical transient spectra & Featureless & Not available & this work \\
 Infrared transient & Yes & Yes & this work, \cite{Burrows2011Nat, Levan2011Sci} \\
 Millimeter $L_\nu$ ($100-170$\,GHz) & ${\sim}10^{33}$ erg s$^{-1}$ Hz$^{-1}$ & ${\sim}10^{32}$ erg s$^{-1}$ Hz$^{-1}$ & \cite{Perley2022GCN.31627.SMAdetection}, this work; \cite{Zauderer2011Nat, Yuan2016} \\
 Radio transient & Yes & Yes & this work; \\
 Lorentz factor & $\Gamma \simeq 12$ & $2 \lesssim \Gamma \lesssim 20 $ & this work; \cite{Bloom2011Sci, Burrows2011Nat,  Zauderer2011Nat}\\
X--ray column density $N_H$ & $< 6.4 \times 10^{21}$\,cm$^{-2}$ & $\simeq 1 \times 10^{22}$\,cm$^{-2}$  & this work, \cite{Burrows2011Nat, Levan2011Sci}\\
 Host galaxy in archival images & No & Yes & this work; \cite{Bloom2011Sci, Burrows2011Nat, Levan2011Sci, Zauderer2011Nat}  \\ 
 Host galaxy luminosity & $M_r > -21.4$\,mag & $M_V \simeq -18.19$\,mag  & this work; \cite{Levan2011Sci, Zauderer2011Nat} \\ 
 Host galaxy star formation rate & $< 135$\,M$_\odot$\,yr$^{-1}$ &  0.5\,M$_{\odot}$\,yr$^{-1}$ & this work; \cite{Levan2011Sci} \\
 
 \multirow{2}{*}{Optical polarization}
  & $P_{\rm lin}\sim 0$\% &  \multirow{2}{*}{$P_{\rm lin} = 7.4\% \pm 3.5\%$} &  \multirow{2}{*}{Cikota et al. (in prep.); \cite{Wiersema2012}} \\
& $P_{\rm circ}\sim 0$\% &  & \\
Radio polarization & Not available & $P_{\rm lin} < 9.7\%$ & \cite{Wiersema2012} \\
 SMBH mass & $< 5\times 10^{8}$\,M$_{\odot}$ & $\lesssim 10^{7}$\,M$_{\odot}$ & this work; \cite{Bloom2011Sci, Burrows2011Nat}\\
SMBH spin & $\gtrsim 0.3$ & $\gtrsim0.7$ & this work\\
Inferred on-axis jetted TDE rate & $0.02 ^{+ 0.04 }_{- 0.01 }$ Gpc$^{-3}$ yr$^{-1}$ & $0.03^{+0.04}_{-0.02}\,$Gpc$^{-3}$\,yr$^{-1}$ & this work;\cite{Sun2015ApJ}\\
    \hline
    \end{tabular}
    \caption{Comparison of observational and inferred properties of \at\ and the well-studied jetted TDE Swift J1644+57. The inferred on-axis jetted TDE rate in the Swift J1644+57 column was calculated using the entire population of X-ray jetted TDEs.}
    \label{tab:comparisonJ1644}
\end{table}

\clearpage

\noindent {\bf References}

%\bibliographystyle{naturemag}
%\bibliography{references, gcns}

\clearpage

\noindent {\bf Affiliations}
\begin{small}
\begin{affiliations}
\label{sec:affiliations}

\item Joint Space-Science Institute, University of Maryland, College Park, MD 20742, USA
\item Department of Astronomy, University of Maryland, College Park, MD 20742, USA
\item Astrophysics Science Division, NASA Goddard Space Flight Center, Mail Code 661, Greenbelt, MD 20771, USA
\item School of Physics and Astronomy, University of Minnesota, Minneapolis, Minnesota 55455, USA
\item Astrophysics Research Institute, Liverpool John Moores University, IC2, Liverpool Science Park, 146 Brownlow Hill, Liverpool L3 5RF, UK
\item Division of Physics, Mathematics and Astronomy, California Institute of Technology, Pasadena, CA 91125, USA
\item Department of Astrophysical Sciences, Princeton University, Princeton, NJ 08544, USA
\item Indian Institute of Technology Bombay, Powai, Mumbai 400076, India
\item Department of Astronomy, University of California, Berkeley, CA 94720, USA
\item Lawrence Berkeley National Laboratory, 1 Cyclotron Road, MS 50B-4206, Berkeley, CA 94720, USA
\item Miller Institute for Basic Research in Science, 468 Donner Lab, Berkeley, CA 94720, USA
\item Artemis, Universit'e C\^ote d'Azur, Observatoire de la C\^ote d'Azur, CNRS, F-06304 Nice, France
\item The Oskar Klein Centre, Department of Physics, Stockholm University, AlbaNova, SE-106 91 Stockholm, Sweden
\item Instituto de Astrofísica de Andalucía, Glorieta de la Astronomía, S/N, 18008, Granada, Spain
\item The Oskar Klein Centre, Department of Astronomy, Stockholm University, AlbaNova, SE-106 91 Stockholm, Sweden
\item Department of Physics and Astronomy, University of Leicester, University Road, LE1 7RH, U.K
\item Space Telescope Science Institute, 3700 San Martin Drive, Baltimore, MD 21218, USA
\item Department of Physics and Astronomy, The Johns Hopkins University, 3400 North Charles Street, Baltimore, MD 21218, USA
\item DARK, Niels Bohr Institute, University of Copenhagen, Jagtvej 128, 2200 Copenhagen
\item Center for Gravitation, Cosmology and Astrophysics, Department of Physics, University of Wisconsin--Milwaukee, P.O.\ Box 413, Milwaukee, WI 53201, USA
\item Indian Institute of Astrophysics, Bangalore 560034, India
\item School of Physics, University of Melbourne, Parkville, Victoria 3010, Australia
\item ARC Centre of Excellence for All Sky Astrophysics in 3 Dimensions (ASTRO 3D)
\item Department of Astronomy and Astrophysics, University of California, Santa Cruz, CA 95064, USA
\item DIRAC Institute, Department of Astronomy, University of Washington, 3910 15th Avenue NE, Seattle, WA 98195, USA
\item Institut de Radioastronomie Millimétrique (IRAM), 300 Rue de la Piscine, F-38406 Saint Martin d’Hères, France
\item Department of Physics \& Astronomy, Louisiana State University, Baton Rouge, LA 70803, USA
\item INAF-Osservatorio Astronomico di Brera, Via E. Bianchi 46, I-23807 Merate (LC), Italy
\item National Centre for Radio Astrophysics, Tata Institute of Fundamental Research, Pune University Campus, Ganeshkhind, Pune 411 007, India
\item DTU Space, National Space Institute, Technical University of Denmark, Elektrovej 327, 2800 Kgs. Lyngby, Denmark
\item Australian Research Council Centre of Excellence for Gravitational Wave Discovery (OzGrav), Swinburne University of Technology, Hawthorn, VIC, 3122, Australia
\item Centre for Astrophysics and Supercomputing, Swinburne University of Technology, Hawthorn, VIC, 3122, Australia
\item Space Science Data Center - Agenzia Spaziale Italiana, via del Politecnico, s.n.c., I-00133, Roma, Italy
\item Center for Interdisciplinary Exploration and Research in Astrophysics (CIERA), Northwestern University, Evanston, IL 60208, USA
\item IPAC, California Institute of Technology, 1200 E. California Blvd, Pasadena, CA 91125, USA
\item Department of Astrophysics, Radboud University, 6525 AJ Nĳmegen, The Netherlands
\item W. M. Keck Observatory, 65-1120 Mamalahoa Highway, Kamuela, HI 96743, USA
\item Center for Data Driven Discovery, California Institute of Technology, Pasadena, CA 91125, USA
\item DST INSPIRE Faculty fellow, Indian Institute of Astrophysics, Bengaluru 560034, India
\item Nikhef, Science Park 105, 1098 XG Amsterdam, The Netherlands
\item Institute for Gravitational and Subatomic Physics (GRASP), Utrecht University, Princetonplein 1, 3584 CC Utrecht, The Netherlands
\item IAASARS, National Observatory of Athens, 15236, Penteli, Greece
\item Department of Astrophysics, Astronomy \& Mechanics, Faculty of Physics, National and Kapodistrian University of Athens, 15784, Athens, Greece
\item Nordic Optical Telescope, Rambla Jose Ana Fernandez Perez, 7, 38711 Brena Baja, Spain
\item Department of Physics and Astronomy, Aarhus University, NyMunkegade 120, DK-8000 Aarhus C, Denmark
\item National Radio Astronomy Observatory, P.O. Box O, Socorro, NM 87801, USA
\item Center for Astrophysics — Harvard \& Smithsonian, 60 Garden St, MS 72, Cambridge, MA 02138, USA
\item School of Physical and Chemical Sciences | Te Kura Matu, University of Canterbury, Private Bag 4800, Christchurch 8140, New Zealand
\item Caltech Optical Observatories, California Institute of Technology, Pasadena, CA 91125, USA
\item Univ Lyon, Univ Claude Bernard Lyon 1, CNRS, IP2I Lyon / IN2P3, IMR 5822, F-69622, Villeurbanne, France
\item Institute for Astronomy, University of Hawaii, 34 Ohia Ku St., Pukalani, HI 96768-8288, USA
\item Astronomical Institute of the Czech Academy of Sciences (ASU-CAS), Fri\v cova 298, 251 65 Ond\v rejov, Czech Republic
\item Department of Astronomy and Astrophysics, University of Toronto, 50 St George St, Toronto, ON, Canada
\item National Optical-Infrared Astronomy Research Laboratory, 950 N. Cherry Avenue, Tucson, AZ 85719, USA
\item GEPI, Observatoire de Paris, PSL University, CNRS, Place Jules Janssen, 92190 Meudon
\item Institut d’Astrophysique de Paris, CNRS-UPMC, UMR7095, 98bis bd Arago, 75014 Paris, France

\end{affiliations}
\end{small}

\begin{addendum}

\item The authors thank the anonymous referees for their thorough review of the manuscript and scientific input. The authors thank Dheeraj R. Pasham, Kevin Burdge, David Cook, Aleksandar Cikota, and Sam Oates. 

M.~W. Coughlin acknowledges support from the National Science Foundation with grant numbers PHY-2010970 and OAC-2117997.
E.~C. Kool acknowledges support from the G.R.E.A.T research environment and the Wenner-Gren Foundations. M. Bulla acknowledges support from the Swedish Research Council (Reg. no. 2020-03330). H. Kumar and T. Ahumada thank the LSSTC Data Science Fellowship Program, which is funded by LSSTC, NSF Cybertraining Grant \#1829740, the Brinson Foundation, and the Moore Foundation; their participation in the program has benefited this work. W. Lu was supported by the Lyman Spitzer, Jr. Fellowship at Princeton University. M.R. has received funding from the European Research Council (ERC) under the European Union's Horizon 2020 research and innovation programme (grant agreement n 759194 - USNAC).
G.~L., P.~C. and M.~P. were supported by a research grant (19054) from VILLUM FONDEN. P.~T.~H.~P. is supported by the research program of the Netherlands Organization for Scientific Research (NWO). 
D.~A. Kann acknowledges support from Spanish
National Research Project RTI2018-098104-J-I00 (GRBPhot).
The material is based on work supported by NASA under award No. 80GSFC17M0002.
A.~J. Nayana acknowledges DST-INSPIRE Faculty Fellowship (IFA20-PH-259) for supporting this research.
This work has been supported by the research project grant “Understanding the Dynamic Universe” funded by the Knut and Alice Wallenberg Foundation under Dnr KAW 2018.0067.

Based on observations obtained with the Samuel Oschin Telescope 48-inch and the 60-inch Telescope at the Palomar Observatory as part of the Zwicky Transient Facility project. ZTF is supported by the National Science Foundation under Grant No. AST-1440341 and a collaboration including Caltech, IPAC, the Weizmann Institute for Science, the Oskar Klein Center at Stockholm University, the University of Maryland, the University of Washington (UW), Deutsches Elektronen-Synchrotron and Humboldt University, Los Alamos National Laboratories, the TANGO Consortium of Taiwan, the University of Wisconsin at Milwaukee, and Lawrence Berkeley National Laboratories. Operations are conducted by Caltech Optical Observatories, IPAC, and UW.  The work is partly based on the observations made with the Gran Telescopio Canarias (GTC), installed in the Spanish Observatorio del Roque de los Muchachos of the Instituto de Astrofisica de Canarias, in the island of La Palma. We acknowledge all co-Is of our GTC proposal.
SED Machine is based upon work supported by the National Science Foundation under Grant No. 1106171.
The ZTF forced-photometry service was funded under the Heising-Simons Foundation grant \#12540303 (PI: Graham).

The research leading to these results has received funding from the European Union Seventh Framework Programme (FP7/2013-2016) under grant agreement No. 312430 (OPTICON). Based on observations collected at the European Southern Observatory under ESO programme
106.21T6.015.

This work made use of data from the GROWTH-India Telescope (GIT) set up by the Indian Institute of Astrophysics (IIA) and the Indian Institute of Technology Bombay (IITB). It is located at the Indian Astronomical Observatory (Hanle), operated by IIA. We acknowledge funding by the IITB alumni batch of 1994, which partially supports operations of the telescope. Telescope technical details are available online.\cite{growth_india}

Based on observations made with the Nordic Optical Telescope, owned in collaboration by the University of Turku and Aarhus University, and operated jointly by Aarhus University, the University of Turku and the University of Oslo, representing Denmark, Finland and Norway, the University of Iceland and Stockholm University at the Observatorio del Roque de los Muchachos, La Palma, Spain, of the Instituto de Astrofisica de Canarias.

The Liverpool Telescope is operated on the island of La Palma by Liverpool John Moores University in the Spanish Observatorio del Roque de los Muchachos of the Instituto de Astrofisica de Canarias with financial support from the UK Science and Technology Facilities Council.

The National Radio Astronomy Observatory is a facility of the National Science Foundation operated under cooperative agreement by Associated Universities, Inc.

Partly based on observations collected at Centro Astron\'omico Hispano en Andaluc\'ia (CAHA) at Calar Alto, operated jointly by Instituto de Astrof\'isica de Andaluc\'ia (CSIC) and Junta de Andaluc\'ia.

The James Clerk Maxwell Telescope is operated by the East Asian Observatory on behalf of The National Astronomical Observatory of
Japan; Academia Sinica Institute of Astronomy and Astrophysics;
the Korea Astronomy and Space Science Institute; the National
Astronomical Research Institute of Thailand; Center for Astronomical
Mega-Science (as well as the National Key R\&D Program of China with
No. 2017YFA0402700).
Additional funding support is provided by the Science and Technology
Facilities Council of the United Kingdom and participating universities
and organizations in the United Kingdom and Canada.
Additional funds for the construction of SCUBA-2 were provided by
the Canada Foundation for Innovation.
The JCMT data reported here were obtained under project M22AP030
(principal investigator D.A.P.).
We thank Jasmin Silva, Alexis-Ann Acohido, Harold Pena, and the
JCMT staff for the prompt support of these observations.
The {\tt Starlink} software is currently supported by the East Asian
Observatory.

The Submillimeter Array is a joint project between the Smithsonian Astrophysical Observatory and the Academia Sinica Institute of Astronomy and Astrophysics and is funded by the Smithsonian Institution and the Academia Sinica.

This work is based on observations carried out under project number W21BK with the IRAM NOEMA Interferometer. IRAM is supported by INSU/CNRS (France), MPG (Germany) and IGN (Spain).

%% Gemini 
Based on observations obtained at the international Gemini Observatory, a program of NSF’s NOIRLab, which is managed by the Association of Universities for Research in Astronomy (AURA) under a cooperative agreement with the National Science Foundation. on behalf of the Gemini Observatory partnership: the National Science Foundation (United States), National Research Council (Canada), Agencia Nacional de Investigaci\'{o}n y Desarrollo (Chile), Ministerio de Ciencia, Tecnolog\'{i}a e Innovaci\'{o}n (Argentina), Minist\'{e}rio da Ci\^{e}ncia, Tecnologia, Inova\c{c}\~{o}es e Comunica\c{c}\~{o}es (Brazil), and Korea Astronomy and Space Science Institute (Republic of Korea). This work was enabled by observations made from the Gemini North telescope, located within the Maunakea Science Reserve and adjacent to the summit of Maunakea. We are grateful for the privilege of observing the Universe from a place that is unique in both its astronomical quality and its cultural significance.

 \item[Competing Interests] The authors declare no competing interests.

\item[Contributions] All the authors contributed to the scientific interpretation of the source and reviewed the manuscript. IA and MWC discovered the source, led the follow-up observations and were the primary writers of the manuscript. DAP conducted radio and sub-mm data acquisition and analysis, LT data acquisition and analysis, and contributed significantly to the source analysis.  
YY conducted the X-ray and ultraviolet data analysis. WL led the theoretical modeling of the source.
AYQH, RAP, KPM, GCA, SB, PC, NT, IAS, MBremer, MKrips, AJN, GP conducted radio and sub-mm data acquisition and analysis.
MMK is a ZTF science group leader, contributed to the source follow-up and interpretation. SRK contributed to the paper writing. SBC was PI for the HST observations and conducted photometric data analysis.  
SA, TA, CF, VRK, KKD, JVR, MJG, ACR, MJL conducted optical and near infrared follow-up observations and data analysis with Keck and P200 telescopes.
AR, JF, FV conducted DECam observations, data processing and analysis. JZ, JC, DD, AMöller, SG, KA, RR-H conducted DECam observations. QW conducted the ATLAS data analysis.
EH, SG, MBulla, JCM, JSB contributed to the source interpretation. 
JS conducted follow-up observations with NOT and P60.
PC, GL, MP, EP conducted follow-up observations with NOT. SS conducted follow-up observations with multiple telescopes and led analysis on the host limits. 
AS-C, JJS, VR worked on rate calculations in the optical and radio, respectively.
LI, VDE, SV, AL, NT conducted VLT spectroscopic observations and data analysis.
SC conducted Swift XRT data analysis.
AdUP conducted near infrared observations, photometry, and spectroscopic line strength analysis. CT, JFAF conducted near infrared observations with GTC. DAK provided GRB light curves and significantly contributed to the data analysis. 
HK, VB conducted GIT observations and data analysis.
EB conducted searches for gamma-ray counterparts to the transient. 
RDS contributed to the multi-messenger interpretation of the transient. PTHP conducted the bayesian optical light curve analysis. DLK contributed to the source interpretation and thoroughly reviewed the paper.
RR, BR, RRL, AAM, MSM, ECB, GN, YS, MR, EK are ZTF and Fritz marshal builders. 
AT simulated the Swift GRB detection analysis.

 \item[Correspondence] Correspondence and requests for materials
should be addressed to Igor Andreoni~(email: andreoni@umd.edu) and Michael Coughlin~(email: cough052@umn.edu).

 \item[Data Availability] Photometry and spectroscopy of AT2022cmc will be made available via the WISeREP public database. Facilities that make all their data available in public archives, either promptly or after a proprietary period, include: Very Large Telescope, Very Large Array, Liverpool Telescope, Blanco Telescope, W.~M. Keck Observatory, Gemini Observatory, Palomar 48-inch/ZTF, the Neutron Star Interior Composition Explorer, and the Neil Gehrels Swift Observatory. Data from the Asteroid Terrestrial-impact Last Alert System  were obtained from a public source.

 \item[Code Availability] The ZTFReST\cite{AndreoniCough2021ztfrest, ztfrest_v1.0.0} code is publicly available. Upon request, the corresponding author will provide code (primarily in python) used to produce the figures.

\end{addendum}

\clearpage
\newpage

\begin{methods}

\section{Identification of \at}
\label{Methods:Identification}

In recent years, the immense growth in size and complexity of data sets produced by modern astronomical facilities, e.g., Refs.~\cite{aLIGO,adVirgo,AaAc2017, Bellm2019, Graham2019PASP,Ivezic2019} has required a revolution in the data science principles applied to facilitate discovery of very rare phenomena like \at. 
For optical astronomy in particular, the advent of time-domain surveys such as ZTF requires techniques developed for parsing, in real time, the $\sim$ 1 million alerts produced every night. This real time aspect is essential, as the rapid evolution of the many systems requires that they are discovered and characterized as fast as possible, or the opportunity to acquire crucial data is lost. It is these multi-wavelength sources for which follow-up (or even coordinated wide-field observations with rapid triggering) are immediately required, and it is these sources that we target with real-time algorithms such as the ZTFReST project\cite{AndreoniCough2021ztfrest, ztfrest_v1.0.0}.

ZTFReST uses ZTF alert packets combined with forced point-spread-function photometry (ForcePhotZTF\cite{Yao2019}) to search for exotic extragalactic transients, including kilonovae from binary neutron star mergers.  The ZTF22aaajecp transient, which was assigned\cite{TNS_2022cmc} the IAU name \at, was identified as being unusual for both its rapid rise ($\sim 0.48$\,mag/day) and subsequent rapid decay ($\sim 1.29$\,mag/day).
This can be compared to, for example, core-collapse supernovae models, which across parameter space show both shallower rises ($\sim 0.13$\,mag/day) and decay rates ($\sim 0.74$\,mag/day) maximized across the parameter space.
The discovery of \at\ by ZTF demonstrates that modern optical telescopes are capable of finding jetted TDEs independently of $\gamma$-ray monitors. The lack of an associated $\gamma$-ray signal shows optical discovery of these events reduces the limitations in their study due to the Malmquist bias in the $\gamma$-ray band. Both optical and gamma-ray identification of jetted TDEs will increase the detection rates and allow for greater understanding of this rare class of transient, analogous to recent advances in understanding stars collapsing to black holes producing GRBs, e.g., Ref.~\cite{AndreoniCough2021ztfrest}.

\section{Comparison between AT2022cmc and other energetic transients}
\label{sec:comparison_transients}
 
We compare \at\ to four known transient classes which exhibit fast optical variability and the existence of radio and X--ray counterparts: (i) kilonovae, (ii) luminous fast blue optical transients (LFBOTs), and (iii)  $\gamma$-ray bursts (GRBs). Blazars are another source class that could potentially generate a multi-wavelength transient like \at, however the spectral energy distribution of \at\ is inconsistent with those observed in blazars\cite{deColle20_jet_TDE_review}. This comparison will be addressed in detail by Yao et al. (in preparation).

% Kilonova scenario
The initially red color and rapid evolution of \at\ resemble the behavior of the optical/infrared kilonova\cite{Me2017} AT2017gfo\cite{CoFo2017} associated with GW170817\cite{AbEA2017b}, the first binary neutron star merger detected in gravitational waves. Indeed, \at\ was discovered by the ZTFReST pipeline, which was designed for enabling real-time discovery of elusive fast transients such as kilonovae and GRB afterglows in optical survey data. However, the luminosity of kilonovae is expected to be orders of magnitude fainter than \at\ due to the low ejecta masses expected, $\sim 0.05 M\odot$\cite{Me2017}. Furthermore, kilonova models evolve from blue to red as the heavier $r$-process synthesized elements are produced, whereas \at\ evolved from red to blue. 

% 18cow scenario
The recently-discovered LFBOTs\cite{Prentice2018,Perley2019,Margutti2019,Coppejans2020,Ho2020_Koala,Perley2021,Yao2021arXiv} have observer-frame light curves similar to \at, as well as X--ray and radio counterparts\cite{Margutti2019,Ho2019cow,Coppejans2020,Ho2020_Koala,Ho2021arXiv,Yao2021arXiv}. However, unlike the prototypical LFBOT AT\,2018cow\cite{Prentice2018,Perley2019,Margutti2019}, the optical light curve of \at\ is much redder, $\gtrsim 100 \times$ brighter in $r$-band at peak, and fades $\sim2\times$ faster at early phases in the rest frame. A long-duration blue component has not been observed in any LFBOT to date. The X--ray ``isotropic equivalent luminosity'' of \at\ is $\gtrsim10,000 \times$ higher than LFBOTs (Fig.\,\ref{fig:sed_X}). Altogether these properties strongly disfavor this scenario.

% GRBs scenario
The observed redshift of \at\ implies that the optical isotropic equivalent luminosity is comparable to the brightest relativistic transients (Fig.\,\ref{fig:sed_X}, left panel). This high luminosity ($M_r \approx -25$\,mag) in addition to the red color at peak and rapid decline, is consistent with synchrotron emission, which arises from charged particles accelerated near to the speed of light. This emission arises in the decelerating blastwave of material identified in cosmological afterglows associated with GRBs, and has been used as a diagnostic to identify these afterglows in ZTF data\cite{AndreoniCough2021ztfrest}. The large isotropic equivalent luminosities and the long-lived nature of the radio/mm and X--ray emission, along with the fast X--ray variability\cite{Pasham2022ATel15232variability}, however, separate \at\ from the class of GRB afterglows  and is in contrast with an off-axis GRB interpretation (however
an extremely long GRB lasting for a few days that mimics jetted TDEs remains a possibility\cite{Quataert2012MNRAS}). Direct multi-wavelength comparisons between \at\ and other energetic transients are shown in Fig.\,\ref{fig:sed_X}. In particular, data for millimeter-band past observations of transients include LGRBs\cite{Sheth2003,Laskar2018,Laskar2019,Perley2014,deUgartePostigo2012_grblc}); LLGRBs\cite{Kulkarni1998,Perley2017-alma}, LFBOTs;\cite{Ho2019cow,Ho2021arXiv}), core-collapse SNe\cite{Weiler2007,Maeda2021,Horesh2013,Corsi2014,Soderberg2010}, and tidal disruption events (TDEs\cite{Zauderer2011Nat,Yuan2016}).

\ref{fig:optical_parspace_compare} shows where \at, in the first few days since discovery, is placed in the optical transient parameter space. The peak luminosity and duration of \at\ separate it well from most transient classes and are consistent with GRB afterglow observations.
Fig.~\ref{fig:sed_X}c, more specifically, shows a comparison between the observer-frame optical light curve of \at\ and GRB afterglows. The light curves are taken from the samples presented in (\cite{Kann2006ApJ,Kann2010ApJ,Kann2011ApJ}, Kann et al., in prep.). They have initially been corrected for all line-of-sight extinction, potential host-galaxy and supernova contribution, and shifted to $z=1$ following the method of (\cite{Kann2006ApJ}). Then, we determined the individual distance modulus $m-M$ based on the intrinsic spectral slope $\beta$ of each afterglow to transform the light curves into absolute magnitudes. A luminous slow component at late time, as seen for \at, has never been observed for GRB afterglows. The light curve of the prototypical LFBOT AT2018cow\cite{Perley2019} is also shown in Fig.~\ref{fig:sed_X}c for comparison.

\section{Relativistic  Evolution of the Radio Source}\label{sec:relativistic_motion}
Our goal for this section is to constrain the shock radius $r$ from the self-absorbed part of the radio spectrum and then constrain the Lorentz factor of the emitting plasma  and hence the beaming angle of the jet. We take the standard approach in modeling the synchrotron afterglow from relativistic jets \cite{kumar15_GRB_review}. See e.g., Ref.~\cite{strubbe09_disk_wind, shiokawa15_disk_formation, kimitake16_circularization, bonnerot16_disk_formation, metzger16_reprocessing, bonnerot21_disk_formation} for extended discussions of  the hydrodynamics of the stellar debris and the accretion disk, which are not discussed here. All quantities (time, frequency, energy, luminosity, etc) in this section are defined in the rest frame of the SMBH or the host galaxy. It is straightforward to convert these quantities to the observer's frame by multiplying the relevant cosmological factors for a given redshift.

Consider a jet with an isotropic equivalent energy of $\Eiso\sim 10^{53}$--$10^{54}\rm\, erg$ (as implied by the X--ray emission) and  a hydrogen number density ahead of the forward shock of $n_0$. From energy conservation, we write
\begin{equation}\label{eq:rdec}
    \Eiso \simeq {4\pi \over 3} n_0 r^3 \Gamma^2 \mp c^2,
\end{equation}
where $\Gamma$ is the Lorentz factor of the shock-heated gas, $\mp$ is the proton mass, and the numerical pre-factor depends on the radial density profile of the medium (here taken to be uniform, but the shock radius depends very weakly on this profile). 
The magnetic field strength in the comoving frame of the shocked region is given by the Rankine-Hugoniot jump conditions
\begin{equation}\label{eq:epsB}
    {B^2\over 8\pi} = 4\Gamma^2 \epsB n_0 \mp c^2,
\end{equation}
where  $\epsB$ is the fraction of the thermal energy shared by the magnetic fields. Combining Eqs.~\ref{eq:rdec} and~\ref{eq:epsB}, we obtain the magnetic field strength
\begin{equation}
    B \simeq 2\sqrt{6} \lrb{\Eiso \epsB\over r^3}^{1/2}.
\end{equation}
The isotropic-equivalent specific luminosity at the self-absorption frequency $\nua$ (defined where the absorption optical depth $\tau(\nua)=1$) is given by
\begin{equation}\label{eq:Lnua}
    L_{\nua} \simeq 4\pi^2 \Gamma^{-2} r^2 I_{\nua},
\end{equation}
where we have considered an emitting area of $\pi \Gamma^{-2} r^2$ as a result of relativistic beaming and the specific intensity on the surface area $I_{\nua}$ to be related to that in the plasma's comoving frame $I'_{\nua'}$ by a Lorentz transform $I_{\nua} \simeq \Gamma^3 I'_{\nua'}$. Since the plasma is optically thick, the intensity in the comoving frame is given by the Planck function and the plasma temperature of the electrons responsible for the absorption is $T_{\rm e}\simeq \ga \me c^2/\kB$ ($\me$ and $\kB$ being electron mass and Boltzmann constant), i.e.,
\begin{equation}
    I_{\nua'}' \simeq 2\nua'^2 \ga \me,
\end{equation}
and the frequency in the comoving frame is given by Lorentz transform $\nua' \simeq \nua/\Gamma$. The electron Lorentz factor $\ga$ is related to the emitting frequency $\nua$ by
\begin{equation}\label{eq:nua}
    \nua \simeq \Gamma \ga^2 {3eB\over 4\pi \me c},
\end{equation}
where $e$ is the charge of the electron.  We plug the expressions for $B$, $\ga$, and $I_{\nua}$ into Eq.~\ref{eq:Lnua} and obtain
\begin{equation}
    L_{\nua} \simeq r^{11/4} \Gamma^{-3/2} \nua^{5/2} \Eiso^{-1/4} \epsB^{-1/4} \times {23.2\me^{3/2} c^{1/2}\over e^{1/2}}. 
\end{equation}
The strong dependence of $L_{\nua}$ on $r$ means that it is possible to constrain the shock radius $r$ using observed values of $L_{\nua}\sim 10^{33}\rm\, erg\,s^{-1}\,Hz^{-1}$ and $\nua\sim 10^{11}\rm\, Hz$. Putting this all together, the result is
\begin{equation}
    r\simeq 0.22\mr{\,pc}\, L_{\nua,33}^{4/11} \nu_{\rm a,11}^{-10/11} \Gamma_1^{6/11} E_{\rm iso,53}^{1/11} \eps_{\rm B,-2}^{1/11},
\end{equation}
where we have used, for notational brevity, $Q = 10^x Q_x$ in cgs units (e.g., $\Gamma = 10\Gamma_1$). We also know that the radio fluxes evolved on a timescale of $t_{\rm var}\simeq r/(2\Gamma^2 c) \sim 1\rm \, day$, and this constrains the Lorentz factor of the emitting plasma
\begin{equation}
    \Gamma\simeq 12 \lrb{t_{\rm var}/\mr{day}}^{11/16} L_{\nua,33}^{1/4} \nu_{\rm a,11}^{-5/8} E_{\rm iso,53}^{1/16} \eps_{\rm B,-2}^{1/16}.
\end{equation}
Thus, this confirms the relativistic jet picture. The shock radius can be plugged back into Eq.~\ref{eq:rdec} to estimate the density of the gas ahead of the shock to be $n_0\sim 0.5\mr{\,cm^{-3}}$ and the magnetic field strength in the comoving frame of the emitting region to be $B\sim 0.3\rm\, G$, under our fiducial values of $L_{\nua,33}=\nu_{\rm a,11}=E_{\rm iso,53}=\eps_{\rm B,-2}=1$; however, the values of $n_0\propto (\Eiso/\epsB)^{1/2}$ and $B\propto (\Eiso\epsB)^{5/16}$ are uncertain by about an order of magnitude, due to their stronger dependence on $E_{\rm iso}$ and $\eps_{\rm B}$.

Next, we address the origin of the emitting plasma. When the jet reaches the deceleration radius, a strong reverse shock (RS), which is mildly relativistic in the comoving frame of the unshocked jet, heats up most of the material inside the jet\cite{metzger12_j1644_afterglow}. About half of the energy is deposited in the RS heated gas and the other half in the gas swept up by the forward shock (FS). In the black hole's rest frame, the FS-heated gas has a specific energy $\Gamma^2 \mp c^2$ per proton (due to bulk plus random motions) whereas the RS-heated gas has $\Gamma \mp c^2$ per proton (mainly due to bulk motion). This means (i) there are many more electrons in the RS-heated region than in the FS-heated region by a factor of $\sim \Gamma$ and (ii) electrons have lower Lorentz factors in the RS-heated region than in the FS-heated region by a factor of $\sim \Gamma$. These low Lorentz factor electrons in the RS-heated region, with a typical Lorentz factor of $\gamma_{\rm RS} \simeq 0.5 \epse \mp/\me \sim 100$ (where $\epse\sim 0.1$ is the fraction of thermal energy in relativistic electrons), dominate the emission and absorption at radio frequencies. Indeed, their characteristic synchrotron frequency is $\nu_{\rm RS}\sim 10^{11}\rm\,Hz$ under our fiducial parameters (see Eq.~\ref{eq:nua}). The high Lorentz factor electrons in the FS region dominate the red component of the optical emission, although electrons in the RS region may also contribute a significant fraction (depending on the Lorentz factor distribution of the shock-accelerated electrons).

Finally, we use the peak isotropic X--ray luminosity $L_{\rm X}\simeq 3\times10^{47}\rm\, erg\,s^{-1}$ to constrain the spin of the SMBH. Under the assumption that the jet is powered by the Penrose-Blandford-\.{Z}najek mechanism \cite{blandford77_BZ_jet}, the maximum jet power is given by $L_{\rm max}\simeq \eta_{\rm BZ}\dot{M}c^2$, where $\dot{M}$ is the accretion rate and the maximum power is achieved when the magnetic fields near the event horizon reach the limiting strength beyond which the magnetic pressure will expel the accreting gas --- the result is a magnetically arrested disk (MAD). In the MAD limit, the jet efficiency is given by \cite{tchkhovskoy10_MAD_efficiency}
\begin{equation}
    \eta_{\rm BZ} \simeq {\kappa_{\rm B}\phi_{\rm B}^2\over 4\pi} f(a), \ f(a) = \Omega_{\rm H}^2 (1 + 1.38\Omega_{\rm H}^2 - 9.2\Omega_{\rm H}^4),
\end{equation}
where $\kappa_{\rm B}\simeq 0.05$ depends weakly on the magnetic field geometry, $\phi_{\rm B}\simeq 50$ is the dimensionless magnetic flux, $\Omega_{\rm H} = a/(2r_{\rm H})$ is the dimensionless angular frequency of the event horizon, $r_{\rm H} = 1 + \sqrt{1-a^2}$ is the radius of the outer event horizon in units of the gravitational radius $GM/c^2$, and $a$ is the dimensionless spin parameter. For a jet Lorentz factor of $\Gamma\simeq 10$, the beaming fraction of the X--ray emission is $f_{\rm b}\simeq 10^{-2}$ (a much smaller beaming factor is unlikely). For a radiative efficiency of 10\%, we infer the beaming-corrected peak jet power to be $L_{\rm jet}\gtrsim 3\times10^{46}\rm\, erg\,s^{-1}$, while the actual peak power may be larger because (i) a fraction of the radiation is likely in the $\gamma$-ray band and (ii) our earliest X--ray observation may have missed the peak of the lightcurve.

Hydrodynamic simulations of TDEs show that the rate at which the stellar debris falls back towards the BH is generally less than $10\,M_\odot \rm \,yr^{-1}$ --- this value corresponds to the peak fallback rate for a $M_*=3M_\odot$ main-sequence star disrupted by a $M=10^6M_\odot$ SMBH\cite{law-smith20_fallback_rate}. Even though the peak fallback rate depends on the masses of the star and SMBH $\dot{M}_{\rm fb,peak}\propto M_*^{1/2} M^{-1/2}$, at highly super-Eddington accretion rates, most of the fallback material is blown away by the radiation pressure instead of accreted by the SMBH \cite{jiang19_super-Eddington_accretion}. Therefore, we generally expect the accretion rate to be $\dot{M}<10\,M_\odot\rm\,yr^{-1}$. Based on the above arguments, we constrain
\begin{equation}
    f(a) \geq {L_{\rm jet}\over \dot{M}c^2} {4\pi\over \kappa_{\rm B}\phi_{\rm B}^2},
\end{equation}
which provides a lower limit on the BH spin parameter $a\gtrsim 0.3$ for $L_{\rm jet} = 3\times10^{46}\rm\, erg\,s^{-1}$ (for the case of \at). We also obtain $a\gtrsim 0.7$ for $L_{\rm jet} = 3\times10^{47}\rm\, erg\,s^{-1}$, as is the case of Swift J1644+57 whose peak X--ray luminosity is 10 times higher.

\section{Redshift}
\label{sec: redshift}

The redshift of \at\ was first determined from the VLT/X-shooter spectrum (Methods section~\ref{Methods:Xshooter}).
The spectrum shows a single emission line that we identify as [OIII]$\lambda$5008 at a redshift of $z=1.1933$, as previously reported via GCN\cite{Tanvir2022GCN.31602.VLTredshift}. At a similar redshift, we detect absorption features of AlIII, FeII, MnII, MgII, MgI and CaII. The average redshift of these features is $z=1.19325\pm0.00024$. However, we notice that there are two velocity components in these features, one at $z=1.19318\pm0.00019$ that dominates the absorption of the AlIII, FeII, MnII, MgII and MgI lines and one at $z=1.19361\pm0.00010$, which dominates in the CaII lines. The velocity difference between these lines is $\sim130$ km s$^{-1}$. \ref{table:EWs} displays the equivalent width measurements of the absorption lines.

\section{Spectral line strength analysis}
\label{sec: line strength}

We have performed a line strength analysis\cite{deUgartePostigo2012} on the VLT/X-shooter spectrum (Methods section~\ref{Methods:Xshooter}), which compares the strength of the absorption features measured in our spectrum with those of a sample of GRB afterglow spectra.
GRBs are typically found to be located well within star-forming host galaxies, and their spectra probe light paths from deep within their hosts. The spectral features imprinted in GRB afterglow spectra have been found to be at hundreds of parsecs or even kpcs from the GRB, so they are probing the overall material in the host galaxy and not necessarily their very local environment. This is similar to what one would expect from the path probed by a jet emitted from the core of an active and similarly star-forming galaxy, but is in contrast to what one sees in the spectra of damped Lyman $\alpha$ (DLA) absorbers in the line of sight of quasars, which probe the outskirts of intervening galaxies and show much weaker features.
\ref{fig:LSD} shows a line strength diagram, in which the average feature strength of the GRB sample is shown with a thick black line and the 1-$\sigma$ deviation in the log-normal space with dotted lines. The features measured in the \at\ spectrum (shown in red) resemble, quite closely, the average strengths seen in GRB spectra. Only the CaII lines show a somewhat lower strength than average, which, as mentioned before, also display a slightly different velocity component which means that they are probably produced in a different region as the rest of the lines (this is commonly seen in GRB spectra). The overall line strength parameter (LSP) compares the line strengths with the sample using a single number. In this case we obtain LSP $=-0.20\pm0.25$ (zero would be the average and $\pm 1$ the $\pm 1\sigma$ deviation), which implies that the lines are just slightly weaker than the average, equivalent to those of the 42nd percentile in the sample.

The fact that the spectral features are similar to the ones seen in GRBs implies that the environment density and composition is probably not unlike the one in which these stellar explosions are produced. 
Since GRBs are known to happen well within star forming galaxies, at a median projected offset of 1.3 kpc from the core of the galaxy \cite{Bloom2002,Blanchard2016}, the observation of a similar environment in the case of AT2022cmc points towards this transient happening well within a galaxy of a similar type.

\section{Host galaxy}
\label{sec:host}

The field of \at\ was observed in $u$ and $r$ bands with the MegaPrime camera at the 3.58m Canada-French-Hawaii Telescope between 2015 and 2016. We retrieved the science-ready level-3 data from the Canadian Astronomy Data Centre. We used aperture photometry at the position of the optical transient (aperture radius: 1.5 $\times$ FWHM of the stellar PSF) to try measure the brightness of the host galaxy. Once an instrumental magnitude was established, it was calibrated against the brightnesses of several stars from a cross-matched SDSS catalogue. The host evaded detection in both bands. Using forced photometry, we measure $>24.19$ and $>24.54$~mag in $u$ and $r$ band; $3\sigma$ confidence; not corrected for Milky-Way extinction), respectively.

From the {\it Swift} X--ray Telescope (XRT)\cite{Burrows2005} data analysis of \at\ (see Methods section~\ref{Methods:Swift}), we estimate the equivalent neutral hydrogen column density of the host galaxy to be 
 $N_H < 6.4 \times 10^{21}$\,cm$^{-2}$ (90\% confidence). The presence of a counterpart in the ultraviolet, first detected with {\it Swift} on 2022 February 23 (day 5.3) with magnitude $UWM2 = 21.30 \pm 0.25$\,mag, provides additional evidence that the host galaxy extinction is significantly lower than in the case of Swift J1644+57 ($A_V \approx 4.5$\,mag, corresponding to an equivalent neutral hydrogen column density of $N_H \approx 1 \times 10^{22}$\,cm$^{-2}$)\cite{Burrows2011Nat}. However, our spectral line strength analysis (Methods section~\ref{sec: line strength} ) yielded results similar to most GRBs from stars collapsing to black holes, which suggests that \at\ happened well within its host galaxy.

To put a limit on the host galaxy properties, we create a possible model for the spectral energy distribution of the host with the software package \texttt{Prospector}\cite{prospector} version 0.3 \cite{Johnson2021a}. \texttt{Prospector} uses the \texttt{Flexible Stellar Population Synthesis} (\texttt{FSPS}) code \cite{Conroy2009a} to generate the underlying physical model and \texttt{python-fsps} \cite{ForemanMackey2014a} to interface with \texttt{FSPS} in python. The \texttt{FSPS} code also accounts for the contribution from the diffuse gas based on the Cloudy models from ref. \cite{Byler2017a}. Furthermore, we assumed a Chabrier initial mass function \cite{Chabrier2003a} and approximated the star formation history (SFH) by a linearly increasing SFH at early times followed by an exponential decline at late times (functional form $t \times \exp\left(-t/\tau\right)$). The model was attenuated with the Calzetti dust model\cite{Calzetti2000a}. The priors were set identical to reference \cite{Schulze2021a}: uniform galaxy mass (5--13 log M$_*$/M${_{\odot}}$), uniform $V$-band optical depth (0--8 $\tau_V$), uniform stellar metallicity ($-2$--0.5 log Z/Z${_{\odot}}$), log-uniform age of the star formation episode (0.001--13.8 $t_{\rm age}$/Gyr), log-uniform e-folding time-scale of the star formation episode (0.1--100 $\tau$/Gyr).

Upper limits on the host galaxy luminosity lead to upper 
limits (at 95\% confidence) on galaxy mass of $\log\,M/M_\odot < 11.2$, 
star-formation rate of 135~$M_\odot\,\rm yr^{-1}$, and an absolute magnitude of $M_r > -21.4$~mag (corrected for Milky Way extinction but not corrected for host attenuation). These upper limits are not strongly constraining, hence deeper imaging is needed in the future.

We use a  galaxy bulge -- black hole mass relation\cite{McConnell2013} and the upper limit on the \at\ galaxy mass to obtain an upper limit on SMBH mass of $M_{\rm BH} < 4.7 \times 10^8$\,M$_{\odot}$.
 The SMBH mass can also be (weakly) constrained based on the Hill's mass argument --- a main-sequence star of less than a few solar masses can be tidally disrupted outside the event horizon of a rapidly spinning SMBH of mass $\lesssim 10^9M_\odot$\cite{kesden12_Hills_mass}.
The upper limit also implies an Eddington luminosity of $L_{\rm Edd} < 6 \times 10^{46} $\,erg\,s$^{-1}$. This Eddington limit is an order of magnitude lower than the NICER soft X--ray isotropic equivalent luminosity\cite{Pasham2022GCN.31601NICERdetection} of $\sim 2.6 \times 10^{47} $\,erg\,s$^{-1}$, which confirms that a jet strongly beamed towards the Earth is potentially responsible for the X--ray emission.

Observations with the Hubble Space Telescope HST  and, possibly, with the James Webb Space Telescope should be able to unveil the faint host galaxy, once the transient has disappeared.

\section{Detectability of GRB 110328A (Swift J1644+57) at $z\sim 1.2$}
\label{Methods:GRB_SwJ1644}
 
The Swift J1644+57 event was first detected as a GRB\cite{Cummings2011GCN}, labelled GRB 110328A, by the Neil Gehrels Swift Observatory (see Methods section~\ref{Methods:Swift}). Placing GRB 110328A at the distance of \at\ would result in an onboard trigger by Swift/BAT in only the most optimal cases. It would require a maximal duration exposure (30~minutes) image trigger with the source near the center of the BAT coded field-of-view, and around the few hour period when GRB 110328A was at its brightest. Assuming a uniform sky distribution of GRB 110328A-like sources at $z=1.19325$ and a normal Swift observing schedule, we find that fewer than 5\% of such events would generate an onboard trigger. However, note that both Swift J2058+05 and Swift J1112-82 were found in automated ground analysis of Swift/BAT data with significantly longer exposures (days), which would allow the discovery of the Swift J1644+47 GRB at $z\sim 1.2$.

\section{Millimeter Survey Rate Predictions}

We consider the rate of AT2022cmc-like transients expected to be detected in two millimeter band surveys: the South Pole Telescope Third Generation (SPT-3G) survey \cite{Benson2014SPIE}, and the Stage-4 CMB experiment\cite{Abazajian2019arXiv} (CMB-S4). SPT-3G is an ongoing 5.5-year survey covering an area of 1500 deg$^2$ (${\sim}3.6\%$ of the sky) at a frequency of 95 GHz. The observations have a single-epoch rms noise of 6 mJy and ${\sim}1/2$ day cadence \cite{Guns2021ApJ}. CMB-S4 will likely begin observations in 2029. Among the surveys it will be performing, CMB-S4 will observe over half the sky over a frequency range of 30-280 GHz and a daily cadence for seven years.  The 93 GHz, single-epoch rms sensitivity is 6 mJy. To determine the rate of AT2022cmc-like events in these surveys, we assume a typical ${\sim}100$ GHz luminosity of ${\sim}10^{32}$ erg s$^{-1}$ Hz$^{-1}$ for at least ${\sim}20$ days. If we define a detected source as one with a $5\sigma$ single-epoch detection (${\sim}30$ mJy for both surveys), AT2022cmc-like events will be detectable to $D_L \sim 1.67$ Gpc in either survey. From the detection of three jetted TDEs in ${\sim}10$ years of {\it Swift} observations\cite{Bloom2011Sci,Burrows2011Nat, Levan2011Sci, Zauderer2011Nat,Cenko2012,Pasham2015,Brown2015},  the on-axis jetted TDE rate is ${\sim}0.03^{+0.04}_{-0.02}$ Gpc$^{-3}$ yr$^{-1}$ \cite{Sun2015ApJ}. Hence, over the full survey duration, we expect a mean of ${\sim}0.05^{+0.05}_{-0.03} (0.9^{+0.9}_{-0.4})$ events in the SPT-3G (CMB-S4) survey. More events will be detected if the single epoch images are stacked. Because the observed 100 GHz lightcurve is ${\sim}$flat for at least twenty days, we conservatively assume ten day bins.  Then we expect a mean of ${\sim}0.5^{+0.5}_{-0.3} (5^{+5}_{-3})$ events in the SPT-3G (CMB-S4) survey. While these predictions are approximate, we generally expect millimeter rates a factor of ${\mathcal{O}}(10)$ higher than those for the Swift J1644+57 \cite{Eftekhari2021arXiv} because the 100 GHz luminosity of AT2022cmc is a factor of ${\sim}10$ higher than that of Swift J1644.

\section{Optical Rates Estimates}

To estimate the rates of \at-like events using ZTF survey data, we use \texttt{simsurvey}\cite{FeNo2019} to simulate \at-like light curves and estimate the efficiency of their recovery with a filter consistent with ZTFReST\cite{AndreoniCough2021ztfrest, ztfrest_v1.0.0}. Using the survey bandpasses and limiting magnitudes calculated for each exposure, we injected light curves uniformly in co-moving volume to a distance of $z=1.2$, consistent with \at's distance. The light curves are reddened by Milky Way extinction.  To flag them as ``recovered,'' we required (i) at least two detections with $>3\sigma$ significance, at least one of which must have $>5 \sigma$ significance, (ii) a measured fade rate faster than 0.3\,mag/day in each band, and (iii) $>3$ hours of time separation.

We provide estimated rates under two assumptions. The first is that \at\ is the only example in our data set and only when the real-time capabilities of ZTFReST were in place, starting in August 2020. This yields an on-axis, jetted TDE rate of $0.02 ^{+ 0.04 }_{- 0.01 }$ Gpc$^{-3}$ yr$^{-1}$ (95\% confidence), showing strong consistency with the established rate\cite{Sun2015ApJ} from {\it Swift}:  ${\sim}0.03^{+0.04}_{-0.02}\,$Gpc$^{-3}$\,yr$^{-1}$. 

Ref.~\cite{AnKo2020} previously reported a here-to-fore unidentified transient, ZTF19aanhtzz/AT2019aacu, which shows some similar properties to \at, including the rapid decay and lack of confirmed host. In this case, the rate estimated would be $ 0.04 ^{+ 0.02 }_{- 0.02 }$ Gpc$^{-3}$ yr$^{-1}$. However, this transient was found during archival searches and no follow-up observations were triggered to look for a potential bright X--ray or radio counterpart. For this reason, we cannot consider ZTF19aanhtzz a jetted TDE candidate. This fact further confirms the need for real-time, data analysis frameworks capable of identifying rapidly evolving transients to enable prompt follow-up.

\section{Optical light curve modelling}
\label{sec:lightcurve_modelling}

To analyze the event, we have proposed a two-component model. The model consists of a time-dependent power-law component and a static blackbody contribution. The spectral flux density $F_{\nu}$ (in the rest-frame) is given by
\begin{equation}
F_{\nu} = F_{\rm pl} \left(\frac{\nu}{10^{15}{\rm Hz}}\right)^{\beta} 10^{\alpha (t - t_0)} + F_{\rm bb}(\nu, T, L_{\rm bb}),
\end{equation}
where $t_0$ is the brightest time of the event, and $F_{\rm pl}$ is the reference spectral flux density for $\nu = 10^{15}$ Hz at $t=t_0$. In the above, $F_{\rm bb}$ is the contribution of a blackbody at temperature $T$ with a luminosity $L_{\rm bb}$.

We have employed Bayesian inference techniques to analyze the optical data from 2022-02-12 onward, with a Gaussian likelihood in AB magnitude space. The blackbody temperature $T$,  luminosity $L_{\rm bb}$ and the reference spectral flux density $F_{\rm pl}$ are assigned log-uniform priors. Therefore, we can assign uniform priors of $\log_{10}(T / 1K) \sim \mathcal{U}(3, 6)$, $\log_{10}(L_{\rm bb} / (1 {\rm erg} / {\rm s})) \sim \mathcal{U}(40, 50)$ and  $\log_{10}(F_{\rm pl} / (1 {\rm erg} / {\rm s} / {\rm Hz})) \sim \mathcal{U}(10, 80)$. The parameters $\beta$ and $\alpha$ are assigned uniform priors, with $\beta \sim \mathcal{U}(-10, 0)$ and $\alpha \sim \mathcal{U}(-100 \ {\rm day}^{-1}, 0 \ {\rm day}^{-1}$). The Bayesian evidence is estimated, and the posterior distribution is sampled with the nested sampling algorithm implemented in \texttt{PyMultinest}\cite{pymultinest, Feroz_2009}.

The blackbody's temperature is inferred to be 30,000$^{+900}_{-800}K$ and the luminosity $L_{\rm bb}$ is inferred to be $10^{45.53\pm0.02}{\rm erg}/{\rm s}$; both are  median values and $90\%$ credible regions (see \ref{fig:BB_PL_parameters} for the parameter estimates). This implies a blackbody photospheric radius of $r_{\rm ph} = [L_{\rm bb}/(4\pi \sigma_{\rm SB} T_{\rm bb}^4)]^{1/2}\simeq 2\times10^{15}\rm\, cm$, where $\sigma_{\rm SB}$ is the Stefan-Boltzmann constant. This large radius suggests that the emission comes from an outflow instead of the surface of an accretion disk --- the Keplerian orbital period at a distance of $2\times10^{15}\rm\,cm$ would be $1.5\mathrm{\,yr} (M/10^6M_\odot)^{-1/2}$ which is too long for any plausible black hole mass $M$.  For the power-law contribution, $F_{\rm pl} = 10^{30.51\pm0.02} \  {\rm erg / s / Hz}$, $\beta = -1.32\pm 0.18$, and $\alpha=-0.48\pm0.02 \ {\rm day}^{-1}$ (again the median values and 90\% credible regions). The estimate of $\beta$ is consistent with the prediction $\beta \in [-1.5, -0.5]$ based on synchrotron afterglow theory.

We have further allowed the value of $\beta$ to be time-dependent.  For the linear case, we assumed $\beta = \beta_0 + \beta_1 (t - t_0)$., with priors $\beta_0 \sim \mathcal{U}(-10, 0)$ and $\beta_1 \sim \mathcal{U}(-10 \ {\rm day}^{-1}, 10 \ {\rm day}^{-1})$. Similarly, for the quadratic case, we assumed $\beta = \beta_0 + \beta_1 (t - t_0) + \beta_2 (t - t_0)^2 / 2$ with priors $\beta_0 \sim \mathcal{U}(-10, 0)$, $\beta_1 \sim \mathcal{U}(-10 \ {\rm day}^{-1}, 10 \ {\rm day}^{-1})$ and $\beta_2 \sim \mathcal{U}(-1 \ {\rm day}^{-2}, 1 \ {\rm day}^{-2})$. However, the resulting logarithms of the  Bayesian evidence (linear in time: $-267.95 \pm 0.18$, quadratic in time: $-270.25 \pm 0.19$) are lower than the time-independent case ($-263.55 \pm 0.17$), where the uncertainty is estimated with the negative relative entropy; see Ref.~\cite{Feroz:2007kg}. Therefore, there is no evidence that $\beta$ varies significantly in time.

\section{A connection between jetted TDEs and the class of luminous featureless TDEs?}
\label{sec:featureless}

A class of TDEs has recently been identified by Ref.\cite{Hammerstein2022arXiv} that are overluminous ($M_r\sim -22$\,mag at peak, \ref{fig:featureless_histo}) and, unlike most TDEs, do not show any broad features in their optical spectra. A physical explanation for the nature of this class of luminous and featureless TDEs is yet to be proposed. Multi-wavelength follow-up data of luminous featureless TDEs are still sparse, therefore the presence of jets cannot be excluded. 

Our observations of \at\ revealed remarkably consistent characteristics between its thermal (blue, slowly evolving) component and the class of luminous featureless TDEs from the Ref.\cite{Hammerstein2022arXiv} sample.
First, broad emission or absorption features are not observed in any optical or near infrared spectra of \at\ (Fig.\,\ref{fig:spectra}), neither during the rapidly evolving initial flare nor in the late-time blue plateau. This is consistent with observations of the jetted TDE candidate Swift J2058+05, whose (low signal-to-noise ratio) optical spectra were dominated by a blue, featureless continuum\cite{Pasham2015}.

Second, taking the time when the thermal component began to dominate in the optical ($\sim 12$ days from the first detection) with luminosity $M_r \sim -22.2$\,mag (correspond to the rest-frame UV band), the luminosity of \at\ falls near to the observed peak luminosities of featureless TDEs that were found to be consistently brighter than TDEs with features\cite{Hammerstein2022arXiv} (\ref{fig:featureless_histo}).

We therefore suggest that a connection likely exists between TDEs that generate relativistic jets and the class of luminous featureless TDEs. Deep radio observations of these transients will be able to probe the presence of a jet at all viewing angles. This connection between jetted TDEs and luminous featureless TDEs, if confirmed, will enable new studies of jet formation in TDEs and system geometry. Understanding these particularly luminous transients may be the only way to map the rate of TDEs as a function of redshift beyond $z\sim0.4$, which represents the approximate limit for spectroscopic classification of $M \sim -20$\,mag transients with large 8-m class optical telescope.

\section{Observations and Data Processing}

\subsection{Palomar 48-inch Samuel Oschin Telescope}
\at\ was discovered using data acquired by the ZTF camera on the 48-inch Samuel Oschin Telescope at Palomar Observatory.  Observations of \at\ were conducted as part of the ZTF public survey, the Caltech high-cadence survey, and the Partnership extragalactic survey\cite{Bellm2019scheduler, Dekany2020PASP}.  The images were processed in real-time through the ZTF reduction and image subtraction pipelines\cite{Masci2019PASP} at the Infrared Processing and Analysis Center (IPAC). PSF forced photometry was obtained via the ZTF forced-photometry service\cite{Masci2019PASP} at IPAC.

\subsection{Liverpool Telescope}
\label{Methods: LT}

Imaging of AT2022cmc using the IO:O camera on the 2m robotic Liverpool Telescope\cite{Steele2004} (LT) was obtained on several occasions beginning from 2022 February 15.  Observations were conducted in SDSS $g$, $r$, $i$, and $z$ filters.  We downloaded images reduced using the standard LT pipeline, and performed our own astrometric alignment and stacking.  Exposures showing major tracking errors or poor transparency due to cloud cover were discarded.  Many exposures suffered from a failure of the IO:O shutter to close at the end of the observation, producing readout streaks across the detector, but the region around the transient was free of contamination and no discernable impact on the quality of relative photometry of nearby stars was observed, so these exposures were retained.   Photometry of the transient was measured with a custom IDL routine using seeing-matched aperture photometry fixed at the transient location, and calibrated relative to a set of SDSS secondary standard stars in the field.  Photometry is presented in \ref{table:photometry}.

\subsection{Hubble Space Telescope}

The location of \at\ was observed with the Hubble Space Telescope (HST) beginning at 2022 March 8 20:12:21 UT ($\sim 25.4$\,d after discovery). The field was imaged with the Wide-Field Camera 3 (WFC3) in the F606W (UVIS) and F160W (IR) filters for 1044\,s and 1059\,s, respectively.
\at\ was well detected in both bands. We measured AB magnitudes of $F606W = 21.82 \pm 0.03$\,mag and $F160W = 22.64 \pm 0.05$\,mag (Fig.\,\ref{fig:lc_optical}; \ref{table:photometry}). 
The source appeared unresolved, without obvious evidence for extended emission directly underneath.

Based on astrometry of the WFC3 images, the coordinates of \at\ could be placed at J2000 right ascension $\alpha$ =
$13^{\rm{h}}34^{\rm{m}}43^{\rm{s}}.201$
and declination $\delta$ = $+33^{\circ} 13' 00''.648$ (see also Methods section~\ref{Methods:VLA}).

\subsection{Very Large Array}
\label{Methods:VLA}

AT2022cmc was observed with the Karl G.\ Jansky Very Large Array (VLA\cite{Perley2011}) on nine occasions between 2022 February 15 and 2022 March 31 under program 2022A-405 (PI: D.\ Perley).  All visits included an integration using the X-band receivers (8--12 GHz); several of them additionally involved observations using other receivers: typically Ku (12--18 GHz) and Ka (30.5--38.5 GHz), although on 2022-02-18 complete frequency coverage from 5--48 GHz using the C, X, Ku, K, Ka, and Q bands was obtained).  All observations used the 3-bit samplers and full polarization.  The target source is within three degrees of the standard calibrator 3C286 (J1331+3030); this source was used as the phase calibrator as well as the flux and bandpass calibrator for all observations.  Most observations were taken during the reconfiguration of the array from BnA to A.

Data were reduced using standard synthesis imaging techniques using the Astronomical Image Processing System (AIPS).  Because the observations were taken with the VLA in a high resolution configuration, standard models of 3C286 were required to derive the antenna delays, bandpasses, and gains.  RFI was removed by flagging amplitudes higher than about 5-sigma.  Less than 1\% of the data were removed by this editing.  

Due to the long baselines, atmospheric phase instabilities can cause significant decorrelation in the image at high frequencies.  Fortunately, the source was bright enough that phase self-calibration could be utilized to remove the atmospheric phase.  This was done by coherently summing over both polarizations and 16 spectral windows (2.048\,GHz bandwidth) for 24 seconds, providing enough SNR to enable a phase solution using a point-source model at the known location of the transient.  This process was required only for the K, Ka, and Q-band observations.  Flux densities were determined with the AIPS task JMFIT and are reported in \ref{table:radio} and \ref{fig:radio_SED}.

The A-configuration Ku observations provide a highly accurate measurement of the source location: 
standard equinox J2000 right ascension $\alpha$ =
$13^{\rm{h}}34^{\rm{m}}43^{\rm{s}}.20232$
and declination $\delta$ = $+33^{\circ} 13' 00''.6565$ 
(uncertainty 0.01$''$).

\subsection{Submillimeter Array}
\label{sec: SMA}

\at\ was regularly observed with the Submillimeter Array (SMA) under standard observing time (project 2021B-S013; PI: Ho) with follow-up observations under Director Discretionary Time/Target of opportunity program (project 2021B-S071; PI: Ho), in the Compact and Extended configurations. Observations were taken during a period of engineering shut-down, so the number of antennas available ranged from 3 to 6, and cover a range of baseline lengths from 16.4\,m to 181.6\,m. The quasars 1310+323 and 1159+292 were used as primary phase and amplitude gain calibrators, respectively, with absolute flux calibration performed by nightly comparison to Ceres or (maser-free) continuum observations of the emission-line star MWC349a. The quasars 1159+292 and/or 3C279 were used for bandpass calibration. Data were calibrated in IDL using the MIR package. Additional analysis and imaging were performed using the MIRIAD package. Given that the target was a point source, and often only 3 antennas were available, fluxes were derived directly from the calibrated visibilities, but the results agree well with flux estimates derived from the dirty and CLEANed images when the data quality and UV-coverage was adequate.

SMA results are summarized in \ref{table:radio}.

\subsection{NOrthern Extended Millimeter Array (NOEMA)}
\label{sec:NOEMA}

We obtained seven epochs of observations of \at\ with NOEMA in extended 11 and 12 antenna A configuration spanning Band 1 (100\,GHz), Band 2 (150\,GHz), and Band 3 (230\,GHz) under the target-of-opportunity program W21BK (PI: Ho); this program is still in progress. The primary flux calibrators were MWC349 and LKHA101, and the time-dependent phase and amplitude calibrators were the quasars J1310+323 and 1315+346. The data reduction was done with the CLIC software (GILDAS package\cite{GILDAS}). 
Dual-polarization UV tables were written for each of the receiver
sidebands. The resulting calibrated UV tables were analyzed in the
MAPPING software (also from the GILDAS package) and point-source UV
plane fits were performed.

NOEMA results are also summarized in \ref{table:radio}.

\subsection{JCMT SCUBA-2 Sub-millimeter Observations}
\label{Methods:JCMT}

Sub-millimeter observations of \at\ were performed simultaneously at 850~{$\mu$}m (350~GHz) and 450~{$\mu$}m (670~GHz) on two nights using the Submillimetre Common-User Bolometer Array 2 (SCUBA-2) continuum camera\cite{jcmt_Holland2013} on the James Clerk Maxwell Telescope
(JCMT) on Mauna Kea, Hawaii.
The SCUBA-2 data were analyzed in the standard manner using the 2021A version of {\tt Starlink}\cite{jcmt_Currie2014}; this used Version 1.7.0 of {\tt SMURF}\cite{jcmt_Chapin2013} and Version 2.6-12 of {\tt KAPPA}. Observations of the SCUBA-2 calibrator Arp~220 on both nights did not show any anomalous behaviours, so the current standard flux conversion factors were used for the flux normalization \cite{jcmt_Mairs2021}.
In the SCUBA-2 Dynamic Interactive Map-Maker, the Blank Field map was
used for the AT2022cmc observations.
The maps were smoothed using a matched filter.
The root mean square background noise was determined in the central 2\arcmin\ of the
map with the source excluded.

The SCUBA-2 observations of AT2022cmc are summarized in
\ref{table:radio}.
These expand on the preliminary results given in Ref.~\cite{jcmt_Smith2022}.
There was a marginal detection of AT2022cmc at 850~{$\mu$}m on both  nights.
This becomes more significant when all the data are combined, giving an 850~{$\mu$}m flux density of $4.9 \pm 1.3$~mJy/beam at a mid-point
of UT 2022 February 21.510.

\at\ was not detected at 450~{$\mu$}m in the individual
night observations or in the combined data; the root mean square measurement for the combined data is $10.5$~mJy/beam at a mid-point of 2022 February 21 12:14 UTC.

JCMT results are also summarized in \ref{table:radio}.

\subsection{upgraded Giant Metrewave Radio Telescope}
\label{sec: GMRT}

The event \at\ was observed with the upgraded Giant Metrewave Radio Telescope (uGMRT) starting 2022 March 13 until 2022 March 26. The observations were taken in uGMRT  band 5 (1000--1450\,MHz) band 4 (550--900\,MHz) and band 3 (250--500\,MHz). The observations were two hours in duration including overheads using a bandwidth of 400 MHz for bands 4 and 5, and of 3 hours in band 3 using a bandwidth of 200 MHz. 3C 286 was used as flux, bandpass and phase calibrator due to its proximity with the event. The Common Astronomy Software Applications (CASA\cite{McMullin2007}) was used for analysing the data. The data were analyzed in three major steps, i.e flagging, calibration and imaging using the procedure laid out in (\cite{Maity2021}).

The source was not detected in any of the bands (\ref{table:radio}), consistent with the expected optically thick evolution at sub-GHz frequencies based on the higher frequency radio data with the VLA. uGMRT results are also summarized in \ref{table:radio}.

\subsection{GROWTH-India Telescope}
The 0.7m GROWTH-India Telescope (GIT), located at the Indian Astronomical Observatory (IAO), Hanle-Ladakh, started observing \at{} at 19:30:26.78 UT on 2022 February 15. The data were acquired in SDSS $g^\prime$, $r^\prime$ and $i^\prime$ bands with multiple 300~sec exposures. Data were downloaded in real time to our data processing unit at IIT Bombay. After a preliminary bias correction and flat fielding, and cosmic-rays removal with the Astro-SCRAPPY\cite{2019ascl.soft07032M} package, all images acquired on the same night were stacked making use of SWarp\cite{2010ascl.soft10068B}. The pipeline performs PSF photometry to obtain the instrumental magnitudes using standard techniques. These magnitudes were calibrated against the PanSTARRS DR1 catalogue\cite{2016arXiv161205560C} by correcting for zero points. Reported photometric uncertainties (\ref{table:photometry}) are $1\sigma$ values.

\subsection{Blanco Telescope}
We conducted photometric observations of \at\ using the Dark Energy Camera (DECam\cite{Flaugher2015}) optical imager mounted at the prime focus of the Blanco telescope at Cerro Tololo Inter-American Observatory (program ID 2022A-679480, PI: Zhang; program ID 2021B-0325, PI: Rest). After standard calibration (bias correction, flat-fielding, and astrometric alignment) was done by the NSF NOIRLab DECam Community Pipeline\cite{Valdes2014}, difference image photometry was obtained using the \texttt{Photpipe} pipeline\cite{Rest2014}.  Data are presented in \ref{table:photometry}.

\subsection{Nordic Optical Telescope}
We obtained a series of $gri$ photometry with the Alhambra Faint Object Spectrograph and Camera (ALFOSC\cite{ALFOSC}) on the 2.56\,m Nordic Optical Telescope (NOT) at the Observatorio del Roque de los Muchachos on La Palma (Spain) (Program ID: 64-501). The data were reduced with PyNOT\cite{PyNOT} that uses standard routines for imaging data. We used aperture photometry to measure the brightness of the transient. Once an instrumental magnitude was established, it was calibrated against the brightness of several stars from a cross-matched SDSS catalogue. Data are presented in \ref{table:photometry}.

The NOT spectrum in Fig.\,\ref{fig:spectra} was obtained with ALFOSC using Grism 4 which covers 3200 -- 9600 $\AA$ at resolution R=360 and was reduced with PypeIt\cite{PypeIt}.

\subsection{Palomar 60-inch telescope}
Photometry was also obtained on the robotic 
Palomar 60-inch telescope (P60; \cite{2006PASP..118.1396C}) equipped with the 
Spectral Energy Distribution Machine (SEDM\cite{2018PASP..130c5003B, Rigault2019}). 
Photometry was produced with an image-subtraction pipeline\cite{Fremling2016}, with template images from the Sloan Digital Sky Survey (SDSS\cite{2014ApJS..211...17A}). This pipeline produces PSF magnitudes, calibrated against SDSS stars in the field. Data are presented in \ref{table:photometry}.

\subsection{Palomar 200-inch telescope}
We obtained one epoch of near-infrared observations from the Wide Infrared Camera on the Palomar 200 in telescope. On 2022 March 12 we performed a set of 18 dithered exposures of 45\,s each in the J\,band (1.25$\mu$m). We use standard optical reduction techniques in Python to reduce and co-add the images, using 2MASS point source catalog for photometric calibration. We measure aperture photometry using photutils.  Data are presented in  \ref{table:photometry}.

\subsection{Asteroid Terrestrial-impact Last Alert System}
We obtained broad-band ``orange" (5560\,\AA--8260\,\AA) and ``cyan'' (4120\,\AA--6570\,\AA) light curves from the ATLAS\cite{Tonry2018_ATLAS} survey. This data is publicly available through the ATLAS Transient Science Server\cite{Smith2020_ATLAS}. Detections of \at\ were obtained only in the orange filter.

\subsection{Very Large Telescope}
\label{Methods:Xshooter}

The X-shooter spectrograph\cite{Vernet2011} installed on the European Southern Observatory (ESO) Very Large Telescope at Paranal Observatory (Chile) observed \at\ on 2022 February 17 via program 106.21T6 (PI: Tanvir). The observations consisted of $4\times1200$\,s in the UVB and VIS arms and $8\times600$ s in the near-infrared arm, using an ABBA nodding pattern. We used a 1$^{\prime\prime}$ slit in the UVB arm a 0.9$^{\prime\prime}$ in the VIS and the 0.9$^{\prime\prime}$ JHslit in the near infrared, designed to block part of the $K$-band spectrum to reduce the noise in the $J$ and $H$ bands. The resulting spectral coverage goes from 3000\,{\AA} to 21000\,{\AA}. The data reduction was performed using the X-shooter pipeline\cite{Modigliani2010} and additional scripts developed within the \textit{Stargate} collaboration\cite{Selsing2019}.  The spectrum is shown in Figure~\ref{fig:spectra}.

\subsection{Gran Telescopio Canarias}

Near Infrared observations were performed using EMIR (Espectr\'ografo Multiobjeto Infra-Rojo\cite{Garzon2006}) on three different epochs using programs GTCMULTIPLE2H-21B (PI: de Ugarte Postigo) and GTCMULTIPLE2H-22A (PI: Th\"one). The data reduction was performed using a self designed pipeline based on shell scripts and IRAF procedures, which includes flat fielding, background correction, bad pixel masking, fine alignment and combination of images. Relative photometry was performed using multiple field stars from the UKIRT photometric catalog.  Data are presented in \ref{table:photometry}.

\subsection{Calar Alto}

We obtained observations of ZTF22aaajecp/\at\ under program 22A-2.2-019 (PI: Kann) on 2022 February 18 from 04:54:04 - 06:04:15 UT with CAFOS (Calar Alto Faint Object Spectrograph) mounted on the 2.2m telescope at the Centro Astron\'omico Hispano-Alem\'an (CAHA), Almeria, Spain\cite{2022GCN.31626....1K}. Observing conditions were good but images were influenced by the bright Moon. Twelve images of 120 s integration time each were taken in the SDSS $r^\prime$ and $i^\prime$ bands. We reduced the images following standard procedures in IRAF (bias subtraction, flatfielding, sky subtraction, shifting \& co-adding). The source is well-detected in both bands in each stacked image. Photometry was performed with respect to field stars from the SDSS photometric catalog. Data are presented in \ref{table:photometry}.

\subsection{W.~M. Keck Observatory}
\label{sec: Keck}
Spectroscopy was obtained with the DEIMOS (DEep Imaging Multi-Object Spectrograph) and LRIS (Low Resolution Imaging Spectrometer) instruments at the W.~M.~Keck Observatory.

DEIMOS spectroscopy covered the wavelength range 5250-8780\,\AA, consisting of 3 exposures of 900\,s each starting at 2022 February 17 15:07 UTC.  DEIMOS data were reduced using the PypeIt\cite{Prochaska2020PypeIt} data reduction pipeline.
The spectrum had low S/N, but the transient was detected over the full wavelength range\cite{Lundquist2022GCN.31612DEIMOS}.  Absorption lines, first identified with VLT/X-shooter, were found and interpreted at 6132\,\AA, 6148\,\AA\ as the Mg II 2796\,\AA, 2803\,\AA\ lines and the absorption lines at 5671\,\AA, 5702\,\AA as the Fe II 2586\,\AA, 2600\,\AA, agreeing with the redshift of 1.193\cite{Tanvir2022GCN.31602.VLTredshift}.  The Ca II 3934\,\AA\ line was detected at 8629\,\AA, however the Ca II 3969\,\AA\ was not identifiable at 8706\,\AA.

Two LRIS spectra were obtained starting on 2022 February 25 14:28:28 and 2022 March 03 11:24:06.19 UTC. The data were reduced using the LPipe\cite{Perley2019PASP} pipeline. The host galaxy lines identified in the X-shooter and DEIMOS spectra could be recognized, but the LRIS spectra appeared to be otherwise featureless.

All spectra are shown in Figure~\ref{fig:spectra}.

\subsection{Gemini Observatory}
We acquired spectroscopic data of AT2022cmc using the long slit mode of the Gemini Multi-Object Spectrographs (GMOS) mounted at the Gemini-North 8-meter telescope in Mauna Kea on the island of Hawaii, under the program GN-2022A-Q-127 (PI: Ho). 

We used both the R400 and B600 gratings, and we obtained $2 \times 450$\,s exposures on each grating. We used the $1''$
slit, starting at 2022 February 15 14:35 UTC. After two independent reductions using DRAGONS (Data Reduction for Astronomy from Gemini Observatory North and South) \cite{dragons} and Pyraf we report a featureless red
continuum throughout our effective spectral coverage, from 3800\,\AA\ to 9100\,\AA\, although the S/N is low blueward of 5500\,\AA \cite{GCN31595}. The spectrum is shown in Figure~\ref{fig:spectra}.

\subsection{Neil Gehrels Swift Observatory}
\label{Methods:Swift}

\at\ was observed by the X--ray Telescope (XRT\cite{Burrows2005}) and the Ultra-Violet/Optical Telescope (UVOT\cite{Roming2005}) on board the Neil Gehrels Swift Observatory under a series of time-of-opportunity (ToO) requests starting on 2022 February 23 03:25:55 UTC. The {Swift} follow-up campaign began later than expected because of a few weeks of emergency downtime of the observatory.

All XRT observations were obtained in the photon-counting mode. 
First, we ran \texttt{ximage} to determine the position of \at\ in each observation. 
To calculate the background-subtracted count rates, we filtered the cleaned event files using a source region with $r_{\rm src} = 30^{\prime\prime}$, and eight background regions with $r_{\rm bkg} = 25^{\prime\prime}$ evenly spaced at $80^{\prime\prime}$ from \at. A log of XRT observations is given in \ref{table:xrt}.

For observations where the XRT net counts are greater than 100, we grouped the spectra to have at least one count per bin, and modeled the 0.3--10\,keV data with an absorbed power-law model, \texttt{tbabs*ztbabs*powerlaw}.
All data were fitted using $C$-statistics via \texttt{cstat}\cite{Cash1979}. We do not find strong evidence of spectral evolution throughout the first seven XRT observations (see \ref{fig:XRT_Gamma_evol}). Assuming $\Gamma=1.53$ and a host galaxy $N_{\rm H} = 1.1\times 10^{21}\,{\rm cm^{-2}}$ (Methods section \ref{Methods:NICER}), the XRT 0.3--10\,keV count rate (in $\rm count\,s^{-1}$) to flux (in $\rm erg\,cm^{-2}\,s^{-1}$) conversion factor is $4.19\times 10^{-11}$.

The first seven UVOT epochs (obsIDs 15023001--15023007) were conducted with $UBV$+All UV filters. Subsequent observations were  conducted with $U$+All UV filters. We measured the UVOT photometry using the \texttt{uvotsource} tool. We used a circular source region with $r_{\rm src} = 5^{\prime\prime}$, and corrected for the enclosed energy within the aperture. We measured the background using four nearby circular source-free regions with $r_{\rm bkg} = 10^{\prime\prime}$. The UVOT photometry is presented in \ref{table:photometry}.

\subsection{Neutron Star Interior Composition Explorer}
\label{Methods:NICER}

\at\ was observed by the Neutron Star Interior Composition Explorer (NICER\cite{Gendreau2016}) under director's discretionary time (DDT) and ToO programs. The NICER observations will be reported in detail by Pasham et al., in press\footnote{\url{http://www.nature.com/articles/s41550-022-01820-x}}. Here we only analyzed the first NICER good time interval obtained on 2022 Feburary 16.% from 19:05:20 to 22:26:09.

We processed the NICER data using \texttt{heasoft} v6.29c. 
We ran \texttt{nicerl2} to obtain the cleaned and screened event files. 
We removed hot detectors. 
Background was computed using the \texttt{nibackgen3C50} tool \cite{Remillard2022} with hbgcut=0.05 and s0cut=2.0. 
Response files were generated with \texttt{nicerarf} and \texttt{nicerrmf}.
The spectrum was rebinned using \texttt{ftgrouppha} with grouptype=optmin and groupscale=50. 
We added systematic errors of 1\% using \texttt{grppha}.

The final spectrum has an effective exposure time of 1560\,s, and the source is above background at 0.25--8\,keV.
We fitted the 0.25--8\,keV data using an absorbed power-law model, \texttt{tbabs*ztbabs*powerlaw} and $\chi^2$-statistics. 
The Galactic column density $N_{\rm H}$ was fixed at $8.88\times 10^{19}\,{\rm cm^{-2}}$ \cite{HI4PI2016}. 
We obtained a good fit with a $\chi^2$/degrees of freedom ($\chi^2$/dof) of 74.91/83. 
The best-fit power-law index is $\Gamma_X = 1.53 \pm 0.03$, 
and host galaxy absorption is $N_{\rm H} = 1.09^{+0.14}_{-0.13} \times 10^{21}\,{\rm cm^{-2}}$. 
The observed 0.25--8\,keV flux is $(3.29 \pm 0.07) \times 10^{-11}\,{\rm erg\,s^{-1}\,cm^{-2}}$.
The inferred absorbed 0.3--10\,keV flux is $(3.75 \pm 0.09) \times  10^{-11}\,{\rm erg\,s^{-1}\,cm^{-2}}$.
Errors are 90\% confidence level for one parameter of interest. 
The data and best-fit model are shown in Fig.\ref{fig:sed_X}a and Fig.\ref{fig:sed_X}d.

\vspace{1in}
\noindent{\bf References}
%\bibliographystyleNew{naturemag}
%\bibliographyNew{references}
%\bibliographystyle{naturemag}
%\bibliography{references, gcns}

\begin{thebibliography}{100}
\expandafter\ifx\csname url\endcsname\relax
  \def\url#1{\texttt{#1}}\fi
\expandafter\ifx\csname urlprefix\endcsname\relax\def\urlprefix{URL }\fi
\providecommand{\bibinfo}[2]{#2}
\providecommand{\eprint}[2][]{\url{#2}}

\bibitem{Rees1988}
\bibinfo{author}{{Rees}, M.~J.}
\newblock \bibinfo{title}{{Tidal disruption of stars by black holes of
  {}10$^{6}$-{}10$^{8}$ solar masses in nearby galaxies}}.
\newblock \emph{\bibinfo{journal}{\nat}} \textbf{\bibinfo{volume}{333}},
  \bibinfo{pages}{523--528} (\bibinfo{year}{1988}).

\bibitem{Bloom2011Sci}
\bibinfo{author}{{Bloom}, J.~S.} \emph{et~al.}
\newblock \bibinfo{title}{{A Possible Relativistic Jetted Outburst from a
  Massive Black Hole Fed by a Tidally Disrupted Star}}.
\newblock \emph{\bibinfo{journal}{Science}} \textbf{\bibinfo{volume}{333}},
  \bibinfo{pages}{203} (\bibinfo{year}{2011}).
%\newblock \eprint{1104.3257}.

\bibitem{Burrows2011Nat}
\bibinfo{author}{{Burrows}, D.~N.} \emph{et~al.}
\newblock \bibinfo{title}{{Relativistic jet activity from the tidal disruption
  of a star by a massive black hole}}.
\newblock \emph{\bibinfo{journal}{\nat}} \textbf{\bibinfo{volume}{476}},
  \bibinfo{pages}{421--424} (\bibinfo{year}{2011}).
%\newblock \eprint{1104.4787}.

\bibitem{Levan2011Sci}
\bibinfo{author}{{Levan}, A.~J.} \emph{et~al.}
\newblock \bibinfo{title}{{An Extremely Luminous Panchromatic Outburst from the
  Nucleus of a Distant Galaxy}}.
\newblock \emph{\bibinfo{journal}{Science}} \textbf{\bibinfo{volume}{333}},
  \bibinfo{pages}{199} (\bibinfo{year}{2011}).
%\newblock \eprint{1104.3356}.

\bibitem{Zauderer2011Nat}
\bibinfo{author}{{Zauderer}, B.~A.} \emph{et~al.}
\newblock \bibinfo{title}{{Birth of a relativistic outflow in the unusual
  gamma-ray transient Swift J164449.3+573451}}.
\newblock \emph{\bibinfo{journal}{\nat}} \textbf{\bibinfo{volume}{476}},
  \bibinfo{pages}{425--428} (\bibinfo{year}{2011}).
%\newblock \eprint{1106.3568}.

\bibitem{Cenko2012}
\bibinfo{author}{{Cenko}, S.~B.} \emph{et~al.}
\newblock \bibinfo{title}{{Swift J2058.4+0516: Discovery of a Possible Second
  Relativistic Tidal Disruption Flare?}}
\newblock \emph{\bibinfo{journal}{\apj}} \textbf{\bibinfo{volume}{753}},
  \bibinfo{pages}{77} (\bibinfo{year}{2012}).
%\newblock \eprint{1107.5307}.

\bibitem{Brown2015}
\bibinfo{author}{{Brown}, G.~C.} \emph{et~al.}
\newblock \bibinfo{title}{{Swift J1112.2-8238: a candidate relativistic tidal
  disruption flare}}.
\newblock \emph{\bibinfo{journal}{\mnras}} \textbf{\bibinfo{volume}{452}},
  \bibinfo{pages}{4297--4306} (\bibinfo{year}{2015}).
%\newblock \eprint{1507.03582}.

\bibitem{Pasham2015}
\bibinfo{author}{{Pasham}, D.~R.} \emph{et~al.}
\newblock \bibinfo{title}{{A Multiwavelength Study of the Relativistic Tidal
  Disruption Candidate Swift J2058.4+0516 at Late Times}}.
\newblock \emph{\bibinfo{journal}{\apj}} \textbf{\bibinfo{volume}{805}},
  \bibinfo{pages}{68} (\bibinfo{year}{2015}).
%\newblock \eprint{1502.01345}.

\bibitem{Yuan2016}
\bibinfo{author}{{Yuan}, Q.}, \bibinfo{author}{{Wang}, Q.~D.},
  \bibinfo{author}{{Lei}, W.-H.}, \bibinfo{author}{{Gao}, H.} \&
  \bibinfo{author}{{Zhang}, B.}
\newblock \bibinfo{title}{{Catching jetted tidal disruption events early in
  millimetre}}.
\newblock \emph{\bibinfo{journal}{\mnras}} \textbf{\bibinfo{volume}{461}},
  \bibinfo{pages}{3375--3384} (\bibinfo{year}{2016}).
%\newblock \eprint{1606.06830}.

\bibitem{Graham2019PASP}
\bibinfo{author}{{Graham}, M.~J.} \emph{et~al.}
\newblock \bibinfo{title}{{The Zwicky Transient Facility: Science Objectives}}.
\newblock \emph{\bibinfo{journal}{\pasp}} \textbf{\bibinfo{volume}{131}},
  \bibinfo{pages}{078001} (\bibinfo{year}{2019}).
%\newblock \eprint{1902.01945}.

\bibitem{Sun2015ApJ}
\bibinfo{author}{{Sun}, H.}, \bibinfo{author}{{Zhang}, B.} \&
  \bibinfo{author}{{Li}, Z.}
\newblock \bibinfo{title}{{Extragalactic High-energy Transients: Event Rate
  Densities and Luminosity Functions}}.
\newblock \emph{\bibinfo{journal}{\apj}} \textbf{\bibinfo{volume}{812}},
  \bibinfo{pages}{33} (\bibinfo{year}{2015}).
%\newblock \eprint{1509.01592}.

\bibitem{AndreoniCough2021ztfrest}
\bibinfo{author}{{Andreoni}, I.} \emph{et~al.}
\newblock \bibinfo{title}{{Fast-transient Searches in Real Time with ZTFReST:
  Identification of Three Optically Discovered Gamma-Ray Burst Afterglows and
  New Constraints on the Kilonova Rate}}.
\newblock \emph{\bibinfo{journal}{\apj}} \textbf{\bibinfo{volume}{918}},
  \bibinfo{pages}{63} (\bibinfo{year}{2021}).
%\newblock \eprint{2104.06352}.

\bibitem{Pasham2022GCN.31601NICERdetection}
\bibinfo{author}{{Pasham}, D.}, \bibinfo{author}{{Gendreau}, K.},
  \bibinfo{author}{{Arzoumanian}, Z.} \& \bibinfo{author}{{Cenko}, B.}
\newblock \bibinfo{title}{{ZTF22aaajecp/AT2022cmc: NICER X-ray detection}}.
\newblock \emph{\bibinfo{journal}{GRB Coordinates Network}}
  \textbf{\bibinfo{volume}{31601}}, \bibinfo{pages}{1} (\bibinfo{year}{2022}).

\bibitem{Perley2022GCN.31592.VLAdetection}
\bibinfo{author}{{Perley}, D.~A.}
\newblock \bibinfo{title}{{ZTF22aaajecp/AT2022cmc: VLA radio detection}}.
\newblock \emph{\bibinfo{journal}{GRB Coordinates Network}}
  \textbf{\bibinfo{volume}{31592}}, \bibinfo{pages}{1} (\bibinfo{year}{2022}).

\bibitem{Perley2022GCN.31627.SMAdetection}
\bibinfo{author}{{Perley}, D.~A.}, \bibinfo{author}{{Ho}, A.~Y.~Q.},
  \bibinfo{author}{{Petitpas}, G.} \& \bibinfo{author}{{Keating}, G.}
\newblock \bibinfo{title}{{ZTF22aaajecb/AT2022cmc: Submillimeter Array
  detection}}.
\newblock \emph{\bibinfo{journal}{GRB Coordinates Network}}
  \textbf{\bibinfo{volume}{31627}}, \bibinfo{pages}{1} (\bibinfo{year}{2022}).

\bibitem{Planck2018}
\bibinfo{author}{{Planck Collaboration}} \emph{et~al.}
\newblock \bibinfo{title}{{Planck 2018 results. VI. Cosmological parameters}}.
\newblock \emph{\bibinfo{journal}{\aap}} \textbf{\bibinfo{volume}{641}},
  \bibinfo{pages}{A6} (\bibinfo{year}{2020}).
%\newblock \eprint{1807.06209}.

\bibitem{Tanvir2022GCN.31602.VLTredshift}
\bibinfo{author}{{Tanvir}, N.~R.} \emph{et~al.}
\newblock \bibinfo{title}{{ZTF22aaajecp/AT2022cmc: VLT/X-shooter redshift}}.
\newblock \emph{\bibinfo{journal}{GRB Coordinates Network}}
  \textbf{\bibinfo{volume}{31602}}, \bibinfo{pages}{1} (\bibinfo{year}{2022}).

\bibitem{GalYam2017hsn}
\bibinfo{author}{{Gal-Yam}, A.}
\newblock \bibinfo{title}{{Observational and Physical Classification of
  Supernovae}}.
\newblock In \bibinfo{editor}{{Alsabti}, A.~W.} \& \bibinfo{editor}{{Murdin},
  P.} (eds.) \emph{\bibinfo{booktitle}{Handbook of Supernovae}},
  \bibinfo{pages}{195} (\bibinfo{year}{2017}).

\bibitem{McConnell2013}
\bibinfo{author}{{McConnell}, N.~J.} \& \bibinfo{author}{{Ma}, C.-P.}
\newblock \bibinfo{title}{{Revisiting the Scaling Relations of Black Hole
  Masses and Host Galaxy Properties}}.
\newblock \emph{\bibinfo{journal}{\apj}} \textbf{\bibinfo{volume}{764}},
  \bibinfo{pages}{184} (\bibinfo{year}{2013}).
%\newblock \eprint{1211.2816}.

\bibitem{lu20_self_intersection}
\bibinfo{author}{{Lu}, W.} \& \bibinfo{author}{{Bonnerot}, C.}
\newblock \bibinfo{title}{{Self-intersection of the fallback stream in tidal
  disruption events}}.
\newblock \emph{\bibinfo{journal}{\mnras}} \textbf{\bibinfo{volume}{492}},
  \bibinfo{pages}{686--707} (\bibinfo{year}{2020}).
%\newblock \eprint{1904.12018}.

\bibitem{blandford77_BZ_jet}
\bibinfo{author}{{Blandford}, R.~D.} \& \bibinfo{author}{{Znajek}, R.~L.}
\newblock \bibinfo{title}{{Electromagnetic extraction of energy from Kerr black
  holes.}}
\newblock \emph{\bibinfo{journal}{\mnras}} \textbf{\bibinfo{volume}{179}},
  \bibinfo{pages}{433--456} (\bibinfo{year}{1977}).

\bibitem{Pasham2022ATel15232variability}
\bibinfo{author}{{Pasham}, D.} \emph{et~al.}
\newblock \bibinfo{title}{{High-cadence NICER X-ray observations of
  AT2022cmc/ZTF22aaajecpc: flux variability and spectral evolution suggest it
  could be a relativistic tidal disruption event}}.
\newblock \emph{\bibinfo{journal}{The Astronomer's Telegram}}
  \textbf{\bibinfo{volume}{15232}}, \bibinfo{pages}{1} (\bibinfo{year}{2022}).

\bibitem{Yao2022ATel15230NuSTAR}
\bibinfo{author}{{Yao}, Y.}, \bibinfo{author}{{Pasham}, D.~R.} \&
  \bibinfo{author}{{Gendreau}, K.~C.}
\newblock \bibinfo{title}{{NuSTAR observation of AT2022cmc, and joint spectral
  fitting with NICER}}.
\newblock \emph{\bibinfo{journal}{The Astronomer's Telegram}}
  \textbf{\bibinfo{volume}{15230}}, \bibinfo{pages}{1} (\bibinfo{year}{2022}).

\bibitem{Tchekhovskoy2014}
\bibinfo{author}{{Tchekhovskoy}, A.}, \bibinfo{author}{{Metzger}, B.~D.},
  \bibinfo{author}{{Giannios}, D.} \& \bibinfo{author}{{Kelley}, L.~Z.}
\newblock \bibinfo{title}{{Swift J1644+57 gone MAD: the case for dynamically
  important magnetic flux threading the black hole in a jetted tidal disruption
  event}}.
\newblock \emph{\bibinfo{journal}{\mnras}} \textbf{\bibinfo{volume}{437}},
  \bibinfo{pages}{2744--2760} (\bibinfo{year}{2014}).
%\newblock \eprint{1301.1982}.

\bibitem{kumar15_GRB_review}
\bibinfo{author}{{Kumar}, P.} \& \bibinfo{author}{{Zhang}, B.}
\newblock \bibinfo{title}{{The physics of gamma-ray bursts \& relativistic
  jets}}.
\newblock \emph{\bibinfo{journal}{\physrep}} \textbf{\bibinfo{volume}{561}},
  \bibinfo{pages}{1--109} (\bibinfo{year}{2015}).
%\newblock \eprint{1410.0679}.

\bibitem{Dai18_viewing_angle}
\bibinfo{author}{{Dai}, L.}, \bibinfo{author}{{McKinney}, J.~C.},
  \bibinfo{author}{{Roth}, N.}, \bibinfo{author}{{Ramirez-Ruiz}, E.} \&
  \bibinfo{author}{{Miller}, M.~C.}
\newblock \bibinfo{title}{{A Unified Model for Tidal Disruption Events}}.
\newblock \emph{\bibinfo{journal}{\apjl}} \textbf{\bibinfo{volume}{859}},
  \bibinfo{pages}{L20} (\bibinfo{year}{2018}).
%\newblock \eprint{1803.03265}.

%\bibitem{stone13_frozen_in}
%\bibinfo{author}{{Stone}, N.}, \bibinfo{author}{{Sari}, R.} \&
%  \bibinfo{author}{{Loeb}, A.}
%\newblock \bibinfo{title}{{Consequences of strong %compression in tidal  disruption events}}.
%\newblock \emph{\bibinfo{journal}{\mnras}} \textbf{\bibinfo{volume}{435}},
  %\bibinfo{pages}{1809--1824}
  %(\bibinfo{year}{2013}).
%\newblock \eprint{1210.3374}.

\bibitem{bonnerot21_disk_formation}
\bibinfo{author}{{Bonnerot}, C.}, \bibinfo{author}{{Lu}, W.} \&
  \bibinfo{author}{{Hopkins}, P.~F.}
\newblock \bibinfo{title}{{First light from tidal disruption events}}.
\newblock \emph{\bibinfo{journal}{\mnras}} \textbf{\bibinfo{volume}{504}},
  \bibinfo{pages}{4885--4905} (\bibinfo{year}{2021}).
%\newblock \eprint{2012.12271}.

\bibitem{Mattila2018Sci}
\bibinfo{author}{{Mattila}, S.} \emph{et~al.}
\newblock \bibinfo{title}{{A dust-enshrouded tidal disruption event with a
  resolved radio jet in a galaxy merger}}.
\newblock \emph{\bibinfo{journal}{Science}} \textbf{\bibinfo{volume}{361}},
  \bibinfo{pages}{482--485} (\bibinfo{year}{2018}).
%\newblock \eprint{1806.05717}.

\bibitem{stone20_TDE_rate}
\bibinfo{author}{{Stone}, N.~C.} \emph{et~al.}
\newblock \bibinfo{title}{{Rates of Stellar Tidal Disruption}}.
\newblock \emph{\bibinfo{journal}{\ssr}} \textbf{\bibinfo{volume}{216}},
  \bibinfo{pages}{35} (\bibinfo{year}{2020}).
%\newblock \eprint{2003.08953}.

\bibitem{deColle20_jet_TDE_review}
\bibinfo{author}{{De Colle}, F.} \& \bibinfo{author}{{Lu}, W.}
\newblock \bibinfo{title}{{Jets from Tidal Disruption Events}}.
\newblock \emph{\bibinfo{journal}{\nar}} \textbf{\bibinfo{volume}{89}},
  \bibinfo{pages}{101538} (\bibinfo{year}{2020}).
%\newblock \eprint{1911.01442}.

\bibitem{AlVe2020}
\bibinfo{author}{Alexander, K.~D.}, \bibinfo{author}{van Velzen, S.},
  \bibinfo{author}{Horesh, A.} \& \bibinfo{author}{Zauderer, B.~A.}
\newblock \bibinfo{title}{Radio properties of tidal disruption events}.
\newblock \emph{\bibinfo{journal}{Space Science Reviews}}
  \textbf{\bibinfo{volume}{216}} (\bibinfo{year}{2020}).
\newblock \urlprefix\url{http://dx.doi.org/10.1007/s11214-020-00702-w}.

\bibitem{Hammerstein2022arXiv}
\bibinfo{author}{{Hammerstein}, E.} \emph{et~al.}
\newblock \bibinfo{title}{{The Final Season Reimagined: 30 Tidal Disruption
  Events from the ZTF-I Survey}}.
\newblock \emph{\bibinfo{journal}{arXiv e-prints}}
  \bibinfo{pages}{arXiv:2203.01461} (\bibinfo{year}{2022}).
%%\newblock \eprint{2203.01461}.

\bibitem{Planck2014dust}
\bibinfo{author}{{Planck Collaboration}} \emph{et~al.}
\newblock \bibinfo{title}{{Planck 2013 results. XI. All-sky model of thermal
  dust emission}}.
\newblock \emph{\bibinfo{journal}{\aap}} \textbf{\bibinfo{volume}{571}},
  \bibinfo{pages}{A11} (\bibinfo{year}{2014}).
%\newblock \eprint{1312.1300}.

\bibitem{Lundquist2022GCN.31612DEIMOS}
\bibinfo{author}{{Lundquist}, M.~J.}, \bibinfo{author}{{Alvarez}, C.~A.} \&
  \bibinfo{author}{{O'Meara}, J.}
\newblock \bibinfo{title}{{ZTF22aaajecp/AT2022cmc: Keck DEIMOS Redshift}}.
\newblock \emph{\bibinfo{journal}{GRB Coordinates Network}}
  \textbf{\bibinfo{volume}{31612}}, \bibinfo{pages}{1} (\bibinfo{year}{2022}).

\bibitem{Wiersema2012}
\bibinfo{author}{{Wiersema}, K.} \emph{et~al.}
\newblock \bibinfo{title}{{Polarimetry of the transient relativistic jet of GRB
  110328/Swift J164449.3+573451}}.
\newblock \emph{\bibinfo{journal}{\mnras}} \textbf{\bibinfo{volume}{421}},
  \bibinfo{pages}{1942--1948} (\bibinfo{year}{2012}).
%\newblock \eprint{1112.3042}.

\bibitem{growth_india}
\bibinfo{title}{{GOWTH India Telescope}}.
\newblock
  \bibinfo{howpublished}{\url{https://sites.google.com/view/growthindia/}}.

\bibitem{ztfrest_v1.0.0}
\bibinfo{author}{{Andreoni}, I.} \& \bibinfo{author}{{Coughlin}, M.}
\newblock \bibinfo{title}{growth-astro/ztfrest: ztfrest}
  (\bibinfo{year}{2022}).
\newblock \urlprefix\url{https://doi.org/10.5281/zenodo.6827348}.

\end{thebibliography}

\begin{thebibliography}{10}
\expandafter\ifx\csname url\endcsname\relax
  \def\url#1{\texttt{#1}}\fi
\expandafter\ifx\csname urlprefix\endcsname\relax\def\urlprefix{URL }\fi
\providecommand{\bibinfo}[2]{#2}
\providecommand{\eprint}[2][]{\url{#2}}
\makeatletter
\addtocounter{\@listctr}{37}
\makeatother


\bibitem{Me2017}
\bibinfo{author}{{Metzger}, B.~D.}
\newblock \bibinfo{title}{{Kilonovae}}.
\newblock \emph{\bibinfo{journal}{Living Reviews in Relativity}}
  \textbf{\bibinfo{volume}{23}}, \bibinfo{pages}{1} (\bibinfo{year}{2019}).
%\newblock \eprint{1910.01617}.

\bibitem{CoFo2017}
\bibinfo{author}{{Coulter et al.}}
\newblock \bibinfo{title}{Swope supernova survey 2017a (sss17a), the optical
  counterpart to a gravitational wave source}.
\newblock \emph{\bibinfo{journal}{Science}} \textbf{\bibinfo{volume}{358}},
  \bibinfo{pages}{1556--1558} (\bibinfo{year}{2017}).

\bibitem{AbEA2017b}
\bibinfo{author}{{Abbott et al.}}
\newblock \bibinfo{title}{Gw170817: Observation of gravitational waves from a
  binary neutron star inspiral}.
\newblock \emph{\bibinfo{journal}{Phys. Rev. Lett.}}
  \textbf{\bibinfo{volume}{119}}, \bibinfo{pages}{161101}
  (\bibinfo{year}{2017}).
\newblock
  \urlprefix\url{https://link.aps.org/doi/10.1103/PhysRevLett.119.161101}.

\bibitem{Prentice2018}
\bibinfo{author}{{Prentice}, S.~J.} \emph{et~al.}
\newblock \bibinfo{title}{{The Cow: Discovery of a Luminous, Hot, and Rapidly
  Evolving Transient}}.
\newblock \emph{\bibinfo{journal}{\apjl}} \textbf{\bibinfo{volume}{865}},
  \bibinfo{pages}{L3} (\bibinfo{year}{2018}).
%\newblock \eprint{1807.05965}.

\bibitem{Perley2019}
\bibinfo{author}{{Perley}, D.~A.} \emph{et~al.}
\newblock \bibinfo{title}{{The fast, luminous ultraviolet transient AT2018cow:
  extreme supernova, or disruption of a star by an intermediate-mass black
  hole?}}
\newblock \emph{\bibinfo{journal}{\mnras}} \textbf{\bibinfo{volume}{484}},
  \bibinfo{pages}{1031--1049} (\bibinfo{year}{2019}).
%\newblock \eprint{1808.00969}.

\bibitem{Margutti2019}
\bibinfo{author}{{Margutti}, R.} \emph{et~al.}
\newblock \bibinfo{title}{{An Embedded X-Ray Source Shines through the
  Aspherical AT 2018cow: Revealing the Inner Workings of the Most Luminous
  Fast-evolving Optical Transients}}.
\newblock \emph{\bibinfo{journal}{\apj}} \textbf{\bibinfo{volume}{872}},
  \bibinfo{pages}{18} (\bibinfo{year}{2019}).
%\newblock \eprint{1810.10720}.

\bibitem{Coppejans2020}
\bibinfo{author}{{Coppejans}, D.~L.} \emph{et~al.}
\newblock \bibinfo{title}{{A Mildly Relativistic Outflow from the Energetic,
  Fast-rising Blue Optical Transient CSS161010 in a Dwarf Galaxy}}.
\newblock \emph{\bibinfo{journal}{\apjl}} \textbf{\bibinfo{volume}{895}},
  \bibinfo{pages}{L23} (\bibinfo{year}{2020}).
%\newblock \eprint{2003.10503}.

\bibitem{Ho2020_Koala}
\bibinfo{author}{{Ho}, A. Y.~Q.} \emph{et~al.}
\newblock \bibinfo{title}{{The Koala: A Fast Blue Optical Transient with
  Luminous Radio Emission from a Starburst Dwarf Galaxy at z = 0.27}}.
\newblock \emph{\bibinfo{journal}{\apj}} \textbf{\bibinfo{volume}{895}},
  \bibinfo{pages}{49} (\bibinfo{year}{2020}).
%\newblock \eprint{2003.01222}.

\bibitem{Perley2021}
\bibinfo{author}{{Perley}, D.~A.} \emph{et~al.}
\newblock \bibinfo{title}{{Real-time discovery of AT2020xnd: a fast, luminous
  ultraviolet transient with minimal radioactive ejecta}}.
\newblock \emph{\bibinfo{journal}{\mnras}} \textbf{\bibinfo{volume}{508}},
  \bibinfo{pages}{5138--5147} (\bibinfo{year}{2021}).
%\newblock \eprint{2103.01968}.

\bibitem{Yao2021arXiv}
\bibinfo{author}{{Yao}, Y.} \emph{et~al.}
\newblock \bibinfo{title}{{The X-ray and Radio Loud Fast Blue Optical Transient
  AT2020mrf: Implications for an Emerging Class of Engine-Driven Massive Star
  Explosions}}.
\newblock \emph{\bibinfo{journal}{arXiv e-prints}}
  \bibinfo{pages}{arXiv:2112.00751} (\bibinfo{year}{2021}).
%\newblock \eprint{2112.00751}.

\bibitem{Ho2019cow}
\bibinfo{author}{{Ho}, A. Y.~Q.} \emph{et~al.}
\newblock \bibinfo{title}{{AT2018cow: A Luminous Millimeter Transient}}.
\newblock \emph{\bibinfo{journal}{\apj}} \textbf{\bibinfo{volume}{871}},
  \bibinfo{pages}{73} (\bibinfo{year}{2019}).
%\newblock \eprint{1810.10880}.

\bibitem{Ho2021arXiv}
\bibinfo{author}{{Ho}, A. Y.~Q.} \emph{et~al.}
\newblock \bibinfo{title}{{Luminous Millimeter, Radio, and X-ray Emission from
  ZTF20acigmel (AT2020xnd)}}.
\newblock \emph{\bibinfo{journal}{arXiv e-prints}}
  \bibinfo{pages}{arXiv:2110.05490} (\bibinfo{year}{2021}).
%\newblock \eprint{2110.05490}.

\bibitem{Quataert2012MNRAS}
\bibinfo{author}{{Quataert}, E.} \& \bibinfo{author}{{Kasen}, D.}
\newblock \bibinfo{title}{{Swift 1644+57: the longest gamma-ray burst?}}
\newblock \emph{\bibinfo{journal}{\mnras}} \textbf{\bibinfo{volume}{419}},
  \bibinfo{pages}{L1--L5} (\bibinfo{year}{2012}).
%\newblock \eprint{1105.3209}.

\bibitem{Sheth2003}
\bibinfo{author}{{Sheth}, K.} \emph{et~al.}
\newblock \bibinfo{title}{{Millimeter Observations of GRB 030329: Continued
  Evidence for a Two-Component Jet}}.
\newblock \emph{\bibinfo{journal}{\apjl}} \textbf{\bibinfo{volume}{595}},
  \bibinfo{pages}{L33--L36} (\bibinfo{year}{2003}).
%\newblock \eprint{astro-ph/0308188}.

\bibitem{Laskar2018}
\bibinfo{author}{{Laskar}, T.} \emph{et~al.}
\newblock \bibinfo{title}{{First ALMA Light Curve Constrains Refreshed Reverse
  Shocks and Jet Magnetization in GRB 161219B}}.
\newblock \emph{\bibinfo{journal}{\apj}} \textbf{\bibinfo{volume}{862}},
  \bibinfo{pages}{94} (\bibinfo{year}{2018}).
%\newblock \eprint{1808.09476}.

\bibitem{Laskar2019}
\bibinfo{author}{{Laskar}, T.} \emph{et~al.}
\newblock \bibinfo{title}{{A Reverse Shock in GRB 181201A}}.
\newblock \emph{\bibinfo{journal}{\apj}} \textbf{\bibinfo{volume}{884}},
  \bibinfo{pages}{121} (\bibinfo{year}{2019}).
%\newblock \eprint{1907.13128}.

\bibitem{Perley2014}
\bibinfo{author}{{Perley}, D.~A.} \emph{et~al.}
\newblock \bibinfo{title}{{The Afterglow of GRB 130427A from 1 to {}10$^{16}$
  GHz}}.
\newblock \emph{\bibinfo{journal}{\apj}} \textbf{\bibinfo{volume}{781}},
  \bibinfo{pages}{37} (\bibinfo{year}{2014}).
%\newblock \eprint{1307.4401}.

\bibitem{deUgartePostigo2012_grblc}
\bibinfo{author}{{de Ugarte Postigo}, A.} \emph{et~al.}
\newblock \bibinfo{title}{{Pre-ALMA observations of GRBs in the mm/submm
  range}}.
\newblock \emph{\bibinfo{journal}{\aap}} \textbf{\bibinfo{volume}{538}},
  \bibinfo{pages}{A44} (\bibinfo{year}{2012}).
%\newblock \eprint{1108.1797}.

\bibitem{Kulkarni1998}
\bibinfo{author}{{Kulkarni}, S.~R.} \emph{et~al.}
\newblock \bibinfo{title}{{Radio emission from the unusual supernova 1998bw and
  its association with the {\ensuremath{\gamma}}-ray burst of 25 April 1998}}.
\newblock \emph{\bibinfo{journal}{Nature}} \textbf{\bibinfo{volume}{395}},
  \bibinfo{pages}{663--669} (\bibinfo{year}{1998}).

\bibitem{Perley2017-alma}
\bibinfo{author}{{Perley}, D.~A.}, \bibinfo{author}{{Schulze}, S.} \&
  \bibinfo{author}{{de Ugarte Postigo}, A.}
\newblock \bibinfo{title}{{GRB 171205A: ALMA observations.}}
\newblock \emph{\bibinfo{journal}{GRB Coordinates Network}}
  \textbf{\bibinfo{volume}{22252}}, \bibinfo{pages}{1} (\bibinfo{year}{2017}).

\bibitem{Weiler2007}
\bibinfo{author}{{Weiler}, K.~W.} \emph{et~al.}
\newblock \bibinfo{title}{{Long-Term Radio Monitoring of SN 1993J}}.
\newblock \emph{\bibinfo{journal}{\apj}} \textbf{\bibinfo{volume}{671}},
  \bibinfo{pages}{1959--1980} (\bibinfo{year}{2007}).
%\newblock \eprint{0709.1136}.

\bibitem{Maeda2021}
\bibinfo{author}{{Maeda}, K.} \emph{et~al.}
\newblock \bibinfo{title}{{The final months of massive star evolution from the
  circumstellar environment around SN Ic 2020oi}}.
\newblock \emph{\bibinfo{journal}{arXiv e-prints}}
  \bibinfo{pages}{arXiv:2106.11618} (\bibinfo{year}{2021}).
%\newblock \eprint{2106.11618}.

\bibitem{Horesh2013}
\bibinfo{author}{{Horesh}, A.} \emph{et~al.}
\newblock \bibinfo{title}{{An early and comprehensive millimetre and centimetre
  wave and X-ray study of SN 2011dh: a non-equipartition blast wave expanding
  into a massive stellar wind}}.
\newblock \emph{\bibinfo{journal}{\mnras}} \textbf{\bibinfo{volume}{436}},
  \bibinfo{pages}{1258--1267} (\bibinfo{year}{2013}).
%\newblock \eprint{1209.1102}.

\bibitem{Corsi2014}
\bibinfo{author}{{Corsi}, A.} \emph{et~al.}
\newblock \bibinfo{title}{{A Multi-wavelength Investigation of the Radio-loud
  Supernova PTF11qcj and its Circumstellar Environment}}.
\newblock \emph{\bibinfo{journal}{\apj}} \textbf{\bibinfo{volume}{782}},
  \bibinfo{pages}{42} (\bibinfo{year}{2014}).
%\newblock \eprint{1307.2366}.

\bibitem{Soderberg2010}
\bibinfo{author}{{Soderberg}, A.~M.} \emph{et~al.}
\newblock \bibinfo{title}{{A relativistic type Ibc supernova without a detected
  {\ensuremath{\gamma}}-ray burst}}.
\newblock \emph{\bibinfo{journal}{\nat}} \textbf{\bibinfo{volume}{463}},
  \bibinfo{pages}{513--515} (\bibinfo{year}{2010}).
%\newblock \eprint{0908.2817}.

\bibitem{Kann2006ApJ}
\bibinfo{author}{{Kann}, D.~A.}, \bibinfo{author}{{Klose}, S.} \&
  \bibinfo{author}{{Zeh}, A.}
\newblock \bibinfo{title}{{Signatures of Extragalactic Dust in Pre-Swift GRB
  Afterglows}}.
\newblock \emph{\bibinfo{journal}{\apj}} \textbf{\bibinfo{volume}{641}},
  \bibinfo{pages}{993--1009} (\bibinfo{year}{2006}).
%\newblock \eprint{astro-ph/0512575}.

\bibitem{Kann2010ApJ}
\bibinfo{author}{{Kann}, D.~A.} \emph{et~al.}
\newblock \bibinfo{title}{{The Afterglows of Swift-era Gamma-ray Bursts. I.
  Comparing pre-Swift and Swift-era Long/Soft (Type II) GRB Optical
  Afterglows}}.
\newblock \emph{\bibinfo{journal}{\apj}} \textbf{\bibinfo{volume}{720}},
  \bibinfo{pages}{1513--1558} (\bibinfo{year}{2010}).
%\newblock \eprint{0712.2186}.

\bibitem{Kann2011ApJ}
\bibinfo{author}{{Kann}, D.~A.} \emph{et~al.}
\newblock \bibinfo{title}{{The Afterglows of Swift-era Gamma-Ray Bursts. II.
  Type I GRB versus Type II GRB Optical Afterglows}}.
\newblock \emph{\bibinfo{journal}{\apj}} \textbf{\bibinfo{volume}{734}},
  \bibinfo{pages}{96} (\bibinfo{year}{2011}).
%\newblock \eprint{0804.1959}.

\bibitem{Fremling2020}
\bibinfo{author}{{Fremling}, C.} \emph{et~al.}
\newblock \bibinfo{title}{{The Zwicky Transient Facility Bright Transient
  Survey. I. Spectroscopic Classification and the Redshift Completeness of
  Local Galaxy Catalogs}}.
\newblock \emph{\bibinfo{journal}{\apj}} \textbf{\bibinfo{volume}{895}},
  \bibinfo{pages}{32} (\bibinfo{year}{2020}).
%\newblock \eprint{1910.12973}.

\bibitem{Perley2020}
\bibinfo{author}{{Perley}, D.~A.} \emph{et~al.}
\newblock \bibinfo{title}{{The Zwicky Transient Facility Bright Transient
  Survey. II. A Public Statistical Sample for Exploring Supernova
  Demographics}}.
\newblock \emph{\bibinfo{journal}{\apj}} \textbf{\bibinfo{volume}{904}},
  \bibinfo{pages}{35} (\bibinfo{year}{2020}).
%\newblock \eprint{2009.01242}.

\bibitem{Ho2021_RET}
\bibinfo{author}{{Ho}, A. Y.~Q.} \emph{et~al.}
\newblock \bibinfo{title}{{The Photometric and Spectroscopic Evolution of
  Rapidly Evolving Extragalactic Transients in ZTF}}.
\newblock \emph{\bibinfo{journal}{arXiv e-prints}}
  \bibinfo{pages}{arXiv:2105.08811} (\bibinfo{year}{2021}).
%\newblock \eprint{2105.08811}.

\bibitem{Ho2022arXiv}
\bibinfo{author}{{Ho}, A. Y.~Q.} \emph{et~al.}
\newblock \bibinfo{title}{{Cosmological Fast Optical Transients with the Zwicky
  Transient Facility: A Search for Dirty Fireballs}}.
\newblock \emph{\bibinfo{journal}{arXiv e-prints}}
  \bibinfo{pages}{arXiv:2201.12366} (\bibinfo{year}{2022}).
%\newblock \eprint{2201.12366}.

\bibitem{Cenko2015}
\bibinfo{author}{{Cenko}, S.~B.} \emph{et~al.}
\newblock \bibinfo{title}{{iPTF14yb: The First Discovery of a Gamma-Ray Burst
  Afterglow Independent of a High-energy Trigger}}.
\newblock \emph{\bibinfo{journal}{\apjl}} \textbf{\bibinfo{volume}{803}},
  \bibinfo{pages}{L24} (\bibinfo{year}{2015}).
%\newblock \eprint{1504.00673}.


\bibitem{Cowperthwaite2017}
\bibinfo{author}{{Cowperthwaite}, P.~S.} \emph{et~al.}
\newblock \bibinfo{title}{{The Electromagnetic Counterpart of the Binary
  Neutron Star Merger LIGO/Virgo GW170817. II. UV, Optical, and Near-infrared
  Light Curves and Comparison to Kilonova Models}}.
\newblock \emph{\bibinfo{journal}{\apjl}} \textbf{\bibinfo{volume}{848}},
  \bibinfo{pages}{L17} (\bibinfo{year}{2017}).
%\newblock \eprint{1710.05840}.

\bibitem{Kasliwal2017}
\bibinfo{author}{{Kasliwal}, M.~M.} \emph{et~al.}
\newblock \bibinfo{title}{{Illuminating gravitational waves: A concordant
  picture of photons from a neutron star merger}}.
\newblock \emph{\bibinfo{journal}{Science}} \textbf{\bibinfo{volume}{358}},
  \bibinfo{pages}{1559--1565} (\bibinfo{year}{2017}).
%\newblock \eprint{1710.05436}.

\bibitem{Drout2017}
\bibinfo{author}{{Drout}, M.~R.} \emph{et~al.}
\newblock \bibinfo{title}{{Light curves of the neutron star merger
  GW170817/SSS17a: Implications for r-process nucleosynthesis}}.
\newblock \emph{\bibinfo{journal}{Science}} \textbf{\bibinfo{volume}{358}},
  \bibinfo{pages}{1570--1574} (\bibinfo{year}{2017}).
%\newblock \eprint{1710.05443}.

\bibitem{Villar2017}
\bibinfo{author}{{Villar}, V.~A.}, \bibinfo{author}{{Berger}, E.},
  \bibinfo{author}{{Metzger}, B.~D.} \& \bibinfo{author}{{Guillochon}, J.}
\newblock \bibinfo{title}{{Theoretical Models of Optical Transients. I. A Broad
  Exploration of the Duration-Luminosity Phase Space}}.
\newblock \emph{\bibinfo{journal}{\apj}} \textbf{\bibinfo{volume}{849}},
  \bibinfo{pages}{70} (\bibinfo{year}{2017}).
%\newblock \eprint{1707.08132}.

\bibitem{strubbe09_disk_wind}
\bibinfo{author}{{Strubbe}, L.~E.} \& \bibinfo{author}{{Quataert}, E.}
\newblock \bibinfo{title}{{Optical flares from the tidal disruption of stars by
  massive black holes}}.
\newblock \emph{\bibinfo{journal}{\mnras}} \textbf{\bibinfo{volume}{400}},
  \bibinfo{pages}{2070--2084} (\bibinfo{year}{2009}).
%\newblock \eprint{0905.3735}.

\bibitem{shiokawa15_disk_formation}
\bibinfo{author}{{Shiokawa}, H.}, \bibinfo{author}{{Krolik}, J.~H.},
  \bibinfo{author}{{Cheng}, R.~M.}, \bibinfo{author}{{Piran}, T.} \&
  \bibinfo{author}{{Noble}, S.~C.}
\newblock \bibinfo{title}{{General Relativistic Hydrodynamic Simulation of
  Accretion Flow from a Stellar Tidal Disruption}}.
\newblock \emph{\bibinfo{journal}{\apj}} \textbf{\bibinfo{volume}{804}},
  \bibinfo{pages}{85} (\bibinfo{year}{2015}).
%\newblock \eprint{1501.04365}.

\bibitem{kimitake16_circularization}
\bibinfo{author}{{Hayasaki}, K.}, \bibinfo{author}{{Stone}, N.} \&
  \bibinfo{author}{{Loeb}, A.}
\newblock \bibinfo{title}{{Circularization of tidally disrupted stars around
  spinning supermassive black holes}}.
\newblock \emph{\bibinfo{journal}{\mnras}} \textbf{\bibinfo{volume}{461}},
  \bibinfo{pages}{3760--3780} (\bibinfo{year}{2016}).
%\newblock \eprint{1501.05207}.

\bibitem{bonnerot16_disk_formation}
\bibinfo{author}{{Bonnerot}, C.}, \bibinfo{author}{{Rossi}, E.~M.},
  \bibinfo{author}{{Lodato}, G.} \& \bibinfo{author}{{Price}, D.~J.}
\newblock \bibinfo{title}{{Disc formation from tidal disruptions of stars on
  eccentric orbits by Schwarzschild black holes}}.
\newblock \emph{\bibinfo{journal}{\mnras}} \textbf{\bibinfo{volume}{455}},
  \bibinfo{pages}{2253--2266} (\bibinfo{year}{2016}).
%\newblock \eprint{1501.04635}.

\bibitem{metzger16_reprocessing}
\bibinfo{author}{{Metzger}, B.~D.} \& \bibinfo{author}{{Stone}, N.~C.}
\newblock \bibinfo{title}{{A bright year for tidal disruptions}}.
\newblock \emph{\bibinfo{journal}{\mnras}} \textbf{\bibinfo{volume}{461}},
  \bibinfo{pages}{948--966} (\bibinfo{year}{2016}).
%\newblock \eprint{1506.03453}.

\bibitem{metzger12_j1644_afterglow}
\bibinfo{author}{{Metzger}, B.~D.}, \bibinfo{author}{{Giannios}, D.} \&
  \bibinfo{author}{{Mimica}, P.}
\newblock \bibinfo{title}{{Afterglow model for the radio emission from the
  jetted tidal disruption candidate Swift J1644+57}}.
\newblock \emph{\bibinfo{journal}{\mnras}} \textbf{\bibinfo{volume}{420}},
  \bibinfo{pages}{3528--3537} (\bibinfo{year}{2012}).
%\newblock \eprint{1110.1111}.

\bibitem{tchkhovskoy10_MAD_efficiency}
\bibinfo{author}{{Tchekhovskoy}, A.}, \bibinfo{author}{{Narayan}, R.} \&
  \bibinfo{author}{{McKinney}, J.~C.}
\newblock \bibinfo{title}{{Black Hole Spin and The Radio Loud/Quiet Dichotomy
  of Active Galactic Nuclei}}.
\newblock \emph{\bibinfo{journal}{\apj}} \textbf{\bibinfo{volume}{711}},
  \bibinfo{pages}{50--63} (\bibinfo{year}{2010}).
%\newblock \eprint{0911.2228}.

\bibitem{law-smith20_fallback_rate}
\bibinfo{author}{{Law-Smith}, J. A.~P.}, \bibinfo{author}{{Coulter}, D.~A.},
  \bibinfo{author}{{Guillochon}, J.}, \bibinfo{author}{{Mockler}, B.} \&
  \bibinfo{author}{{Ramirez-Ruiz}, E.}
\newblock \bibinfo{title}{{Stellar Tidal Disruption Events with Abundances and
  Realistic Structures (STARS): Library of Fallback Rates}}.
\newblock \emph{\bibinfo{journal}{\apj}} \textbf{\bibinfo{volume}{905}},
  \bibinfo{pages}{141} (\bibinfo{year}{2020}).
%\newblock \eprint{2007.10996}.

\bibitem{jiang19_super-Eddington_accretion}
\bibinfo{author}{{Jiang}, Y.-F.}, \bibinfo{author}{{Stone}, J.~M.} \&
  \bibinfo{author}{{Davis}, S.~W.}
\newblock \bibinfo{title}{{Super-Eddington Accretion Disks around Supermassive
  Black Holes}}.
\newblock \emph{\bibinfo{journal}{\apj}} \textbf{\bibinfo{volume}{880}},
  \bibinfo{pages}{67} (\bibinfo{year}{2019}).
%\newblock \eprint{1709.02845}.

\bibitem{deUgartePostigo2012}
\bibinfo{author}{{de Ugarte Postigo}, A.} \emph{et~al.}
\newblock \bibinfo{title}{{The distribution of equivalent widths in long GRB
  afterglow spectra}}.
\newblock \emph{\bibinfo{journal}{\aap}} \textbf{\bibinfo{volume}{548}},
  \bibinfo{pages}{A11} (\bibinfo{year}{2012}).
%\newblock \eprint{1209.0891}.

\bibitem{Bloom2002}
\bibinfo{author}{{Bloom}, J.~S.}, \bibinfo{author}{{Kulkarni}, S.~R.} \&
  \bibinfo{author}{{Djorgovski}, S.~G.}
\newblock \bibinfo{title}{{The Observed Offset Distribution of Gamma-Ray Bursts
  from Their Host Galaxies: A Robust Clue to the Nature of the Progenitors}}.
\newblock \emph{\bibinfo{journal}{\aj}} \textbf{\bibinfo{volume}{123}},
  \bibinfo{pages}{1111--1148} (\bibinfo{year}{2002}).
%\newblock \eprint{astro-ph/0010176}.

\bibitem{Blanchard2016}
\bibinfo{author}{{Blanchard}, P.~K.}, \bibinfo{author}{{Berger}, E.} \&
  \bibinfo{author}{{Fong}, W.-f.}
\newblock \bibinfo{title}{{The Offset and Host Light Distributions of Long
  Gamma-Ray Bursts: A New View From HST Observations of Swift Bursts}}.
\newblock \emph{\bibinfo{journal}{\apj}} \textbf{\bibinfo{volume}{817}},
  \bibinfo{pages}{144} (\bibinfo{year}{2016}).
%\newblock \eprint{1509.07866}.

\bibitem{Burrows2005}
\bibinfo{author}{{Burrows}, D.~N.} \emph{et~al.}
\newblock \bibinfo{title}{{The Swift X-Ray Telescope}}.
\newblock \emph{\bibinfo{journal}{\ssr}} \textbf{\bibinfo{volume}{120}},
  \bibinfo{pages}{165--195} (\bibinfo{year}{2005}).
%\newblock \eprint{astro-ph/0508071}.

\bibitem{prospector}
\bibinfo{title}{{Prospector}}.
\newblock \bibinfo{howpublished}{\url{https://github.com/bd-j/prospector}}.

\bibitem{Johnson2021a}
\bibinfo{author}{{Johnson}, B.~D.}, \bibinfo{author}{{Leja}, J.},
  \bibinfo{author}{{Conroy}, C.} \& \bibinfo{author}{{Speagle}, J.~S.}
\newblock \bibinfo{title}{{Stellar Population Inference with Prospector}}.
\newblock \emph{\bibinfo{journal}{\apjs}} \textbf{\bibinfo{volume}{254}},
  \bibinfo{pages}{22} (\bibinfo{year}{2021}).
%\newblock \eprint{2012.01426}.

\bibitem{Conroy2009a}
\bibinfo{author}{{Conroy}, C.}, \bibinfo{author}{{Gunn}, J.~E.} \&
  \bibinfo{author}{{White}, M.}
\newblock \bibinfo{title}{{The Propagation of Uncertainties in Stellar
  Population Synthesis Modeling. I. The Relevance of Uncertain Aspects of
  Stellar Evolution and the Initial Mass Function to the Derived Physical
  Properties of Galaxies}}.
\newblock \emph{\bibinfo{journal}{\apj}} \textbf{\bibinfo{volume}{699}},
  \bibinfo{pages}{486--506} (\bibinfo{year}{2009}).
%\newblock \eprint{0809.4261}.

\bibitem{ForemanMackey2014a}
\bibinfo{author}{{Foreman-Mackey}, D.}, \bibinfo{author}{{Sick}, J.} \&
  \bibinfo{author}{{Johnson}, B.}
\newblock \bibinfo{title}{{Python-Fsps: Python Bindings To Fsps (V0.1.1)}}
  (\bibinfo{year}{2014}).

\bibitem{Byler2017a}
\bibinfo{author}{{Byler}, N.}, \bibinfo{author}{{Dalcanton}, J.~J.},
  \bibinfo{author}{{Conroy}, C.} \& \bibinfo{author}{{Johnson}, B.~D.}
\newblock \bibinfo{title}{{Nebular Continuum and Line Emission in Stellar
  Population Synthesis Models}}.
\newblock \emph{\bibinfo{journal}{\apj}} \textbf{\bibinfo{volume}{840}},
  \bibinfo{pages}{44} (\bibinfo{year}{2017}).
%\newblock \eprint{1611.08305}.

\bibitem{Chabrier2003a}
\bibinfo{author}{{Chabrier}, G.}
\newblock \bibinfo{title}{{Galactic Stellar and Substellar Initial Mass
  Function}}.
\newblock \emph{\bibinfo{journal}{\pasp}} \textbf{\bibinfo{volume}{115}},
  \bibinfo{pages}{763} (\bibinfo{year}{2003}).
%\newblock \eprint{astro-ph/0304382}.

\bibitem{Calzetti2000a}
\bibinfo{author}{{Calzetti}, D.} \emph{et~al.}
\newblock \bibinfo{title}{{The Dust Content and Opacity of Actively
  Star-forming Galaxies}}.
\newblock \emph{\bibinfo{journal}{\apj}} \textbf{\bibinfo{volume}{533}},
  \bibinfo{pages}{682--695} (\bibinfo{year}{2000}).
%\newblock \eprint{astro-ph/9911459}.

\bibitem{Schulze2021a}
\bibinfo{author}{{Schulze}, S.} \emph{et~al.}
\newblock \bibinfo{title}{{The Palomar Transient Factory Core-collapse
  Supernova Host-galaxy Sample. I. Host-galaxy Distribution Functions and
  Environment Dependence of Core-collapse Supernovae}}.
\newblock \emph{\bibinfo{journal}{\apjs}} \textbf{\bibinfo{volume}{255}},
  \bibinfo{pages}{29} (\bibinfo{year}{2021}).
%\newblock \eprint{2008.05988}.

\bibitem{kesden12_Hills_mass}
\bibinfo{author}{{Kesden}, M.}
\newblock \bibinfo{title}{{Tidal-disruption rate of stars by spinning
  supermassive black holes}}.
\newblock \emph{\bibinfo{journal}{\prd}} \textbf{\bibinfo{volume}{85}},
  \bibinfo{pages}{024037} (\bibinfo{year}{2012}).
%\newblock \eprint{1109.6329}.

\bibitem{Cummings2011GCN}
\bibinfo{author}{{Cummings}, J.~R.} \emph{et~al.}
\newblock \bibinfo{title}{{GRB 110328A: Swift detection of a burst.}}
\newblock \emph{\bibinfo{journal}{GRB Coordinates Network}}
  \textbf{\bibinfo{volume}{11823}}, \bibinfo{pages}{1} (\bibinfo{year}{2011}).

\bibitem{Benson2014SPIE}
\bibinfo{author}{{Benson}, B.~A.} \emph{et~al.}
\newblock \bibinfo{title}{{SPT-3G: a next-generation cosmic microwave
  background polarization experiment on the South Pole telescope}}.
\newblock In \bibinfo{editor}{{Holland}, W.~S.} \&
  \bibinfo{editor}{{Zmuidzinas}, J.} (eds.)
  \emph{\bibinfo{booktitle}{Millimeter, Submillimeter, and Far-Infrared
  Detectors and Instrumentation for Astronomy VII}}, vol.
  \bibinfo{volume}{9153} of \emph{\bibinfo{series}{Society of Photo-Optical
  Instrumentation Engineers (SPIE) Conference Series}}, \bibinfo{pages}{91531P}
  (\bibinfo{year}{2014}).
%\newblock \eprint{1407.2973}.

\bibitem{Abazajian2019arXiv}
\bibinfo{author}{{Abazajian}, K.} \emph{et~al.}
\newblock \bibinfo{title}{{CMB-S4 Science Case, Reference Design, and Project
  Plan}}.
\newblock \emph{\bibinfo{journal}{arXiv e-prints}}
  \bibinfo{pages}{arXiv:1907.04473} (\bibinfo{year}{2019}).
%\newblock \eprint{1907.04473}.

\bibitem{Guns2021ApJ}
\bibinfo{author}{{Guns}, S.} \emph{et~al.}
\newblock \bibinfo{title}{{Detection of Galactic and Extragalactic
  Millimeter-wavelength Transient Sources with SPT-3G}}.
\newblock \emph{\bibinfo{journal}{\apj}} \textbf{\bibinfo{volume}{916}},
  \bibinfo{pages}{98} (\bibinfo{year}{2021}).
%\newblock \eprint{2103.06166}.

\bibitem{Eftekhari2021arXiv}
\bibinfo{author}{{Eftekhari}, T.} \emph{et~al.}
\newblock \bibinfo{title}{{Extragalactic Millimeter Transients in the Era of
  Next Generation CMB Surveys}}.
\newblock \emph{\bibinfo{journal}{arXiv e-prints}}
  \bibinfo{pages}{arXiv:2110.05494} (\bibinfo{year}{2021}).
%\newblock \eprint{2110.05494}.

\bibitem{FeNo2019}
\bibinfo{author}{Feindt, U.} \emph{et~al.}
\newblock \bibinfo{title}{simsurvey: estimating transient discovery rates for
  the zwicky transient facility}.
\newblock \emph{\bibinfo{journal}{Journal of Cosmology and Astroparticle
  Physics}} \textbf{\bibinfo{volume}{2019}}, \bibinfo{pages}{005–005}
  (\bibinfo{year}{2019}).
\newblock \urlprefix\url{http://dx.doi.org/10.1088/1475-7516/2019/10/005}.

\bibitem{AnKo2020}
\bibinfo{author}{Andreoni, I.} \emph{et~al.}
\newblock \bibinfo{title}{Constraining the kilonova rate with zwicky transient
  facility searches independent of gravitational wave and short gamma-ray burst
  triggers}.
\newblock \emph{\bibinfo{journal}{The Astrophysical Journal}}
  \textbf{\bibinfo{volume}{904}}, \bibinfo{pages}{155} (\bibinfo{year}{2020}).
\newblock \urlprefix\url{https://doi.org/10.3847%2F1538-4357%2Fabbf4c}.

\bibitem{pymultinest}
\bibinfo{author}{{Buchner, J.}} \emph{et~al.}
\newblock \bibinfo{title}{X-ray spectral modelling of the agn obscuring region
  in the cdfs: Bayesian model selection and catalogue}.
\newblock \emph{\bibinfo{journal}{A\&A}} \textbf{\bibinfo{volume}{564}},
  \bibinfo{pages}{A125} (\bibinfo{year}{2014}).
\newblock \urlprefix\url{https://doi.org/10.1051/0004-6361/201322971}.

\bibitem{Feroz_2009}
\bibinfo{author}{Feroz, F.}, \bibinfo{author}{Hobson, M.~P.} \&
  \bibinfo{author}{Bridges, M.}
\newblock \bibinfo{title}{Multinest: an efficient and robust bayesian inference
  tool for cosmology and particle physics}.
\newblock \emph{\bibinfo{journal}{Monthly Notices of the Royal Astronomical
  Society}} \textbf{\bibinfo{volume}{398}}, \bibinfo{pages}{1601–1614}
  (\bibinfo{year}{2009}).
\newblock \urlprefix\url{http://dx.doi.org/10.1111/j.1365-2966.2009.14548.x}.

\bibitem{Feroz:2007kg}
\bibinfo{author}{Feroz, F.} \& \bibinfo{author}{Hobson, M.~P.}
\newblock \bibinfo{title}{{Multimodal nested sampling: an efficient and robust
  alternative to MCMC methods for astronomical data analysis}}.
\newblock \emph{\bibinfo{journal}{Mon. Not. Roy. Astron. Soc.}}
  \textbf{\bibinfo{volume}{384}}, \bibinfo{pages}{449} (\bibinfo{year}{2008}).
%\newblock \eprint{0704.3704}.

\bibitem{aLIGO}
\bibinfo{author}{{Aasi et al}}.
\newblock \bibinfo{title}{Advanced ligo}.
\newblock \emph{\bibinfo{journal}{Classical and Quantum Gravity}}
  \textbf{\bibinfo{volume}{32}}, \bibinfo{pages}{074001}
  (\bibinfo{year}{2015}).

\bibitem{adVirgo}
\bibinfo{author}{{Acernese et al}}.
\newblock \bibinfo{title}{Advanced {V}irgo}.
\newblock \emph{\bibinfo{journal}{Classical and Quantum Gravity}}
  \textbf{\bibinfo{volume}{32}}, \bibinfo{pages}{024001}
  (\bibinfo{year}{2015}).

\bibitem{AaAc2017}
\bibinfo{author}{Aartsen, M.} \emph{et~al.}
\newblock \bibinfo{title}{The {IceCube} neutrino observatory: instrumentation
  and online systems}.
\newblock \emph{\bibinfo{journal}{Journal of Instrumentation}}
  \textbf{\bibinfo{volume}{12}}, \bibinfo{pages}{P03012--P03012}
  (\bibinfo{year}{2017}).
\newblock
  \urlprefix\url{https://doi.org/10.1088%2F1748-0221%2F12%2F03%2Fp03012}.

\bibitem{Bellm2019}
\bibinfo{author}{{Bellm}, E.~C.} \emph{et~al.}
\newblock \bibinfo{title}{{The Zwicky Transient Facility: System Overview,
  Performance, and First Results}}.
\newblock \emph{\bibinfo{journal}{\pasp}} \textbf{\bibinfo{volume}{131}},
  \bibinfo{pages}{018002} (\bibinfo{year}{2019}).
%\newblock \eprint{1902.01932}.

\bibitem{Ivezic2019}
\bibinfo{author}{{Ivezi{\'c}}, {\v{Z}}.} \emph{et~al.}
\newblock \bibinfo{title}{{LSST: From Science Drivers to Reference Design and
  Anticipated Data Products}}.
\newblock \emph{\bibinfo{journal}{\apj}} \textbf{\bibinfo{volume}{873}},
  \bibinfo{pages}{111} (\bibinfo{year}{2019}).
%\newblock \eprint{0805.2366}.

\bibitem{Yao2019}
\bibinfo{author}{{Yao}, Y.} \emph{et~al.}
\newblock \bibinfo{title}{{ZTF Early Observations of Type Ia Supernovae. I.
  Properties of the 2018 Sample}}.
\newblock \emph{\bibinfo{journal}{\apj}} \textbf{\bibinfo{volume}{886}},
  \bibinfo{pages}{152} (\bibinfo{year}{2019}).
%\newblock \eprint{1910.02967}.

\bibitem{TNS_2022cmc}
\bibinfo{author}{{Andreoni}, I.}
\newblock \bibinfo{title}{{ZTF Transient Discovery Report for 2022-02-14}}.
\newblock \emph{\bibinfo{journal}{Transient Name Server Discovery Report}}
  \textbf{\bibinfo{volume}{2022-397}}, \bibinfo{pages}{1}
  (\bibinfo{year}{2022}).

\bibitem{Bellm2019scheduler}
\bibinfo{author}{{Bellm}, E.~C.} \emph{et~al.}
\newblock \bibinfo{title}{{The Zwicky Transient Facility: Surveys and
  Scheduler}}.
\newblock \emph{\bibinfo{journal}{\pasp}} \textbf{\bibinfo{volume}{131}},
  \bibinfo{pages}{068003} (\bibinfo{year}{2019}).
%\newblock \eprint{1905.02209}.

\bibitem{Dekany2020PASP}
\bibinfo{author}{{Dekany}, R.} \emph{et~al.}
\newblock \bibinfo{title}{{The Zwicky Transient Facility: Observing System}}.
\newblock \emph{\bibinfo{journal}{\pasp}} \textbf{\bibinfo{volume}{132}},
  \bibinfo{pages}{038001} (\bibinfo{year}{2020}).
%\newblock \eprint{2008.04923}.

\bibitem{Masci2019PASP}
\bibinfo{author}{{Masci}, F.~J.} \emph{et~al.}
\newblock \bibinfo{title}{{The Zwicky Transient Facility: Data Processing,
  Products, and Archive}}.
\newblock \emph{\bibinfo{journal}{\pasp}} \textbf{\bibinfo{volume}{131}},
  \bibinfo{pages}{018003} (\bibinfo{year}{2019}).
%\newblock \eprint{1902.01872}.

\bibitem{Steele2004}
\bibinfo{author}{{Steele}, I.~A.} \emph{et~al.}
\newblock \bibinfo{title}{{The Liverpool Telescope: performance and first
  results}}.
\newblock In \bibinfo{editor}{{Oschmann}, J., Jacobus~M.} (ed.)
  \emph{\bibinfo{booktitle}{Ground-based Telescopes}}, vol.
  \bibinfo{volume}{5489} of \emph{\bibinfo{series}{Society of Photo-Optical
  Instrumentation Engineers (SPIE) Conference Series}},
  \bibinfo{pages}{679--692} (\bibinfo{year}{2004}).

\bibitem{Perley2011}
\bibinfo{author}{{Perley}, R.~A.}, \bibinfo{author}{{Chandler}, C.~J.},
  \bibinfo{author}{{Butler}, B.~J.} \& \bibinfo{author}{{Wrobel}, J.~M.}
\newblock \bibinfo{title}{{The Expanded Very Large Array: A New Telescope for
  New Science}}.
\newblock \emph{\bibinfo{journal}{\apjl}} \textbf{\bibinfo{volume}{739}},
  \bibinfo{pages}{L1} (\bibinfo{year}{2011}).
%\newblock \eprint{1106.0532}.

\bibitem{Eftekhari2018}
\bibinfo{author}{{Eftekhari}, T.}, \bibinfo{author}{{Berger}, E.},
  \bibinfo{author}{{Zauderer}, B.~A.}, \bibinfo{author}{{Margutti}, R.} \&
  \bibinfo{author}{{Alexander}, K.~D.}
\newblock \bibinfo{title}{{Radio Monitoring of the Tidal Disruption Event Swift
  J164449.3+573451. III. Late-time Jet Energetics and a Deviation from
  Equipartition}}.
\newblock \emph{\bibinfo{journal}{\apj}} \textbf{\bibinfo{volume}{854}},
  \bibinfo{pages}{86} (\bibinfo{year}{2018}).
%\newblock \eprint{1710.07289}.

\bibitem{GILDAS}
\bibinfo{title}{{GILDAS package}}.
\newblock \bibinfo{howpublished}{\url{https://www.iram.fr}}.

\bibitem{jcmt_Holland2013}
\bibinfo{author}{Holland, W.~S.} \emph{et~al.}
\newblock \bibinfo{title}{{SCUBA-2: the 10 000 pixel bolometer camera on the
  James Clerk Maxwell Telescope}}.
\newblock \emph{\bibinfo{journal}{Monthly Notices of the Royal Astronomical
  Society}} \textbf{\bibinfo{volume}{430}}, \bibinfo{pages}{2513--2533}
  (\bibinfo{year}{2013}).
\newblock \urlprefix\url{https://doi.org/10.1093/mnras/sts612}.
\newblock
  \eprint{https://academic.oup.com/mnras/article-pdf/430/4/2513/3809877/sts612.pdf}.

\bibitem{jcmt_Currie2014}
\bibinfo{author}{{Currie}, M.~J.} \emph{et~al.}
\newblock \bibinfo{title}{{Starlink Software in 2013}}.
\newblock In \bibinfo{editor}{{Manset}, N.} \& \bibinfo{editor}{{Forshay}, P.}
  (eds.) \emph{\bibinfo{booktitle}{Astronomical Data Analysis Software and
  Systems XXIII}}, vol. \bibinfo{volume}{485} of
  \emph{\bibinfo{series}{Astronomical Society of the Pacific Conference
  Series}}, \bibinfo{pages}{391} (\bibinfo{year}{2014}).

\bibitem{jcmt_Chapin2013}
\bibinfo{author}{Chapin, E.~L.} \emph{et~al.}
\newblock \bibinfo{title}{{SCUBA-2: iterative map-making with the
  Sub-Millimetre User Reduction Facility}}.
\newblock \emph{\bibinfo{journal}{Monthly Notices of the Royal Astronomical
  Society}} \textbf{\bibinfo{volume}{430}}, \bibinfo{pages}{2545--2573}
  (\bibinfo{year}{2013}).
\newblock \urlprefix\url{https://doi.org/10.1093/mnras/stt052}.
\newblock
  \eprint{https://academic.oup.com/mnras/article-pdf/430/4/2545/3828956/stt052.pdf}.

\bibitem{jcmt_Mairs2021}
\bibinfo{author}{Mairs, S.} \emph{et~al.}
\newblock \bibinfo{title}{A decade of {SCUBA}-2: A comprehensive guide to
  calibrating 450 $\upmu$m and 850 $\upmu$m continuum data at the {JCMT}}.
\newblock \emph{\bibinfo{journal}{The Astronomical Journal}}
  \textbf{\bibinfo{volume}{162}}, \bibinfo{pages}{191} (\bibinfo{year}{2021}).
\newblock \urlprefix\url{https://doi.org/10.3847/1538-3881/ac18bf}.

\bibitem{jcmt_Smith2022}
\bibinfo{author}{{Smith}, I.~A.}, \bibinfo{author}{{Perley}, D.~A.} \&
  \bibinfo{author}{{Tanvir}, N.~R.}
\newblock \bibinfo{title}{{ZTF22aaajecp/AT2022cmc: JCMT SCUBA-2 sub-mm
  observations}}.
\newblock \emph{\bibinfo{journal}{GRB Coordinates Network}}
  \textbf{\bibinfo{volume}{31602}}, \bibinfo{pages}{1} (\bibinfo{year}{2022}).

\bibitem{McMullin2007}
\bibinfo{author}{{McMullin}, J.~P.}, \bibinfo{author}{{Waters}, B.},
  \bibinfo{author}{{Schiebel}, D.}, \bibinfo{author}{{Young}, W.} \&
  \bibinfo{author}{{Golap}, K.}
\newblock \bibinfo{title}{{CASA Architecture and Applications}}.
\newblock In \bibinfo{editor}{{Shaw}, R.~A.}, \bibinfo{editor}{{Hill}, F.} \&
  \bibinfo{editor}{{Bell}, D.~J.} (eds.) \emph{\bibinfo{booktitle}{Astronomical
  Data Analysis Software and Systems XVI}}, vol. \bibinfo{volume}{376} of
  \emph{\bibinfo{series}{Astronomical Society of the Pacific Conference
  Series}}, \bibinfo{pages}{127} (\bibinfo{year}{2007}).

\bibitem{Maity2021}
\bibinfo{author}{{Maity}, B.} \& \bibinfo{author}{{Chandra}, P.}
\newblock \bibinfo{title}{{1000 Days of the Lowest-frequency Emission from the
  Low-luminosity GRB 171205A}}.
\newblock \emph{\bibinfo{journal}{\apj}} \textbf{\bibinfo{volume}{907}},
  \bibinfo{pages}{60} (\bibinfo{year}{2021}).
%\newblock \eprint{2012.05166}.

\bibitem{2019ascl.soft07032M}
\bibinfo{author}{{McCully}, C.} \& \bibinfo{author}{{Tewes}, M.}
\newblock \bibinfo{title}{{Astro-SCRAPPY: Speedy Cosmic Ray Annihilation
  Package in Python}} (\bibinfo{year}{2019}).
%\newblock \eprint{1907.032}.

\bibitem{2010ascl.soft10068B}
\bibinfo{author}{{Bertin}, E.}
\newblock \bibinfo{title}{{SWarp: Resampling and Co-adding FITS Images
  Together}} (\bibinfo{year}{2010}).
%\newblock \eprint{1010.068}.

\bibitem{2016arXiv161205560C}
\bibinfo{author}{{Chambers}, K.~C.} \emph{et~al.}
\newblock \bibinfo{title}{{The Pan-STARRS1 Surveys}}.
\newblock \emph{\bibinfo{journal}{arXiv e-prints}}
  \bibinfo{pages}{arXiv:1612.05560} (\bibinfo{year}{2016}).
%\newblock \eprint{1612.05560}.

\bibitem{Flaugher2015}
\bibinfo{author}{{Flaugher}, B.} \emph{et~al.}
\newblock \bibinfo{title}{{The Dark Energy Camera}}.
\newblock \emph{\bibinfo{journal}{\aj}} \textbf{\bibinfo{volume}{150}},
  \bibinfo{pages}{150} (\bibinfo{year}{2015}).
%\newblock \eprint{1504.02900}.

\bibitem{Valdes2014}
\bibinfo{author}{{Valdes}, F.}, \bibinfo{author}{{Gruendl}, R.} \&
  \bibinfo{author}{{DES Project}}.
\newblock \bibinfo{title}{{The DECam Community Pipeline}}.
\newblock In \bibinfo{editor}{{Manset}, N.} \& \bibinfo{editor}{{Forshay}, P.}
  (eds.) \emph{\bibinfo{booktitle}{Astronomical Data Analysis Software and
  Systems XXIII}}, vol. \bibinfo{volume}{485} of
  \emph{\bibinfo{series}{Astronomical Society of the Pacific Conference
  Series}}, \bibinfo{pages}{379} (\bibinfo{year}{2014}).

\bibitem{Rest2014}
\bibinfo{author}{{Rest}, A.} \emph{et~al.}
\newblock \bibinfo{title}{{Cosmological Constraints from Measurements of Type
  Ia Supernovae Discovered during the First 1.5 yr of the Pan-STARRS1 Survey}}.
\newblock \emph{\bibinfo{journal}{\apj}} \textbf{\bibinfo{volume}{795}},
  \bibinfo{pages}{44} (\bibinfo{year}{2014}).
%\newblock \eprint{1310.3828}.

\bibitem{ALFOSC}
\bibinfo{title}{{ALFOSC}}.
\newblock
  \bibinfo{howpublished}{\url{http://www.not.iac.es/instruments/alfosc}}.

\bibitem{PyNOT}
\bibinfo{title}{{PyNOT}}.
\newblock \bibinfo{howpublished}{\url{https://github.com/jkrogager/PyNOT}}.

\bibitem{PypeIt}
\bibinfo{author}{{J. Xavier Prochaska}} \emph{et~al.}
\newblock \bibinfo{title}{pypeit/pypeit: Release 1.0.0} (\bibinfo{year}{2020}).
\newblock \urlprefix\url{https://zenodo.org/record/3743493}.

\bibitem{2006PASP..118.1396C}
\bibinfo{author}{{Cenko}, S.~B.} \emph{et~al.}
\newblock \bibinfo{title}{{The Automated Palomar 60 Inch Telescope}}.
\newblock \emph{\bibinfo{journal}{\pasp}} \textbf{\bibinfo{volume}{118}},
  \bibinfo{pages}{1396--1406} (\bibinfo{year}{2006}).
%\newblock \eprint{astro-ph/0608323}.

\bibitem{2018PASP..130c5003B}
\bibinfo{author}{{Blagorodnova}, N.} \emph{et~al.}
\newblock \bibinfo{title}{{The SED Machine: A Robotic Spectrograph for Fast
  Transient Classification}}.
\newblock \emph{\bibinfo{journal}{\pasp}} \textbf{\bibinfo{volume}{130}},
  \bibinfo{pages}{035003} (\bibinfo{year}{2018}).
%\newblock \eprint{1710.02917}.

\bibitem{Rigault2019}
\bibinfo{author}{{Rigault}, M.} \emph{et~al.}
\newblock \bibinfo{title}{{Fully automated integral field spectrograph pipeline
  for the SEDMachine: pysedm}}.
\newblock \emph{\bibinfo{journal}{\aap}} \textbf{\bibinfo{volume}{627}},
  \bibinfo{pages}{A115} (\bibinfo{year}{2019}).
%\newblock \eprint{1902.08526}.

\bibitem{Fremling2016}
\bibinfo{author}{Fremling, C.} \emph{et~al.}
\newblock \bibinfo{title}{Ptf12os and iptf13bvn}.
\newblock \emph{\bibinfo{journal}{Astronomy \& Astrophysics}}
  \textbf{\bibinfo{volume}{593}}, \bibinfo{pages}{A68} (\bibinfo{year}{2016}).
\newblock \urlprefix\url{http://dx.doi.org/10.1051/0004-6361/201628275}.

\bibitem{2014ApJS..211...17A}
\bibinfo{author}{{Ahn}, C.~P.} \emph{et~al.}
\newblock \bibinfo{title}{{The Tenth Data Release of the Sloan Digital Sky
  Survey: First Spectroscopic Data from the SDSS-III Apache Point Observatory
  Galactic Evolution Experiment}}.
\newblock \emph{\bibinfo{journal}{\apjs}} \textbf{\bibinfo{volume}{211}},
  \bibinfo{pages}{17} (\bibinfo{year}{2014}).
%\newblock \eprint{1307.7735}.

\bibitem{Tonry2018_ATLAS}
\bibinfo{author}{{Tonry}, J.~L.} \emph{et~al.}
\newblock \bibinfo{title}{{ATLAS: A High-cadence All-sky Survey System}}.
\newblock \emph{\bibinfo{journal}{\pasp}} \textbf{\bibinfo{volume}{130}},
  \bibinfo{pages}{064505} (\bibinfo{year}{2018}).
%\newblock \eprint{1802.00879}.

\bibitem{Smith2020_ATLAS}
\bibinfo{author}{{Smith}, K.~W.} \emph{et~al.}
\newblock \bibinfo{title}{{Design and Operation of the ATLAS Transient Science
  Server}}.
\newblock \emph{\bibinfo{journal}{\pasp}} \textbf{\bibinfo{volume}{132}},
  \bibinfo{pages}{085002} (\bibinfo{year}{2020}).
%\newblock \eprint{2003.09052}.

\bibitem{Vernet2011}
\bibinfo{author}{{Vernet}, J.} \emph{et~al.}
\newblock \bibinfo{title}{{X-shooter, the new wide band intermediate resolution
  spectrograph at the ESO Very Large Telescope}}.
\newblock \emph{\bibinfo{journal}{\aap}} \textbf{\bibinfo{volume}{536}},
  \bibinfo{pages}{A105} (\bibinfo{year}{2011}).
%\newblock \eprint{1110.1944}.

\bibitem{Modigliani2010}
\bibinfo{author}{{Modigliani}, A.} \emph{et~al.}
\newblock \bibinfo{title}{{The X-shooter pipeline}}.
\newblock In \emph{\bibinfo{booktitle}{Observatory Operations: Strategies,
  Processes, and Systems III}}, vol. \bibinfo{volume}{7737} of
  \emph{\bibinfo{series}{\procspie}}, \bibinfo{pages}{773728}
  (\bibinfo{year}{2010}).

\bibitem{Selsing2019}
\bibinfo{author}{{Selsing}, J.} \emph{et~al.}
\newblock \bibinfo{title}{{The X-shooter GRB afterglow legacy sample
  (XS-GRB)}}.
\newblock \emph{\bibinfo{journal}{\aap}} \textbf{\bibinfo{volume}{623}},
  \bibinfo{pages}{A92} (\bibinfo{year}{2019}).
%\newblock \eprint{1802.07727}.

\bibitem{Garzon2006}
\bibinfo{author}{{Garz{\'o}n}, F.} \emph{et~al.}
\newblock \bibinfo{title}{{EMIR: the GTC NIR multi-object
  imager-spectrograph}}.
\newblock In \bibinfo{editor}{{McLean}, I.~S.} \& \bibinfo{editor}{{Iye}, M.}
  (eds.) \emph{\bibinfo{booktitle}{Society of Photo-Optical Instrumentation
  Engineers (SPIE) Conference Series}}, vol. \bibinfo{volume}{6269} of
  \emph{\bibinfo{series}{Society of Photo-Optical Instrumentation Engineers
  (SPIE) Conference Series}}, \bibinfo{pages}{626918} (\bibinfo{year}{2006}).

\bibitem{2022GCN.31626....1K}
\bibinfo{author}{{Kann}, D.~A.} \emph{et~al.}
\newblock \bibinfo{title}{{ZTF22aaajecp/AT 2022cmc: CAHA 2.2m/CAFOS detection,
  luminous transient}}.
\newblock \emph{\bibinfo{journal}{GRB Coordinates Network}}
  \textbf{\bibinfo{volume}{31626}}, \bibinfo{pages}{1} (\bibinfo{year}{2022}).

\bibitem{Prochaska2020PypeIt}
\bibinfo{author}{{Prochaska}, J.} \emph{et~al.}
\newblock \bibinfo{title}{{PypeIt: The Python Spectroscopic Data Reduction
  Pipeline}}.
\newblock \emph{\bibinfo{journal}{The Journal of Open Source Software}}
  \textbf{\bibinfo{volume}{5}}, \bibinfo{pages}{2308} (\bibinfo{year}{2020}).
%\newblock \eprint{2005.06505}.

\bibitem{Perley2019PASP}
\bibinfo{author}{{Perley}, D.~A.}
\newblock \bibinfo{title}{{Fully Automated Reduction of Longslit Spectroscopy
  with the Low Resolution Imaging Spectrometer at the Keck Observatory}}.
\newblock \emph{\bibinfo{journal}{\pasp}} \textbf{\bibinfo{volume}{131}},
  \bibinfo{pages}{084503} (\bibinfo{year}{2019}).
%\newblock \eprint{1903.07629}.

\bibitem{dragons}
\bibinfo{author}{{Labrie}, K.}, \bibinfo{author}{{Cardenes}, R.},
  \bibinfo{author}{{Anderson}, K.}, \bibinfo{author}{{Simpson}, C.} \&
  \bibinfo{author}{{Turner}, J.}
\newblock \bibinfo{title}{{DRAGONS: One Pipeline to Rule them All}}.
\newblock In \bibinfo{editor}{{Ballester}, P.}, \bibinfo{editor}{{Ibsen}, J.},
  \bibinfo{editor}{{Solar}, M.} \& \bibinfo{editor}{{Shortridge}, K.} (eds.)
  \emph{\bibinfo{booktitle}{Astronomical Data Analysis Software and Systems
  XXVII}}, vol. \bibinfo{volume}{522} of \emph{\bibinfo{series}{Astronomical
  Society of the Pacific Conference Series}}, \bibinfo{pages}{583}
  (\bibinfo{year}{2020}).

\bibitem{GCN31595}
\bibinfo{author}{{Ahumada}, T.} \emph{et~al.}
\newblock \bibinfo{title}{{ZTF22aaajecp/AT2022cmc: GMOS-N spectroscopy}}.
\newblock \emph{\bibinfo{journal}{GRB Coordinates Network}}
  \textbf{\bibinfo{volume}{31595}}, \bibinfo{pages}{1} (\bibinfo{year}{2022}).

\bibitem{Roming2005}
\bibinfo{author}{{Roming}, P. W.~A.} \emph{et~al.}
\newblock \bibinfo{title}{{The Swift Ultra-Violet/Optical Telescope}}.
\newblock \emph{\bibinfo{journal}{\ssr}} \textbf{\bibinfo{volume}{120}},
  \bibinfo{pages}{95--142} (\bibinfo{year}{2005}).
%\newblock \eprint{astro-ph/0507413}.

\bibitem{Cash1979}
\bibinfo{author}{{Cash}, W.}
\newblock \bibinfo{title}{{Parameter estimation in astronomy through
  application of the likelihood ratio.}}
\newblock \emph{\bibinfo{journal}{\apj}} \textbf{\bibinfo{volume}{228}},
  \bibinfo{pages}{939--947} (\bibinfo{year}{1979}).

\bibitem{Gendreau2016}
\bibinfo{author}{{Gendreau}, K.~C.} \emph{et~al.}
\newblock \bibinfo{title}{{The Neutron star Interior Composition Explorer
  (NICER): design and development}}.
\newblock In \emph{\bibinfo{booktitle}{Space Telescopes and Instrumentation
  2016: Ultraviolet to Gamma Ray}}, vol. \bibinfo{volume}{9905} of
  \emph{\bibinfo{series}{Society of Photo-Optical Instrumentation Engineers
  (SPIE) Conference Series}}, \bibinfo{pages}{99051H} (\bibinfo{year}{2016}).

\bibitem{Remillard2022}
\bibinfo{author}{{Remillard}, R.~A.} \emph{et~al.}
\newblock \bibinfo{title}{{An Empirical Background Model for the NICER X-Ray
  Timing Instrument}}.
\newblock \emph{\bibinfo{journal}{\aj}} \textbf{\bibinfo{volume}{163}},
  \bibinfo{pages}{130} (\bibinfo{year}{2022}).
%\newblock \eprint{2105.09901}.

\bibitem{HI4PI2016}
\bibinfo{author}{{HI4PI Collaboration}} \emph{et~al.}
\newblock \bibinfo{title}{{HI4PI: A full-sky H I survey based on EBHIS and
  GASS}}.
\newblock \emph{\bibinfo{journal}{\aap}} \textbf{\bibinfo{volume}{594}},
  \bibinfo{pages}{A116} (\bibinfo{year}{2016}).
%\newblock \eprint{1610.06175}.
\end{thebibliography}

\end{methods}

%\clearpage

\begin{extended_data}

\renewcommand{\thefigure}{\arabic{figure}~Extended~Data}
\renewcommand{\thefigure}{Extended Data Figure \arabic{figure}}
\renewcommand{\figurename}{}
\setcounter{figure}{0}

\renewcommand{\thetable}{\arabic{table}~Extended~Data}
\renewcommand{\thetable}{Extended Data Table \arabic{table}}
\renewcommand{\tablename}{}
\setcounter{table}{0}

\begin{table}
\begin{center}
\caption{Equivalent widths measured in the X-shooter spectrum, in observer frame.}
\label{table:EWs}
\begin{tabular}{ccccc} 
\hline\hline
Observed $\lambda$ [\AA]& Feature & $z$ & EW {\AA} \\ 
\hline
4067.87 & AlIII 1854.72	& 1.19325		& 1.37 $\pm$ 0.23 \\ 
4084.84 & AlIII 1862.78	& 1.19287		& 1.01 $\pm$ 0.25 \\ 
5140.88 & FeII 2344.21 	& 1.19301		& 3.68 $\pm$ 0.35 \\ 
5225.70 & FeII 2382.77  & 1.19312		& 3.17 $\pm$ 0.32 \\ 
5673.57 & FeII 2586.65 	& 1.19340		& 2.54 $\pm$ 0.27 \\ 
5703.60 & FeII 2600.17 	& 1.19355 		& 4.03 $\pm$ 0.37 \\ 
5716.34 & MnII 2606.46 	& 1.19314 		& 1.49 $\pm$ 0.28 \\
6132.95 & MgII 2796.35 	& 1.19320	    & 3.98 $\pm$ 0.30 \\ 
6148.52 & MgII 2803.53 	& 1.19314	    & 4.68 $\pm$ 0.34  \\ 
6256.81 & MgI 2852.96 	& 1.19309	    & 2.60 $\pm$ 0.18 \\
8631.28 & CaII 3934.78 	& 1.19359       & 1.83 $\pm$ 0.17 \\
8707.82 & CaII 3969.59 	& 1.19363	    & 1.14 $\pm$ 0.14 \\
\hline
\end{tabular}
\end{center}
\end{table}

\begin{figure*}[ht]
    \centering
    \includegraphics[width=0.9\columnwidth]{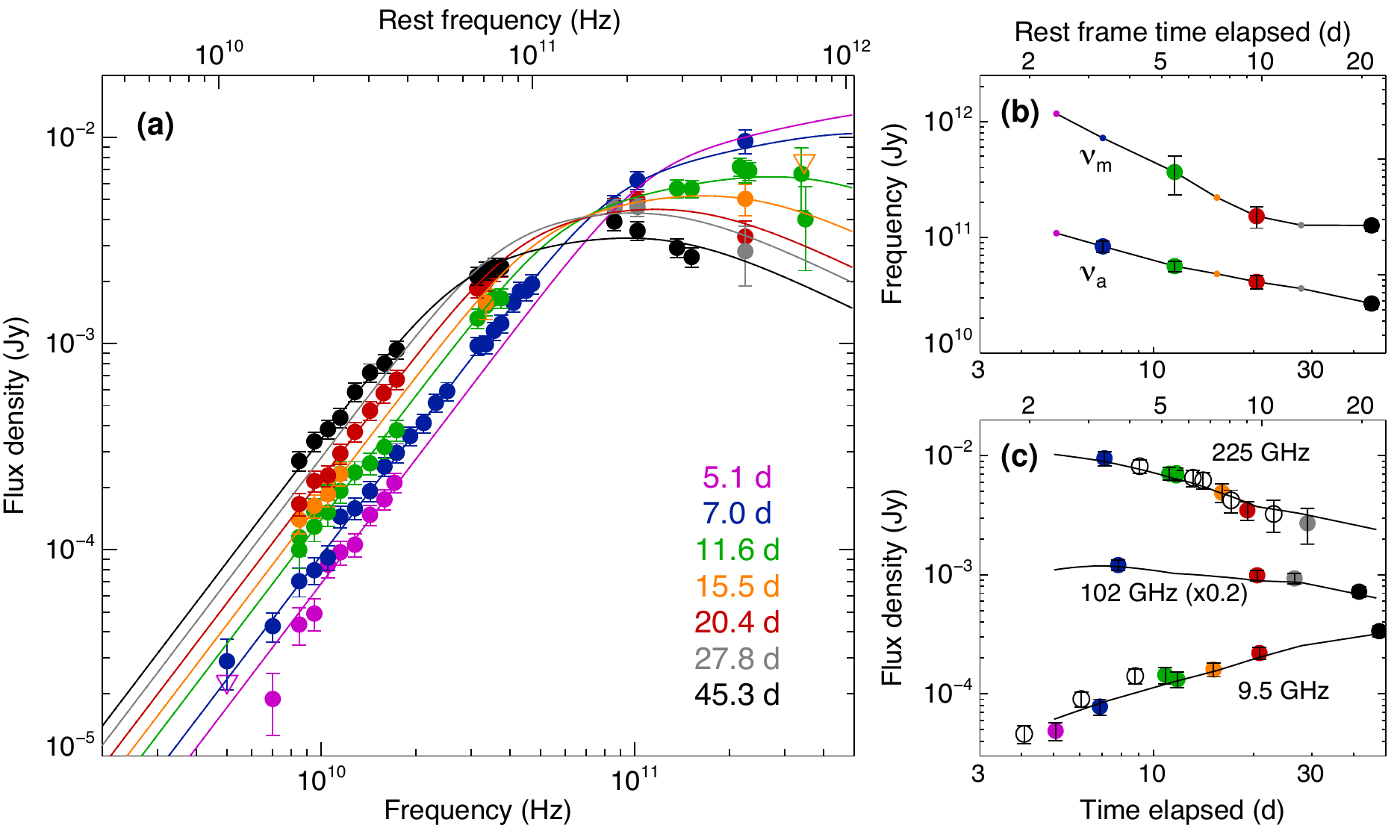}
    \caption{Time-dependent long-wavelength spectral evolution of AT2022cmc from observations with the VLA, NOEMA, SMA, JCMT, and ATCA.
(a): Co-eval energy distributions for AT2022cmc.  Measurements are shown as circles with error bars (a 10\% systematic component has been included) color-coded by observation epoch.  A synchrotron broken power-law model has been fit to the data assuming a spectral index ($F_\nu \propto \nu^\alpha$) of $\alpha=+2$ at low frequencies  ($\nu < \nu_a$), $\alpha=1/3$ at mid-frequencies ($\nu_a < \nu < \nu_m$), and $\alpha=-1$ at high frequencies ($\nu_m < \nu$).}  
For the SEDs at 7.0, 11.6, 20.4, and 45.3 days (observer-frame) the model is fit with all parameters free to vary; for the remaining epoch the break frequencies are fixed based on a plausible extrapolation/interpolation of the other epochs and only the flux scale is fit.
(b): Evolution of the spectral break frequencies.  Larger circles with error bars show measured break frequencies; the remaining points are interpolated.
(c): Light curves at 9.5, 102, and 235 GHz with predictions of the interpolated SED model overplotted.  (Unfilled circles show additional measurements not used in the co-eval SEDs.)
The general evolution of the SED and light curve are very similar to what was seen in Swift J1644\cite{Eftekhari2018}, with a low-frequency SED that remains self-absorbed out to late times. \label{fig:radio_SED}.
\end{figure*}

\begin{figure}[htbp!]
    \centering
    \includegraphics{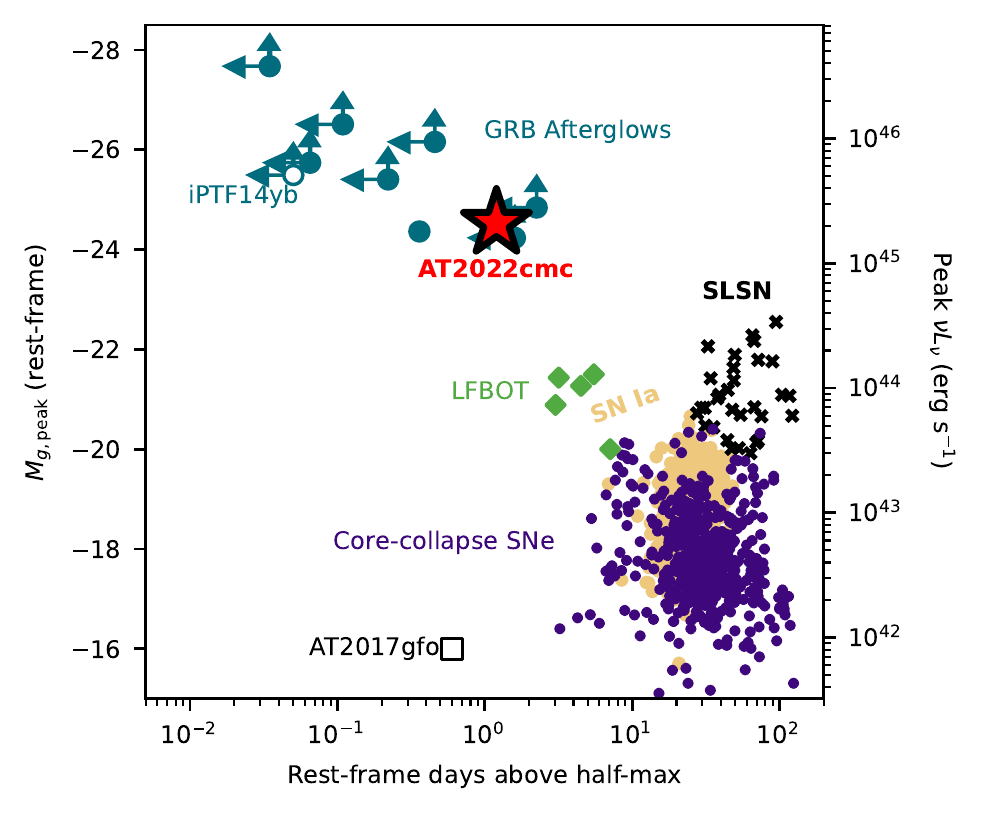}
    \caption{Duration and luminosity of optical transients compared to \at, including superluminous SNe (SLSN), Type Ia SNe (SN Ia), core-collapse SNe \cite{Fremling2020,Perley2020,Ho2021_RET}, luminous fast blue optical transients(LFBOTs) \cite{Prentice2018,Perley2019,Perley2021,Ho2020_Koala,Ho2021_RET,Yao2021arXiv,Ho2022arXiv}, GRB afterglows \cite{Cenko2015,Ho2022arXiv}, and the kilonova AT\,2017gfo \cite{CoFo2017,Cowperthwaite2017,Kasliwal2017,Drout2017,Villar2017}.  \label{fig:optical_parspace_compare}}
\end{figure}

\begin{figure}[t]
 \centering
\includegraphics[width=\textwidth]{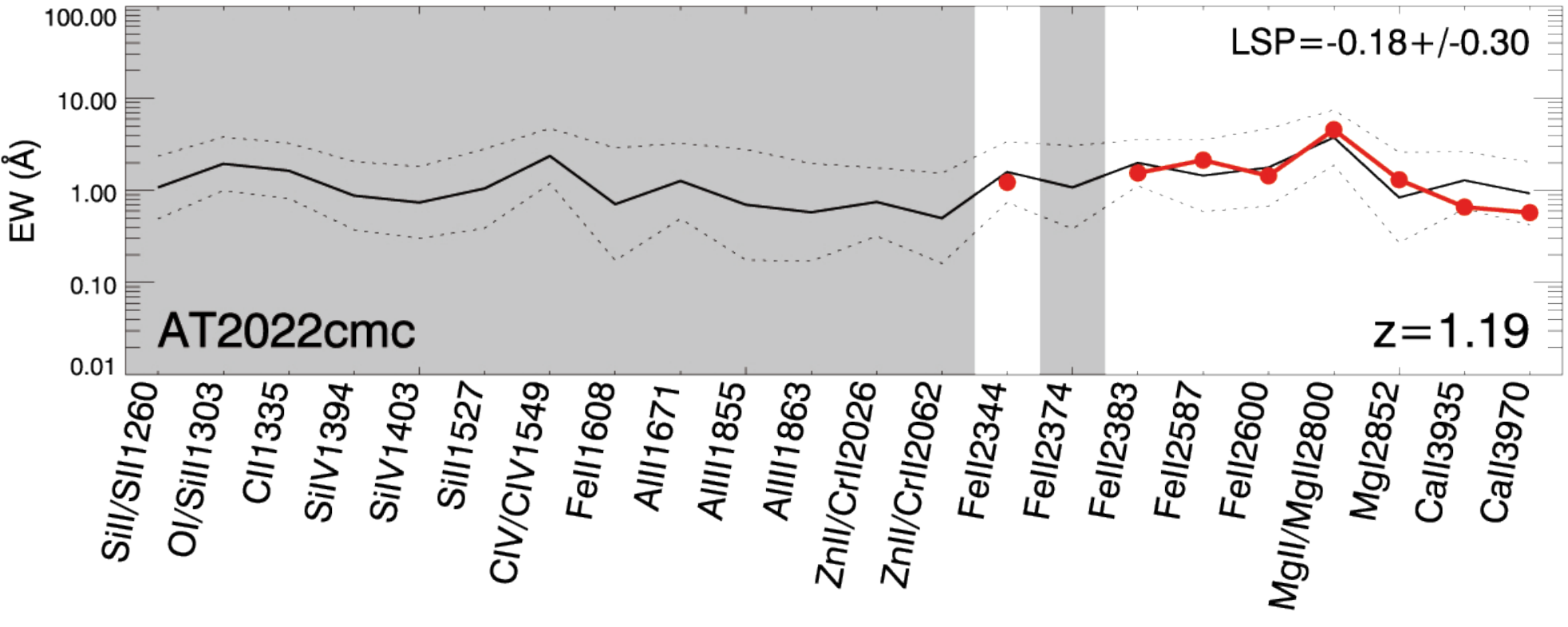}
  \caption{Line strength diagram comparing the equivalent widths (EWs) of the absorption features measured in the X-shooter spectrum of AT2022cmc (in red) with a sample of GRB afterglow spectra. The thick black line marks the average strength of the sample and the dotted lines the standard deviation in log normal space. The shaded features are those for which we cannot provide reliable measurements because they fall outside the spectral range of our data, or because they are in a region of the spectrum affected by a very low signal to noise ratio or by telluric features. The features seen in the line of sight of AT2022cmc have very similar strength as those of a typical GRB.}
 \label{fig:LSD}
\end{figure}

\begin{figure}[htbp!]
    \centering
    \includegraphics[width=\columnwidth]{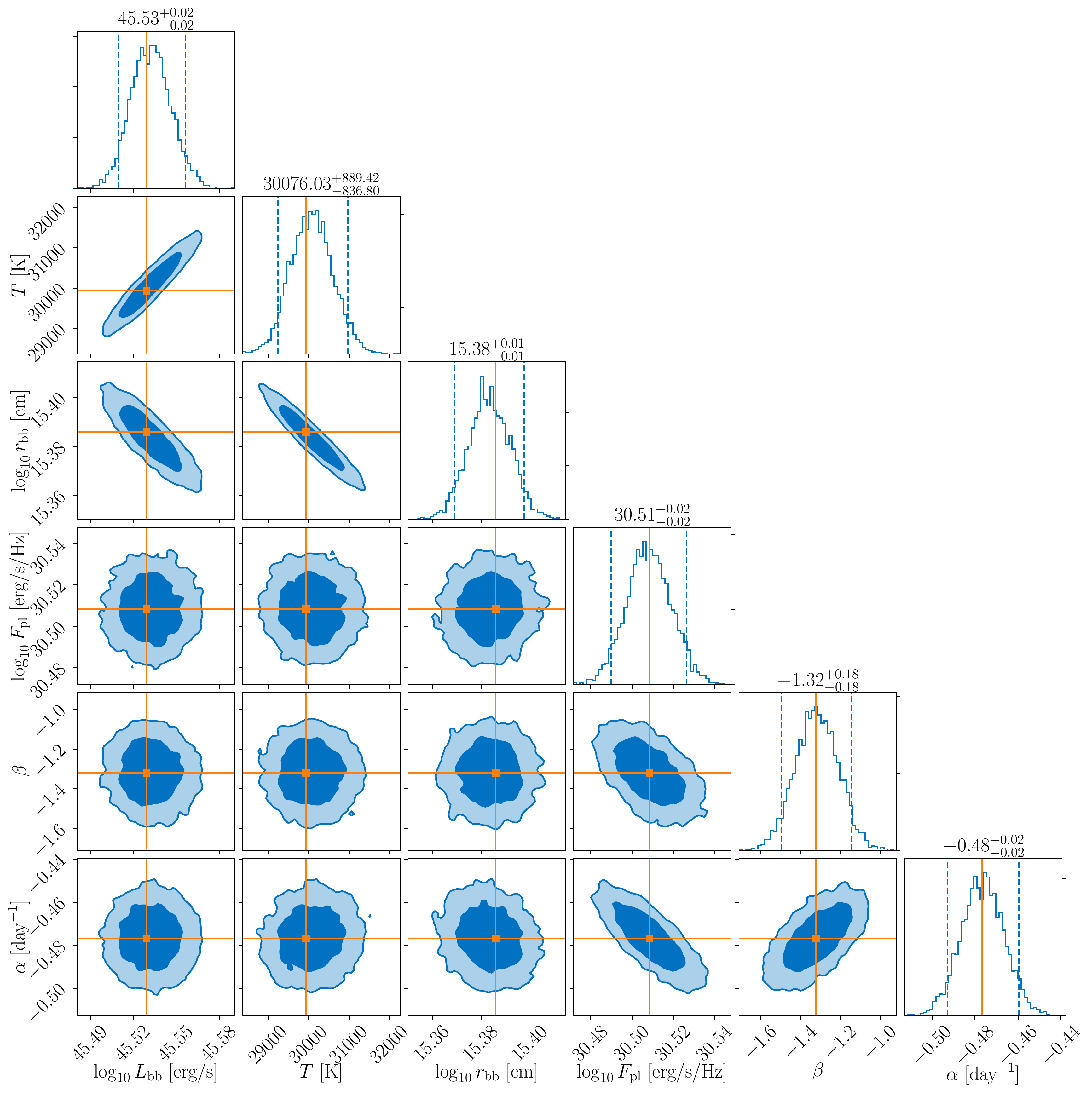}
    \caption{Marginalized histograms for the optical light curve modelling discussed in Section \ref{sec:lightcurve_modelling}. The parameter estimates given correspond to median and 90\% Bayesian credible intervals, as marked by the blue dashed vertical and horizontal lines. The best-fit (maximum likelihood) parameters are marked with the orange lines. The $68\%$ ($95\%$) credible regions are colored in dark (light) blue. \label{fig:BB_PL_parameters}}
\end{figure}

\begin{figure}[htbp!]
    \centering
    \includegraphics[width=0.8\columnwidth]{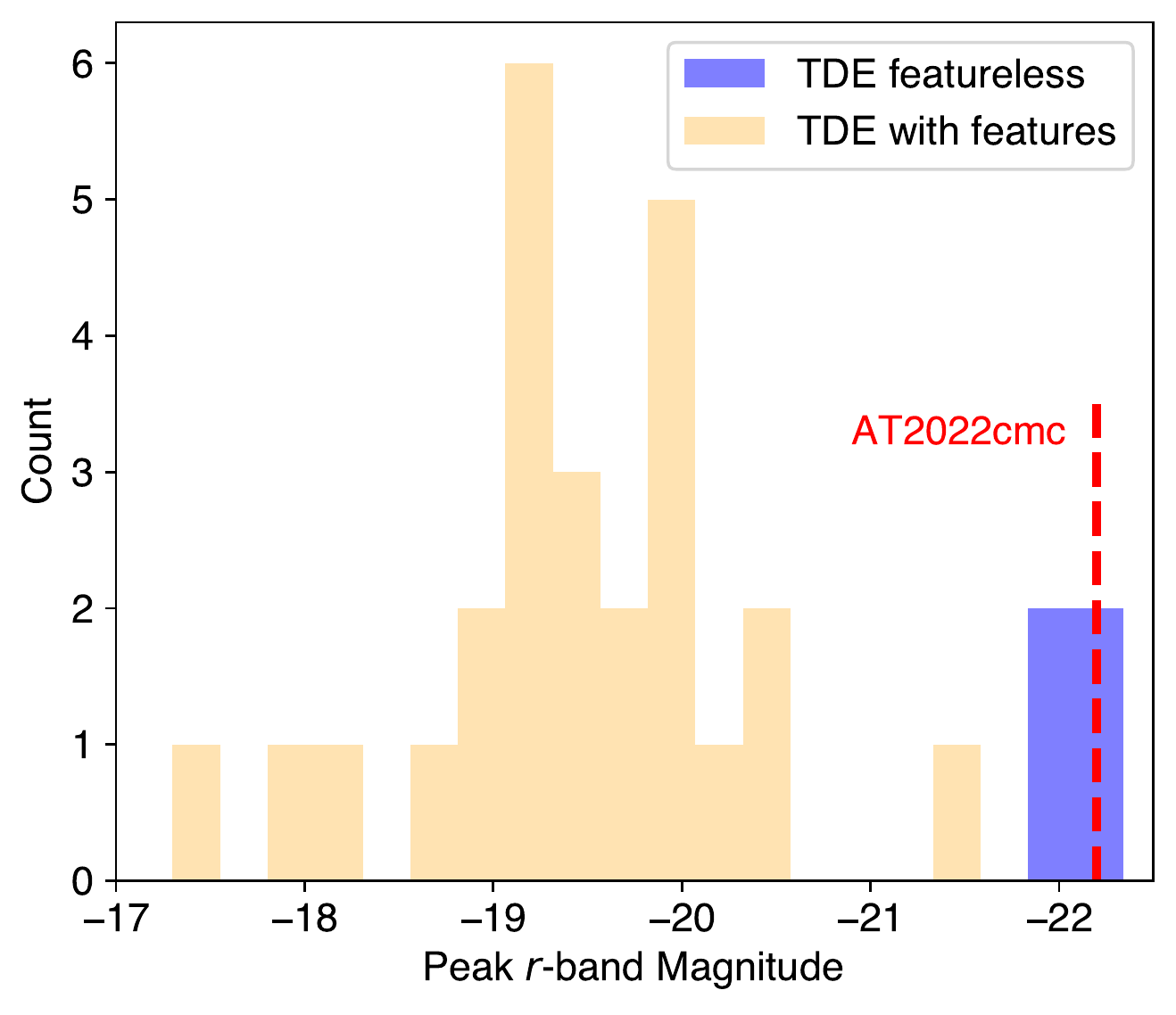}
    \caption{Distribution of the peak absolute magnitudes ($r$-band) for a population of TDEs\cite{Hammerstein2022arXiv}. Featureless TDEs are consistently brighter than TDEs that show broad features in their optical spectra. The absolute magnitude of \at\ when the slow/blue component dominates falls in the ballpark of featureless TDE peak luminosities, which supports a possible connection between TDEs with relativistic jets and the class of featureless TDEs.     \label{fig:featureless_histo}}
\end{figure}

\begin{figure}[htbp!]
    \centering
    \includegraphics[width=0.75\columnwidth]{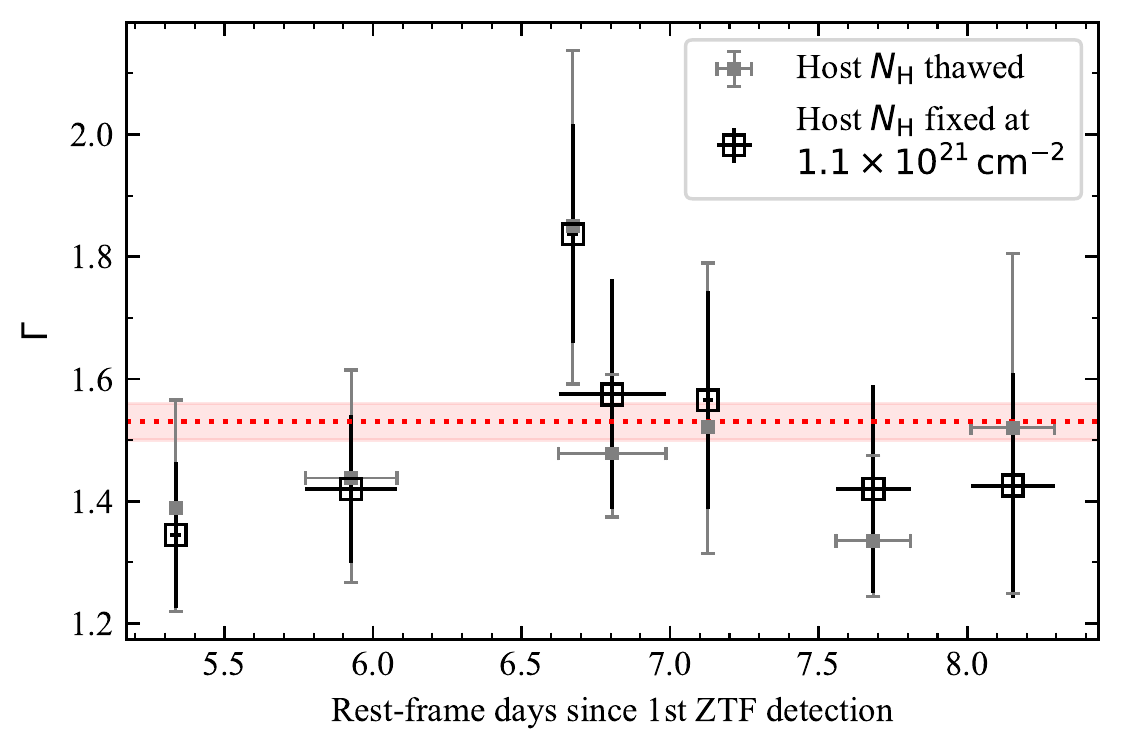}
    \caption{{\bf Evolution of the power-law photon index $\Gamma_X$ in the first seven XRT observations.} All measurements are consistent with the best-fit $\Gamma_X$ in the first NICER observation (section \ref{Methods:NICER}), as marked by the horizontal dotted line. \label{fig:XRT_Gamma_evol}}
\end{figure}

\begin{table*}
\centering
\begin{tabular}{ccccccc}
\hline\hline
obsID & Start Date & $\Delta t$ & Exp. & Net Count Rate & Obs. Flux & Obs. Luminosity \\
      & (UT)   & (days) &     (s) & ($\rm count\,s^{-1}$)  & ($10^{-13}\,{\rm erg\,s^{-1}\,cm^{-2}}$) & ($10^{45}\,{\rm erg\,s^{-1}}$) \\ 
\hline
15023001 & 2022-02-23.15 & 5.34 & 2629 & $0.155 \pm 0.008$ & $65.03 \pm 3.32$ & $55.48 \pm 2.83$ \\
15023002 & 2022-02-24.44 & 5.93 & 3096 & $0.129 \pm 0.007$ & $53.98 \pm 2.78$ & $46.05 \pm 2.37$ \\
15023003 & 2022-02-26.37 & 6.80 & 2737 & $0.064 \pm 0.005$ & $26.95 \pm 2.12$ & $22.99 \pm 1.81$ \\
15023004 & 2022-02-26.08 & 6.67 & 2829 & $0.064 \pm 0.005$ & $26.95 \pm 2.12$ & $22.99 \pm 1.81$ \\
15023005 & 2022-02-27.08 & 7.13 & 2599 & $0.079 \pm 0.006$ & $33.31 \pm 2.41$ & $28.42 \pm 2.06$ \\
15023006 & 2022-02-28.30 & 7.68 & 2694 & $0.080 \pm 0.006$ & $33.46 \pm 2.37$ & $28.55 \pm 2.02$ \\
15023007 & 2022-03-01.33 & 8.15 & 2654 & $0.066 \pm 0.005$ & $27.58 \pm 2.19$ & $23.53 \pm 1.87$ \\
15023008 & 2022-03-05.26 & 9.95 & 664 & $0.029 \pm 0.008$ & $12.27 \pm 3.16$ & $10.47 \pm 2.69$ \\
15023009 & 2022-03-07.46 & 10.95 & 2634 & $0.030 \pm 0.004$ & $12.74 \pm 1.51$ & $10.87 \pm 1.29$ \\
15023010 & 2022-03-10.61 & 12.39 & 2829 & $0.016 \pm 0.003$ & $6.89 \pm 1.11$ & $5.88 \pm 0.94$ \\
15023011 & 2022-03-13.63 & 13.76 & 1485 & $0.017 \pm 0.004$ & $6.95 \pm 1.56$ & $5.93 \pm 1.33$ \\
\hline
\end{tabular}
\caption{ \textbf{XRT observations of \at.} 
$\Delta t$ is rest-frame days since the first ZTF detection epoch.
The count rate, flux, and luminosity are given in the observer frame 0.3--10\,keV. 
The uncertainties are represented by the 68\% confidence intervals, assuming Poisson symmetrical errors. } 
\label{table:xrt}
\end{table*}

\end{extended_data}

\clearpage

\begin{supplement}

\renewcommand{\thefigure}{Supplementary Information Figure~\arabic{figure}}
\renewcommand{\figurename}{}
\setcounter{figure}{0}

\renewcommand{\thetable}{Supplementary Information Table~\arabic{table}}
 \renewcommand{\tablename}{}
\setcounter{table}{0}

\begin{center}
\begin{longtable}{llllll}
%\begin{table}%{llllll}
\caption{Infrared/Optical/Ultraviolet Photometry table, not corrected for the Galactic extinction. The second column reports the rest frame time from the first ZTF detection. }
%\begin{tabular}{llllll}
\label{table:photometry} \\
\hline
\hline
Date UT & $\Delta t$ & Filter & Mag & eMag & Instrument \\ 
 & (days) & & (AB) &  (AB) \\
\hline
2022-01-21 08:57 & -9.6093 & $ZTF_r$ & $>19.8$ & - & ZTF \\
2022-01-21 09:54 & -9.5913 & $ZTF_g$ & $>19.7$ & - & ZTF \\
2022-02-11 10:42 & 0.0000 & $ZTF_r$ & 20.71 & 0.17 & ZTF \\
2022-02-11 11:08 & 0.0081 & $ZTF_g$ & 20.91 & 0.17 & ZTF \\
2022-02-12 09:58 & 0.4422 & $ZTF_r$ & 19.08 & 0.09 & ZTF \\
2022-02-12 09:59 & 0.4424 & $ZTF_r$ & 19.15 & 0.11 & ZTF \\
2022-02-12 10:21 & 0.4493 & $ZTF_i$ & 18.88 & 0.09 & ZTF \\
2022-02-12 11:27 & 0.4702 & $ZTF_g$ & 19.33 & 0.06 & ZTF \\
2022-02-12 11:28 & 0.4706 & $o$ & 19.12 & 0.19 & ATLAS \\
2022-02-12 11:33 & 0.4721 & $o$ & 18.97 & 0.18 & ATLAS \\
2022-02-12 11:39 & 0.4739 & $o$ & 18.89 & 0.17 & ATLAS \\
2022-02-12 11:46 & 0.4762 & $o$ & 19.29 & 0.28 & ATLAS \\
2022-02-12 11:57 & 0.4797 & $ZTF_r$ & 19.10 & 0.04 & ZTF \\
2022-02-12 12:03 & 0.4817 & $ZTF_g$ & 19.53 & 0.06 & ZTF \\
2022-02-12 12:34 & 0.4913 & $ZTF_r$ & 19.08 & 0.04 & ZTF \\
2022-02-13 09:47 & 0.8945 & $ZTF_r$ & 19.69 & 0.10 & ZTF \\
2022-02-13 10:20 & 0.9048 & $ZTF_g$ & 19.81 & 0.12 & ZTF \\
2022-02-13 12:15 & 0.9413 & $o$ & 19.35 & 0.26 & ATLAS \\
2022-02-13 12:19 & 0.9427 & $o$ & 19.32 & 0.26 & ATLAS \\
2022-02-13 12:21 & 0.9433 & $o$ & 19.30 & 0.28 & ATLAS \\
2022-02-13 12:29 & 0.9457 & $o$ & 19.55 & 0.30 & ATLAS \\
2022-02-14 09:36 & 1.3472 & $ZTF_r$ & 19.87 & 0.12 & ZTF \\
2022-02-14 09:39 & 1.3480 & $ZTF_r$ & 19.78 & 0.11 & ZTF \\
2022-02-14 12:50 & 1.4085 & $ZTF_g$ & 20.38 & 0.20 & ZTF \\
2022-02-14 12:52 & 1.4090 & $ZTF_g$ & 20.06 & 0.14 & ZTF \\
2022-02-15 00:53 & 1.6375 & $r$ & 20.27 & 0.13 & IO:O \\
2022-02-15 00:55 & 1.6382 & $i$ & 20.24 & 0.17 & IO:O \\
2022-02-15 00:59 & 1.6392 & $z$ & 19.79 & 0.10 & IO:O \\
2022-02-15 02:12 & 1.6624 & $g$ & 20.47 & 0.15 & IO:O \\
2022-02-15 02:16 & 1.6636 & $r$ & 20.49 & 0.09 & IO:O \\
2022-02-15 02:20 & 1.6651 & $i$ & 20.10 & 0.08 & IO:O \\
2022-02-15 02:25 & 1.6666 & $z$ & 19.87 & 0.09 & IO:O \\
2022-02-15 03:15 & 1.6825 & $g$ & 20.68 & 0.14 & IO:O \\
2022-02-15 03:20 & 1.6840 & $r$ & 20.35 & 0.10 & IO:O \\
2022-02-15 03:26 & 1.6858 & $i$ & 20.11 & 0.08 & IO:O \\
2022-02-15 03:32 & 1.6876 & $z$ & 19.99 & 0.11 & IO:O \\
2022-02-15 05:56 & 1.7335 & $g$ & 20.60 & 0.14 & IO:O \\
2022-02-15 06:02 & 1.7353 & $r$ & 20.41 & 0.10 & IO:O \\
2022-02-15 06:08 & 1.7371 & $i$ & 20.15 & 0.07 & IO:O \\
2022-02-15 06:14 & 1.7389 & $z$ & 20.02 & 0.12 & IO:O \\
2022-02-15 07:53 & 1.7704 & $ZTF_i$ & $>19.4$ & - & ZTF \\
2022-02-15 08:43 & 1.7863 & $g$ & 20.43 & 0.08 & DECam \\
2022-02-15 08:43 & 1.7864 & $ZTF_g$ & $>19.4$ & - & ZTF \\
2022-02-15 08:44 & 1.7866 & $r$ & 20.53 & 0.05 & DECam \\
2022-02-15 08:45 & 1.7869 & $i$ & 20.34 & 0.06 & DECam \\
2022-02-15 09:50 & 1.8076 & $ZTF_r$ & $>19.7$ & - & ZTF \\
2022-02-15 20:11 & 2.0043 & $r$ & 20.67 & 0.09 & GITCamera \\
2022-02-15 23:43 & 2.0714 & $r$ & 20.68 & 0.08 & GITCamera \\
2022-02-16 01:21 & 2.1024 & $g$ & $>21.0$ & - & IO:O \\
2022-02-16 01:32 & 2.1056 & $r$ & $>20.8$ & - & IO:O \\
2022-02-16 01:41 & 2.1086 & $i$ & $>20.5$ & - & IO:O \\
2022-02-16 01:50 & 2.1115 & $z$ & $>19.8$ & - & IO:O \\
2022-02-16 08:59 & 2.2473 & $g$ & 21.16 & 0.28 & DECam \\
2022-02-16 09:00 & 2.2476 & $r$ & 21.06 & 0.13 & DECam \\
2022-02-16 09:01 & 2.2479 & $i$ & 20.93 & 0.11 & DECam \\
2022-02-16 09:02 & 2.2482 & $z$ & 21.04 & 0.09 & DECam \\
2022-02-16 21:03 & 2.4764 & $r$ & 20.86 & 0.07 & GITCamera \\
2022-02-17 08:52 & 2.7012 & $g$ & 21.09 & 0.22 & DECam \\
2022-02-17 08:53 & 2.7015 & $r$ & 21.15 & 0.15 & DECam \\
2022-02-17 08:54 & 2.7018 & $i$ & 21.28 & 0.19 & DECam \\
2022-02-17 08:55 & 2.7021 & $z$ & 21.15 & 0.14 & DECam \\
2022-02-17 08:57 & 2.7026 & $z$ & 21.38 & 0.15 & DECam \\
2022-02-17 21:55 & 2.9491 & $r$ & 21.16 & 0.10 & GITCamera \\
2022-02-18 01:16 & 3.0128 & $J$ & 21.21 & 0.05 & EMIR \\
2022-02-18 01:24 & 3.0152 & $i$ & 21.47 & 0.24 & IO:O \\
2022-02-18 01:27 & 3.0161 & $H$ & 21.36 & 0.06 & EMIR \\
2022-02-18 01:42 & 3.0208 & $Ks$ & 20.64 & 0.13 & EMIR \\
2022-02-18 04:28 & 3.0736 & $i$ & 21.00 & 0.15 & IO:O \\
2022-02-18 04:54 & 3.0816 & $r$ & 21.26 & 0.08 & CAFOS \\
2022-02-18 05:36 & 3.0949 & $i$ & 21.12 & 0.12 & CAFOS \\
2022-02-18 08:52 & 3.1571 & $g$ & 21.19 & 0.15 & DECam \\
2022-02-18 08:53 & 3.1576 & $g$ & 21.11 & 0.13 & DECam \\
2022-02-18 08:55 & 3.1580 & $i$ & 21.14 & 0.11 & DECam \\
2022-02-18 08:56 & 3.1584 & $i$ & 21.34 & 0.13 & DECam \\
2022-02-18 08:58 & 3.1589 & $z$ & 21.00 & 0.09 & DECam \\
2022-02-18 08:59 & 3.1594 & $z$ & 21.39 & 0.16 & DECam \\
2022-02-19 09:00 & 3.6155 & $g$ & 21.50 & 0.17 & DECam \\
2022-02-19 09:01 & 3.6160 & $g$ & 21.69 & 0.23 & DECam \\
2022-02-19 09:03 & 3.6165 & $i$ & 21.17 & 0.09 & DECam \\
2022-02-19 09:04 & 3.6170 & $i$ & 21.25 & 0.09 & DECam \\
2022-02-19 09:06 & 3.6175 & $z$ & 21.51 & 0.14 & DECam \\
2022-02-19 09:08 & 3.6180 & $z$ & 21.38 & 0.14 & DECam \\
2022-02-20 09:03 & 4.0726 & $i$ & 21.68 & 0.18 & DECam \\
2022-02-20 09:07 & 4.0737 & $z$ & 21.74 & 0.21 & DECam \\
2022-02-21 08:51 & 4.5249 & $i$ & 21.72 & 0.23 & DECam \\
2022-02-21 08:53 & 4.5254 & $i$ & 21.35 & 0.24 & DECam \\
2022-02-21 08:55 & 4.5259 & $z$ & 21.55 & 0.24 & DECam \\
2022-02-22 08:51 & 4.9808 & $i$ & 21.82 & 0.20 & DECam \\
2022-02-22 08:53 & 4.9814 & $i$ & 21.90 & 0.19 & DECam \\
2022-02-23 03:25 & 5.3337 & $UVW1$ & 21.54 & 0.33 & UVOT \\
2022-02-23 03:28 & 5.3345 & $U$ & $>20.7$ & - & UVOT \\
2022-02-23 03:29 & 5.3349 & $B$ & $>19.9$ & - & UVOT \\
2022-02-23 03:32 & 5.3359 & $UVW2$ & 22.28 & 0.34 & UVOT \\
2022-02-23 03:37 & 5.3372 & $V$ & $>19.0$ & - & UVOT \\
2022-02-23 03:41 & 5.3385 & $UVM2$ & 21.42 & 0.19 & UVOT \\
2022-02-23 07:24 & 5.4091 & $g$ & 21.55 & 0.05 & DECam \\
2022-02-23 07:29 & 5.4108 & $r$ & 21.59 & 0.07 & DECam \\
2022-02-23 07:33 & 5.4120 & $r$ & 21.60 & 0.07 & DECam \\
2022-02-23 07:37 & 5.4132 & $i$ & 21.59 & 0.08 & DECam \\
2022-02-23 07:41 & 5.4144 & $i$ & 21.61 & 0.08 & DECam \\
2022-02-23 07:44 & 5.4156 & $z$ & 21.69 & 0.13 & DECam \\
2022-02-23 20:21 & 5.6553 & $r$ & 21.46 & 0.14 & GITCamera \\
2022-02-23 22:31 & 5.6964 & $g$ & 21.47 & 0.11 & GITCamera \\
2022-02-24 07:26 & 5.8657 & $g$ & 21.53 & 0.06 & DECam \\
2022-02-24 07:28 & 5.8664 & $r$ & 21.61 & 0.05 & DECam \\
2022-02-24 07:30 & 5.8670 & $i$ & 21.53 & 0.08 & DECam \\
2022-02-24 07:32 & 5.8677 & $z$ & 21.98 & 0.22 & DECam \\
2022-02-24 08:06 & 5.8784 & $UVW1$ & $>21.5$ & - & UVOT \\
2022-02-24 08:08 & 5.8793 & $U$ & $>20.6$ & - & UVOT \\
2022-02-24 08:10 & 5.8796 & $B$ & $>19.8$ & - & UVOT \\
2022-02-24 08:13 & 5.8806 & $UVW2$ & 22.17 & 0.32 & UVOT \\
2022-02-24 08:17 & 5.8818 & $V$ & $>19.0$ & - & UVOT \\
2022-02-24 08:20 & 5.8828 & $UVM2$ & 21.28 & 0.20 & UVOT \\
2022-02-25 21:25 & 6.5874 & $r$ & 21.71 & 0.12 & GITCamera \\
2022-02-26 01:42 & 6.6689 & $UVW1$ & 21.39 & 0.29 & UVOT \\
2022-02-26 01:44 & 6.6696 & $U$ & 20.58 & 0.32 & UVOT \\
2022-02-26 01:46 & 6.6701 & $B$ & $>20.0$ & - & UVOT \\
2022-02-26 01:50 & 6.6714 & $UVW2$ & 22.38 & 0.34 & UVOT \\
2022-02-26 01:53 & 6.6725 & $V$ & $>19.1$ & - & UVOT \\
2022-02-26 02:00 & 6.6745 & $UVM2$ & 21.55 & 0.20 & UVOT \\
2022-02-26 08:00 & 6.7887 & $r$ & 21.38 & 0.09 & SEDM \\
2022-02-26 08:11 & 6.7921 & $g$ & 21.70 & 0.11 & SEDM \\
2022-02-26 08:22 & 6.7955 & $i$ & 21.60 & 0.18 & SEDM \\
2022-02-26 08:50 & 6.8045 & $UVW1$ & 21.12 & 0.25 & UVOT \\
2022-02-26 08:52 & 6.8050 & $U$ & $>20.7$ & - & UVOT \\
2022-02-26 08:53 & 6.8052 & $B$ & $>19.9$ & - & UVOT \\
2022-02-26 08:54 & 6.8057 & $UVW2$ & 22.19 & 0.30 & UVOT \\
2022-02-26 08:57 & 6.8066 & $V$ & $>19.0$ & - & UVOT \\
2022-02-26 08:59 & 6.8072 & $UVM2$ & 21.65 & 0.23 & UVOT \\
2022-02-27 01:37 & 7.1233 & $UVW1$ & 21.20 & 0.25 & UVOT \\
2022-02-27 01:39 & 7.1240 & $U$ & $>20.7$ & - & UVOT \\
2022-02-27 01:40 & 7.1244 & $B$ & $>19.9$ & - & UVOT \\
2022-02-27 01:45 & 7.1257 & $UVW2$ & 22.17 & 0.30 & UVOT \\
2022-02-27 01:48 & 7.1267 & $V$ & $>19.0$ & - & UVOT \\
2022-02-27 01:54 & 7.1286 & $UVM2$ & 21.69 & 0.25 & UVOT \\
2022-02-27 06:21 & 7.2132 & $r$ & 21.36 & 0.16 & SEDM \\
2022-02-27 06:27 & 7.2151 & $g$ & $>21.4$ & - & SEDM \\
2022-02-27 06:33 & 7.2169 & $i$ & $>21.2$ & - & SEDM \\
2022-02-27 19:59 & 7.4724 & $r$ & 21.63 & 0.09 & GITCamera \\
2022-02-27 22:04 & 7.5120 & $g$ & 21.59 & 0.09 & GITCamera \\
2022-02-28 02:07 & 7.5888 & $g$ & 21.54 & 0.06 & ALFOSC \\
2022-02-28 02:19 & 7.5925 & $r$ & 21.61 & 0.04 & ALFOSC \\
2022-02-28 02:34 & 7.5972 & $i$ & 21.78 & 0.06 & ALFOSC \\
2022-02-28 06:02 & 7.6632 & $g$ & $>19.8$ & - & SEDM \\
2022-02-28 06:13 & 7.6666 & $i$ & $>19.8$ & - & SEDM \\
2022-02-28 07:02 & 7.6822 & $UVW1$ & 20.80 & 0.19 & UVOT \\
2022-02-28 07:05 & 7.6830 & $U$ & $>20.7$ & - & UVOT \\
2022-02-28 07:06 & 7.6834 & $B$ & $>19.9$ & - & UVOT \\
2022-02-28 07:09 & 7.6844 & $UVW2$ & 21.93 & 0.26 & UVOT \\
2022-02-28 07:13 & 7.6858 & $V$ & $>19.0$ & - & UVOT \\
2022-02-28 07:17 & 7.6871 & $UVM2$ & 21.49 & 0.21 & UVOT \\
2022-03-01 07:44 & 8.1514 & $UVW1$ & 21.51 & 0.31 & UVOT \\
2022-03-01 07:46 & 8.1523 & $U$ & $>20.7$ & - & UVOT \\
2022-03-01 07:48 & 8.1527 & $B$ & $>19.9$ & - & UVOT \\
2022-03-01 07:51 & 8.1537 & $UVW2$ & $>22.4$ & - & UVOT \\
2022-03-01 07:55 & 8.1550 & $V$ & $>19.0$ & - & UVOT \\
2022-03-01 07:59 & 8.1563 & $UVM2$ & 21.42 & 0.20 & UVOT \\
2022-03-02 03:03 & 8.5185 & $g$ & 21.73 & 0.10 & IO:O \\
2022-03-02 03:07 & 8.5199 & $r$ & 21.71 & 0.13 & IO:O \\
2022-03-02 03:11 & 8.5212 & $i$ & 21.75 & 0.17 & IO:O \\
2022-03-02 03:16 & 8.5225 & $z$ & $>21.7$ & - & IO:O \\
2022-03-03 05:29 & 9.0209 & $Ks$ & 22.77 & 0.28 & EMIR \\
2022-03-04 06:10 & 9.4896 & $g$ & 21.59 & 0.03 & ALFOSC \\
2022-03-04 06:21 & 9.4933 & $r$ & 21.73 & 0.04 & ALFOSC \\
2022-03-04 06:33 & 9.4970 & $i$ & 21.87 & 0.07 & ALFOSC \\
2022-03-05 06:16 & 9.9477 & $UVM2$ & $>21.2$ & - & UVOT \\
2022-03-05 06:19 & 9.9487 & $UVW1$ & $>20.9$ & - & UVOT \\
2022-03-05 06:21 & 9.9493 & $U$ & $>20.2$ & - & UVOT \\
2022-03-05 06:24 & 9.9501 & $UVW2$ & $>21.4$ & - & UVOT \\
2022-03-06 02:09 & 10.3253 & $g$ & 22.03 & 0.12 & IO:O \\
2022-03-06 02:13 & 10.3266 & $r$ & 21.92 & 0.15 & IO:O \\
2022-03-06 02:17 & 10.3279 & $i$ & 21.64 & 0.17 & IO:O \\
2022-03-06 02:21 & 10.3293 & $z$ & $>21.5$ & - & IO:O \\
2022-03-07 02:22 & 10.7856 & $g$ & 21.80 & 0.07 & IO:O \\
2022-03-07 02:28 & 10.7875 & $r$ & 21.72 & 0.08 & IO:O \\
2022-03-07 02:35 & 10.7895 & $i$ & 21.74 & 0.13 & IO:O \\
2022-03-08 05:08 & 11.2940 & $r$ & $>20.4$ & - & SEDM \\
2022-03-08 05:18 & 11.2974 & $g$ & $>20.8$ & - & SEDM \\
2022-03-08 05:29 & 11.3008 & $i$ & $>20.5$ & - & SEDM \\
2022-03-08 20:04 & 11.5777 & $r$ & 21.74 & 0.10 & GITCamera \\
2022-03-08 20:12 & 11.5803 & $F606W$ & $21.82$ & 0.03 & WFC3 \\
2022-03-08 20:12 & 11.5803 & $F160W$ & $22.64$ & 0.05 & WFC3 \\
2022-03-09 02:07 & 11.6929 & $g$ & 21.68 & 0.08 & ALFOSC \\
2022-03-09 02:19 & 11.6966 & $r$ & 21.76 & 0.05 & ALFOSC \\
2022-03-09 02:31 & 11.7002 & $i$ & 21.92 & 0.08 & ALFOSC \\
2022-03-09 20:13 & 12.0365 & $g$ & $>21.8$ & - & GITCamera \\
2022-03-10 01:21 & 12.1342 & $g$ & 21.95 & 0.14 & IO:O \\
2022-03-10 01:27 & 12.1362 & $r$ & 21.93 & 0.16 & IO:O \\
2022-03-10 01:33 & 12.1381 & $i$ & 21.73 & 0.16 & IO:O \\
2022-03-10 14:38 & 12.3865 & $UVM2$ & 21.99 & 0.30 & UVOT \\
2022-03-10 14:41 & 12.3874 & $UVW1$ & $>21.9$ & - & UVOT \\
2022-03-10 14:42 & 12.3879 & $U$ & $>21.1$ & - & UVOT \\
2022-03-10 14:44 & 12.3886 & $UVW2$ & 22.08 & 0.26 & UVOT \\
2022-03-11 18:53 & 12.9235 & $r$ & 21.49 & 0.08 & GITCamera \\
2022-03-11 20:38 & 12.9566 & $g$ & 21.75 & 0.13 & GITCamera \\
2022-03-12 06:05 & 13.1360 & $r$ & 21.57 & 0.14 & SEDM \\
2022-03-12 06:16 & 13.1395 & $g$ & $>21.6$ & - & SEDM \\
2022-03-12 15:17 & 13.3109 & $J$ & $>22.1$ & - & WIRC \\
2022-03-13 03:47 & 13.5483 & $r$ & 21.81 & 0.06 & ALFOSC \\
2022-03-13 03:58 & 13.5520 & $i$ & 21.88 & 0.05 & ALFOSC \\
2022-03-13 04:44 & 13.5664 & $Ks$ & 23.11 & 0.32 & EMIR \\
2022-03-13 05:53 & 13.5883 & $g$ & 21.75 & 0.05 & ALFOSC \\
2022-03-13 06:05 & 13.5920 & $r$ & 21.82 & 0.03 & ALFOSC \\
2022-03-13 06:16 & 13.5957 & $i$ & 21.89 & 0.04 & ALFOSC \\
2022-03-13 15:03 & 13.7624 & $UVM2$ & 21.72 & 0.36 & UVOT \\
2022-03-13 15:05 & 13.7632 & $UVW1$ & $>21.4$ & - & UVOT \\
2022-03-13 15:06 & 13.7636 & $U$ & $>20.7$ & - & UVOT \\
2022-03-13 15:09 & 13.7643 & $UVW2$ & $>22.1$ & - & UVOT \\
2022-03-16 17:59 & 15.1862 & $UVM2$ & $>22.2$ & - & UVOT \\
2022-03-16 18:04 & 15.1878 & $UVW1$ & $>21.8$ & - & UVOT \\
2022-03-16 18:07 & 15.1887 & $U$ & 20.98 & 0.31 & UVOT \\
2022-03-22 20:08 & 17.9631 & $r$ & $>21.9$ & - & GITCamera \\
2022-03-22 21:55 & 17.9969 & $g$ & $>21.6$ & - & GITCamera \\
2022-03-30 02:02 & 21.2672 & $g$ & 22.20 & 0.20 & IO:O \\
2022-03-30 02:09 & 21.2692 & $r$ & 21.91 & 0.15 & IO:O \\
2022-03-30 02:15 & 21.2711 & $i$ & 21.98 & 0.18 & IO:O \\
2022-03-30 02:21 & 21.2731 & $z$ & 21.66 & 0.30 & IO:O \\
2022-03-30 02:43 & 21.2800 & $g$ & 22.09 & 0.06 & ALFOSC \\
2022-03-30 03:00 & 21.2853 & $r$ & 22.12 & 0.06 & ALFOSC \\
2022-03-30 03:16 & 21.2907 & $i$ & 22.18 & 0.07 & ALFOSC \\
\hline
%\end{tabular}
\end{longtable}
%\end{table}
\end{center}

\clearpage

\begin{center}
%\begin{table*}
\begin{longtable}{cccccc}
%\begin{tabular}{cccccc}
\caption{Radio observations.
$\Delta t$ is observer-frame days since the first ZTF detection epoch, calculated at the observation midpoint.  $\nu$ indicates the central frequency.}
\label{table:radio}\\
\hline\hline
Facility & UT date & $\Delta t$ & $\nu$ & $F_\nu$ &  RMS\\
 & & (days) & (GHz) & ($\mu$Jy) &  ($\mu$Jy) \\
\hline
VLA  &2022-02-15 12:30& 4.07&  8.5&      33&   7\\
VLA  &2022-02-15 12:30& 4.07&  9.5&      46&   7\\
VLA  &2022-02-15 12:30& 4.07& 10.5&      51&   8\\
VLA  &2022-02-15 12:30& 4.07& 11.5&      68&   9\\
VLA  &2022-02-16 12:19& 5.07&  8.5&      43&   8\\
VLA  &2022-02-16 12:19& 5.07&  9.5&      49&   7\\
VLA  &2022-02-16 12:19& 5.07& 10.5&      84&   8\\
VLA  &2022-02-16 12:19& 5.07& 11.5&      97&   9\\
VLA  &2022-02-16 13:07& 5.10& 12.8&     106&   9\\
VLA  &2022-02-16 13:07& 5.10& 14.3&     148&   8\\
VLA  &2022-02-16 13:07& 5.10& 15.9&     176&   8\\
VLA  &2022-02-16 13:07& 5.10& 17.1&     212&   8\\
VLA  &2022-02-16 13:33& 5.12&  5.0&$<$   23&   8\\
VLA  &2022-02-16 13:33& 5.12&  7.0&      19&   6\\
VLA  &2022-02-17 12:01& 6.06&  8.5&      73&  11\\
VLA  &2022-02-17 12:01& 6.06&  9.5&      90&  10\\
VLA  &2022-02-17 12:01& 6.06& 10.5&     137&  10\\
VLA  &2022-02-17 12:01& 6.06& 11.5&     141&  12\\
VLA  &2022-02-18 06:58& 6.84& 41.0&    1540&  50\\
VLA  &2022-02-18 06:58& 6.84& 43.0&    1750&  60\\
VLA  &2022-02-18 06:58& 6.84& 45.0&    1760&  70\\
VLA  &2022-02-18 06:58& 6.84& 47.0&    1900&  90\\
VLA  &2022-02-18 07:22& 6.86& 31.5&     955&  21\\
VLA  &2022-02-18 07:22& 6.86& 33.5&     969&  24\\
VLA  &2022-02-18 07:22& 6.86& 35.5&    1127&  24\\
VLA  &2022-02-18 07:22& 6.86& 37.5&    1226&  31\\
VLA  &2022-02-18 07:50& 6.88& 19.2&     349&  12\\
VLA  &2022-02-18 07:50& 6.88& 21.2&     404&  13\\
VLA  &2022-02-18 07:50& 6.88& 23.2&     506&  13\\
VLA  &2022-02-18 07:50& 6.88& 25.2&     576&  14\\
VLA  &2022-02-18 07:55& 6.88&  8.5&      69&   9\\
VLA  &2022-02-18 07:55& 6.88&  9.5&      78&   9\\
VLA  &2022-02-18 07:55& 6.88& 10.5&      90&   9\\
VLA  &2022-02-18 07:55& 6.88& 11.5&     142&  10\\
VLA  &2022-02-18 08:15& 6.90& 12.8&     156&  11\\
VLA  &2022-02-18 08:15& 6.90& 14.3&     189&  10\\
VLA  &2022-02-18 08:15& 6.90& 15.9&     249&   9\\
VLA  &2022-02-18 08:15& 6.90& 17.4&     290&  12\\
VLA  &2022-02-18 08:48& 6.92&  5.0&      28&   7\\
VLA  &2022-02-18 08:48& 6.92&  7.0&      42&   5\\
SMA  &2022-02-18 13:16& 7.11& 225 &    9540& 820\\
NOEMA&2022-02-19 06:24& 7.82&  86 &    4759&  73\\
NOEMA&2022-02-19 06:24& 7.82& 102 &    6034& 118\\
VLA  &2022-02-20 05:48& 8.80&  8.5&     109&  16\\
VLA  &2022-02-20 05:48& 8.80&  9.5&     141&  14\\
VLA  &2022-02-20 05:48& 8.80& 10.5&     184&  17\\
VLA  &2022-02-20 05:48& 8.80& 11.5&     230&  18\\
SMA  &2022-02-20 12:33& 9.08& 225 &    8172& 610\\
SMA  &2022-02-20 12:33& 9.08& 340 &    6490&2600\\
JCMT &2022-02-20 12:44& 9.08& 350 &    5900&2000\\
JCMT &2022-02-20 12:44& 9.08& 667 &$<$76200&2540\\
VLA  &2022-02-22 06:42&10.83&  8.5&     108&  19\\
VLA  &2022-02-22 06:42&10.83&  9.5&     144&  18\\
VLA  &2022-02-22 06:42&10.83& 10.5&     196&  19\\
VLA  &2022-02-22 06:42&10.83& 11.5&     220&  22\\
JCMT &2022-02-22 11:46&11.04& 350 &    4300&1700\\
JCMT &2022-02-22 11:46&11.04& 667 &$<$34500&1150\\
SMA  &2022-02-22 14:01&11.14& 225 &    7074& 490\\
SMA  &2022-02-22 14:01&11.14& 340 &    7056&2140\\
NOEMA&2022-02-23 03:13&11.69& 216 &    7142&  47\\
NOEMA&2022-02-23 03:13&11.69& 232 &    6835&  68\\
VLA  &2022-02-23 05:16&11.77& 31.5&    1350&  44\\
VLA  &2022-02-23 05:16&11.77& 33.5&    1555&  55\\
VLA  &2022-02-23 05:16&11.77& 35.5&    1687&  49\\
VLA  &2022-02-23 05:16&11.77& 37.5&    1692&  68\\
VLA  &2022-02-23 05:23&11.78&  8.5&     102&  16\\
VLA  &2022-02-23 05:23&11.78&  9.5&     132&  15\\
VLA  &2022-02-23 05:23&11.78& 10.5&     155&  15\\
VLA  &2022-02-23 05:23&11.78& 11.5&     197&  15\\
VLA  &2022-02-23 05:34&11.79& 12.8&     242&  18\\
VLA  &2022-02-23 05:34&11.79& 14.3&     268&  18\\
VLA  &2022-02-23 05:34&11.79& 15.9&     322&  19\\
VLA  &2022-02-23 05:34&11.79& 17.4&     388&  23\\
NOEMA&2022-02-23 21:14&12.44& 136 &    5464&  55\\
NOEMA&2022-02-23 21:14&12.44& 152 &    5446&  70\\
SMA  &2022-02-24 13:29&13.12& 225 &    6538& 810\\
SMA  &2022-02-24 13:29&13.12& 347 &    4262&1890\\
SMA  &2022-02-25 12:38&14.08& 225 &    6227& 780\\
SMA  &2022-02-25 12:38&14.08& 347 &$<$ 9430&3130\\
VLA  &2022-02-26 13:54&15.13&  8.5&     137&  15\\
VLA  &2022-02-26 13:54&15.13&  9.5&     160&  14\\
VLA  &2022-02-26 13:54&15.13& 10.5&     182&  17\\
VLA  &2022-02-26 13:54&15.13& 11.5&     228&  19\\
ATCA &2022-02-26 17:31&15.28& 33.5&    1560& 220\\
SMA  &2022-02-27 12:09&16.06& 225 &    4905& 730\\
SMA  &2022-02-27 12:09&16.06& 347 &$<$ 7246&2050\\
SMA  &2022-02-28 13:05&17.10& 225 &    4226& 810\\
SMA  &2022-03-02 13:49&19.13& 225 &    3485& 510\\
NOEMA&2022-03-03 22:12&20.48&  86 &    4596&  53\\
NOEMA&2022-03-03 22:12&20.48& 102 &    4953&  57\\
VLA  &2022-03-04 05:43&20.79& 31.4&    1890&  32\\
VLA  &2022-03-04 05:43&20.79& 33.5&    2046&  36\\
VLA  &2022-03-04 05:43&20.79& 35.6&    2278&  40\\
VLA  &2022-03-04 05:43&20.79& 37.5&    2389&  42\\
VLA  &2022-03-04 05:49&20.80&  8.5&     170&  12\\
VLA  &2022-03-04 05:49&20.80&  9.5&     220&  11\\
VLA  &2022-03-04 05:49&20.80& 10.5&     234&  13\\
VLA  &2022-03-04 05:49&20.80& 11.5&     299&  14\\
VLA  &2022-03-04 05:59&20.80& 12.8&     381&  13\\
VLA  &2022-03-04 05:59&20.80& 14.3&     483&  12\\
VLA  &2022-03-04 05:59&20.80& 15.8&     587&  15\\
VLA  &2022-03-04 05:59&20.80& 17.4&     684&  16\\
SMA  &2022-03-06 09:48&22.96& 225 &    3243& 930\\
NOEMA&2022-03-09 23:25&26.53&  86 &    4549&  74\\
NOEMA&2022-03-09 23:25&26.53& 102 &    4661&  76\\
SMA  &2022-03-12 11:13&29.02& 225 &    2712& 860\\
uGMRT & 2022-03-13 23:35 & 30.54 & 0.402 & $<98.7$ & 32.9 \\
NOEMA&2022-03-24 22:14&41.48&  86 &    3914&  30\\
NOEMA&2022-03-24 22:14&41.48& 102 &    3609&  34\\
uGMRT & 2022-03-25 00:08 & 41.56 & 1.264 & $<47.7$ & 15.9 \\
NOEMA&2022-03-25 00:45&41.58& 136 &    3045&  41\\
NOEMA&2022-03-25 00:45&41.58& 152 &    2750&  51\\
uGMRT & 2022-03-25 23:27 & 42.53 & 0.750 & $<67.8$ & 22.6 \\
VLA  &2022-03-31 04:08&47.73& 31.4&    2130&  30\\
VLA  &2022-03-31 04:08&47.73& 33.5&    2260&  30\\
VLA  &2022-03-31 04:08&47.73& 35.6&    2350&  40\\
VLA  &2022-03-31 04:08&47.73& 37.5&    2360&  40\\
VLA  &2022-03-31 04:13&47.73&  8.5&     270&  12\\
VLA  &2022-03-31 04:13&47.73&  9.5&     336&  11\\
VLA  &2022-03-31 04:13&47.73& 10.5&     385&  12\\
VLA  &2022-03-31 04:13&47.73& 11.5&     438&  14\\
VLA  &2022-03-31 04:23&47.74& 12.8&     583&  12\\
VLA  &2022-03-31 04:23&47.74& 14.3&     724&  12\\
VLA  &2022-03-31 04:23&47.74& 15.9&     801&  14\\
VLA  &2022-03-31 04:23&47.74& 17.4&     935&  15\\
\hline
%\end{tabular}
%\end{table*}
\end{longtable}
\end{center}

\end{supplement}

%\bibliographystyleNew{naturemag}
%\bibliographyNew{references, gcns}

\end{document}